\documentclass[10pt,a4paper]{article}
\usepackage[left=1in,right=1in,top=1in,bottom=1in]{geometry}
\usepackage{amsmath}
\usepackage{hyperref}
\usepackage{float}
\usepackage{graphicx} % For including images

\newcommand{\Title}[1]{\Large \textbf{#1}}
\newcommand{\Author}[1]{\large #1}
\newcommand{\Affiliation}[1]{\textit{#1}}
\newcommand{\Email}[1]{\href{mailto:#1}{#1}}
\newcommand{\Date}[1]{\small #1}

\begin{document}

% Title Section
\begin{center}
   \Title{Monte Carlo Study of TeV-Scale String Resonances in
Photon-Jet Scattering} \\
   \vspace{0.5em}
   \Author{Kyle Drury$^{1}$}  \\
   \vspace{1em}
   \Affiliation{$^1$Department of Physics and Astronomy, McMaster University, Hamilton, Ontario, L8S 4L8, Canada} \\
   \vspace{1em}
   \Email{Email: druryk5@mcmaster.ca} \\
   \vspace{1em}
   \Date{August 15, 2023}
\end{center}

\begin{abstract}
STRINGS is a Monte Carlo (MC) event generator for simulating the production and decay of first and second string resonances in proton-proton collisions \cite{vakilipourtakalou2018montecarloeventgenerator}. STRINGS can also interface with other programs such as Pythia \cite{Sj_strand_2015, Sj_strand_2006} using the Les Houches Accord \cite{boos2001genericuserprocessinterface, ALWALL2007300} to produce more accurate data. In this paper, we validate STRINGS for the simulation of 2-parton $\rightarrow$ $\gamma$-parton scattering events by comparing to previous literature \cite{lyons2021, Anchordoqui_2014, Anchordoqui_2008}. After validation, we produce MC samples of resonances using $M_s = \{ 5.0, 5.5, 6.0, 6.5, 7.0 \}$ TeV at $\sqrt{s} = \{13, 13.6\}$ TeV with STRINGS and Pythia, and analyze the kinematic data. To accurately reproduce previous results close to resonance, it is necessary to introduce a scaling factor of $\approx 0.53$. With this correction, the resonance structure is as expected.
\end{abstract}

\section{Introduction}

The Standard Model (SM) of particle physics is the most successful theory explaining observed phenomena in the universe. However, it fails to satisfy outstanding issues such as the Hierarchy problem \cite{Arkani_Hamed_1998}. String theory is a proposed framework that grapples with these issues by describing the fundamental particles of the SM in terms of one-dimensional strings. These strings may be open or closed, and interact with each other in ten-dimensional spacetime.

If the six unperceived spatial dimensions are sufficiently large, the scale of the interactions between strings (the string scale) $M_{s}$ should be on the order of a few TeV \cite{Antoniadis_1998}. Using the Large Hadron Collider (LHC) at CERN to study string resonances generated by proton-proton ($pp$) collisions, we can determine if these interactions are being driven by theories predicted by the SM or low-scale string theory. This paper focuses on parton-photon scattering processes $gg \rightarrow g\gamma$ and $gq \rightarrow q\gamma$. These interactions manifest at the LHC as $\gamma$ + jet.

In Section 2, the D-brane model and the Hierarchy Problem are reviewed; Section 3 discusses the kinematics of $pp$ collisions simulated by STRINGS, Pythia, and those detected by ATLAS at the LHC; in Section 4 we validate STRINGS production of the interactions mentioned above; data collection and analysis is then carried out in Section 5. We discuss our findings in Section 6.

\section{Review of String Theory Phenomenology}

\subsection{The Hierarchy Problem}

Despite the successes of the SM, there is a nagging problem that it cannot come to terms with; there appear to be two different fundamental energy scales of nature. The electroweak scale (EWS) $m_{EW} \sim 1$  TeV, and the Planck scale $M_{Pl} \sim 10^{15}$ TeV. The Planck scale is the energy at which the gravitational interaction's relative strength approaches that of the other three fundamental forces \cite{Arkani_Hamed_1998}. Although physicists have proposed new ideas to reconcile this issue, one possibility is that our current understanding of gravity is incomplete. This idea is bolstered when one considers the fact that the EWS has been experimentally confirmed as a fundamental constant, and accurately measured down to distances of $\sim m_{EW}^{-1}$. In contrast, gravity has been accurately probed only on the order of centimeters. Therefore, we may formulate a new theory altering the fundamental strength of gravity.

\subsection{String Theory}

The D-brane model, which underpins our current understanding of low-scale string theory, posits that fermions are manifestations of open strings with their ends attached to stacks of D$p$-branes ($p$-dimensional D-brane; a brane is a multidimensional object), while bosons stretch between individual D$p$-branes in the same stack.

\begin{figure}[H]
   \centering
   \includegraphics[width=0.55\textwidth]{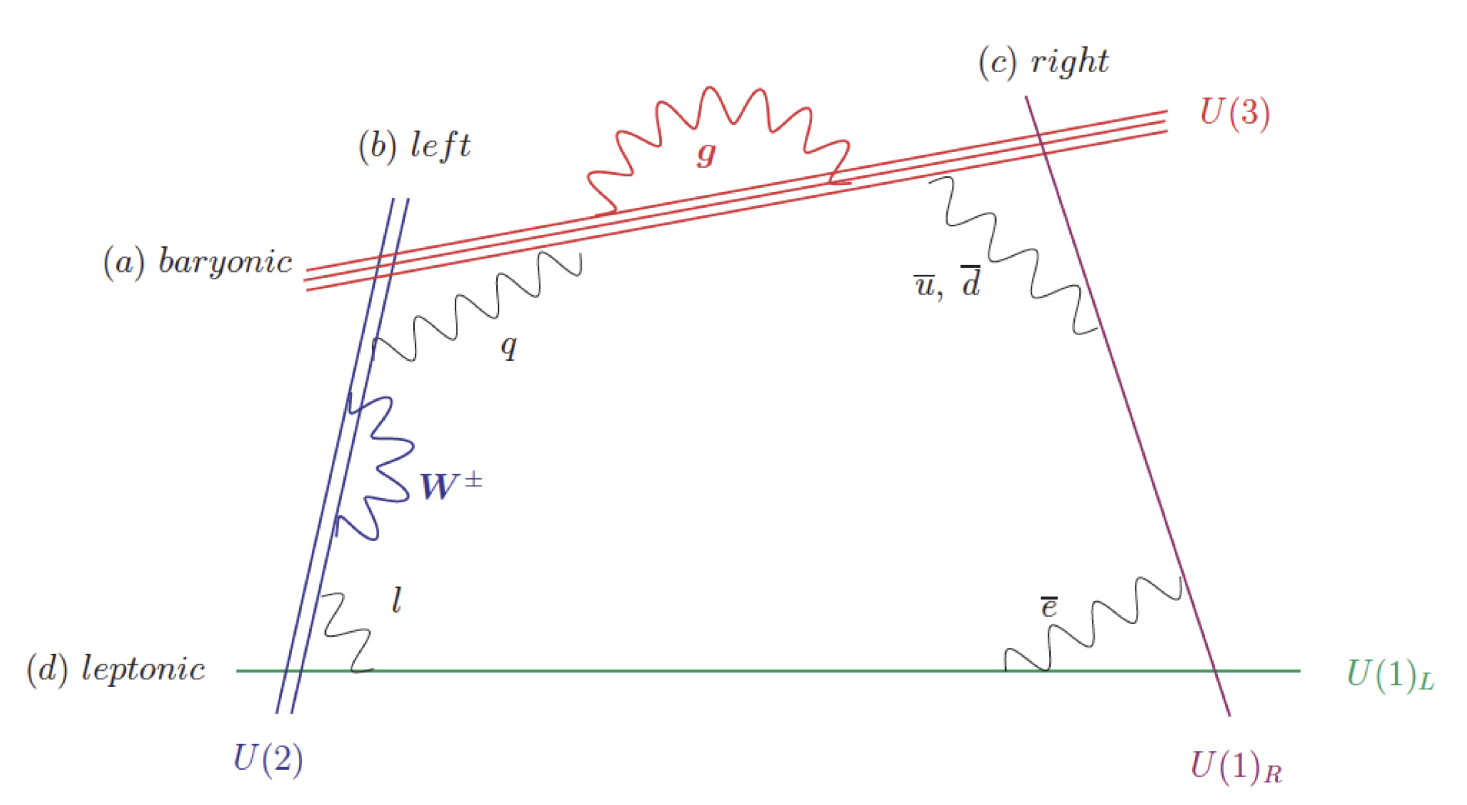}
   \caption{Bosons stretch between the layers of stacks of D-branes. Fermions have endpoints that are attached to different stacks \cite{L_st_2009}.}
   \label{fig:1}
\end{figure}

Since the 6 extra spatial dimensions predicted by string theory cannot be detected, they must be compactified. The D-brane model states that the undiscovered boson that mediates gravity, the so-called graviton, is a manifestation of a closed string that can propagate through all 9 spatial dimensions, completely untethered. Meanwhile, the other bosons are open strings that stretch between D-branes, restricted in their movement by the Dirichlet boundary condition. Therefore gravity might seem fundamentally weaker than the other forces due to its effects 'leaking' into the compactified spatial dimensions unseen on the macroscopic scale.

\begin{figure}[H]
   \centering
   \includegraphics[width=.7\textwidth]{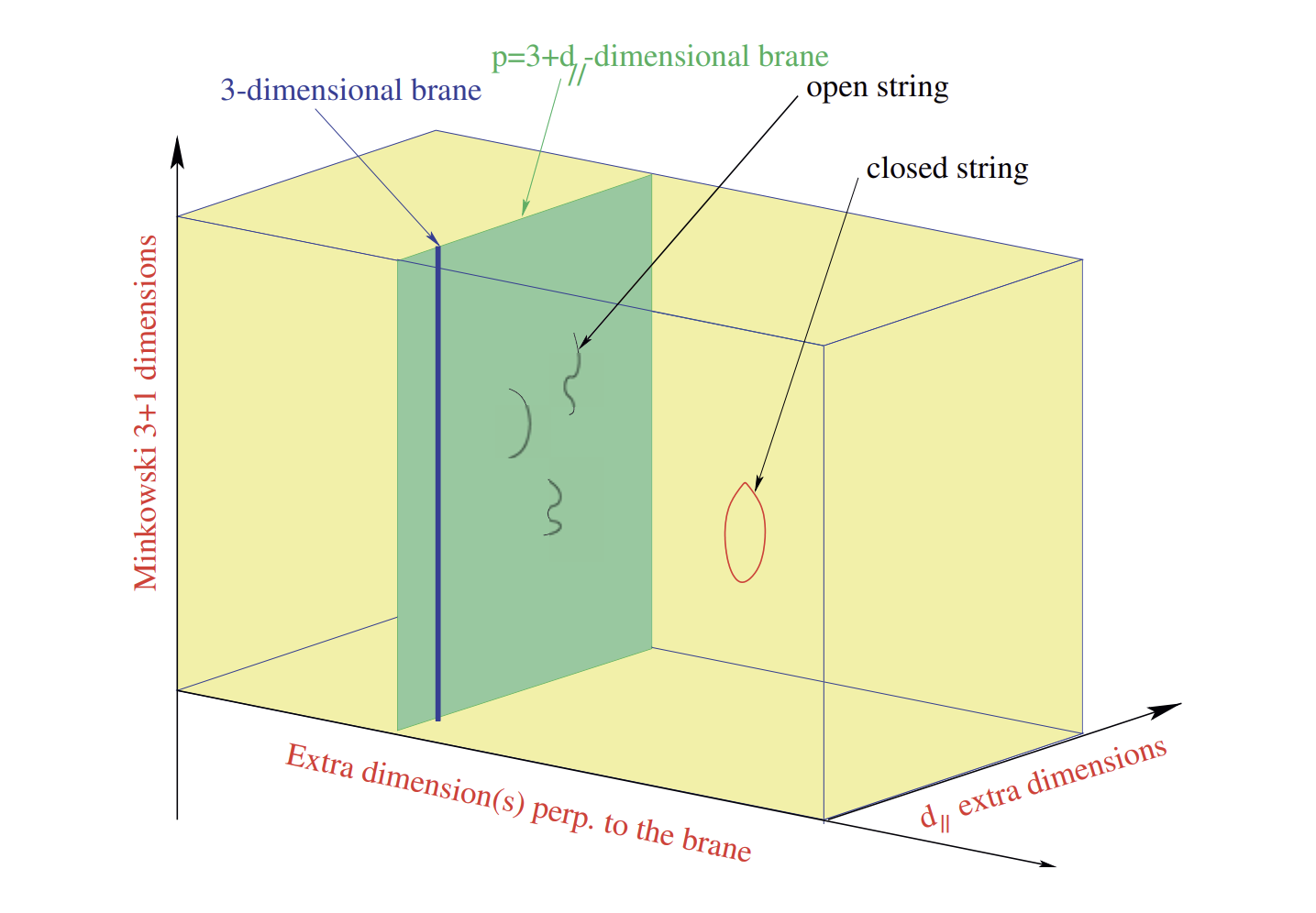}
   \caption{D-brane model diagram with a closed-string graviton propagating in the $d_{\perp}$ direction \cite{L_st_2009}.}
   \label{fig:2}
\end{figure}

\subsection{Extra Dimensions and Compactification}

Since the 4D Planck scale can be derived from physical constants, we can use it in calculations. We imagine the simplest possible compactification of six extra dimensions where each of them has a radius $R$, and the volume of the extra dimensions is given by

\begin{equation}
	V_{6}=(2 \pi R)^6 \tag{2.1} \label{eq:2.1}
\end{equation}

\noindent and the effective 4D Planck scale is related to the fundamental 10D Planck scale by

\begin{equation}
   M_{Pl4}^2 \sim M_{Pl10}^8 R^6 , \tag{2.2} \label{eq:2.2}
\end{equation}

\noindent We set $M_{Pl4}$ to the calculated $10^{15}$ TeV and the fundamental $M_{Pl10}$ to 1 TeV to get rid of the Hierarchy problem. Solving for $R$ yields

\begin{equation}
	R \sim 10^5 \, \, \textup{TeV} . \tag{2.3} \label{eq:2.3}
\end{equation}

The 4D Planck scale can be represented in terms of the volume $V_6$, the string coupling $g_s$, and the string scale $M_s$, the energy scale where stringy effects are prevalent:

\begin{equation}
	M_{Pl10}^2 = \frac{8}{g_s^2} M_s^8 R^6 . \tag{2.4} \label{eq:2.4}
\end{equation}

\noindent Inserting $R \sim 10^5$ TeV yields

\begin{equation}
	M_s \sim g_s^{1/4} \, \, \textup{TeV} . \tag{2.5} \label{eq:2.5}
\end{equation}

\noindent The string coupling is described in \cite{Anchordoqui_2014} as

\begin{equation}
	g_s = \sqrt{4 \pi \alpha_s} , \tag{2.6} \label{eq:2.6}
\end{equation}

\noindent where $\alpha_s$ denotes the gauge coupling:

\begin{equation}
\frac{1}{\alpha_i (M)} = \frac{1}{\alpha_i (M_Z)} - \frac{b_i}{2 \pi} \ln \left[ \frac{M}{M_Z} \right] ; \,\, i = 2,3,Y \qquad \textup{with} \,\,\, b_2 = -19/6, \, b_3 = -7, \, b_Y = 41/6 , \tag{2.7} \label{eq:2.7}
\end{equation}

\noindent where $M_Z = 92.1 \, \textup{GeV}$ is the weak scale or the mass of the weak force-mediating $Z$ boson. We use the calculated values of the couplings at the $M_Z$ pole; $\alpha_3(M_Z) = 0.118 \pm 0.003$, $\alpha_2(M_Z) = 0.0388$, $\alpha_y(M_Z) = 0.01014$ \cite{beringer2012}. These values may be referred to as "running couplings" as they are functions of the energy scale $M$. In this study, the coupling $\alpha_3 = \alpha_s$ is used, which represents the QCD running coupling \cite{patrignani2016}. This makes $g_s \sim \, 1$, so it follows that $M_s$ is on the order of a few TeV, an energy level attainable at the LHC.

\subsection{Scattering Amplitudes and Cross-sections}

This publication by Anchordoqui et al. focused on scattering processes that result in a photon and a jet (a jet results from the creation of a single color-charged particle emerging from a scattering event), of which there are two \cite{Anchordoqui_2014}:

\begin{equation}
   |\mathcal{M}(gg \xrightarrow{} g \gamma)|^2 = \frac{5 g^4 Q^2}{3 M_{s}^4} \left[ \frac{M_{s}^8}{(\hat{s}-M_{s}^2)^{2}+(M_{s} \Gamma_{g*}^{J=0})^2} + \frac{\hat{t}^4 + \hat{u}^4}{(\hat{s} - M_{s}^2)^2 + (M_{s}\Gamma_{g^*}^{J=2})^2} \right] , \tag{2.8} \label{eq:2.8}
\end{equation}

\begin{equation}
    |\mathcal{M}(gq \xrightarrow{} q \gamma)|^2 = \frac{-g^4 Q^2}{3 M_{s}^2} \left[ \frac{\hat{u} M_{s}^4 }{(\hat{s}-M_{s})^{2}+(M_{s} \Gamma_{q*}^{J=\frac{1}{2}})^2} + \frac{\hat{u}^{3}}{(\hat{s}-M_{s}^2)^2 + (M_{s} \Gamma_{q*}^{J=\frac{3}{2}})^2}\right] . \tag{2.9} \label{eq:2.9}
\end{equation}

\noindent Each $\Gamma$ represents a resonance width, also given in \cite{Anchordoqui_2014}:

\begin{equation}
	\Gamma_{g^{*}}^{J=0} = \frac{g^2}{4 \pi} M_{s} \frac{3}{4}, \qquad \Gamma_{g^{*}}^{J=2} = \frac{g^2}{4\pi} M_{s} \frac{9}{20}, \qquad \Gamma_{q^{*}}^{J=\frac{1}{2}} = \frac{g^2}{4 \pi} M_{s} \frac{3}{8}, \qquad \Gamma_{q^{*}}^{J=\frac{3}{2}} = \frac{g^2}{4 \pi} M_{s} \frac{3}{16}, \tag{2.10} \label{eq:2.10}
\end{equation}

\noindent where $Q^2$ is given by \cite{Anchordoqui_2008} as

\begin{equation}
	Q^2 = \frac{1}{6} \kappa^2 \cos^2 \theta_W \approx 2.55 \times 10^{-3}, \tag{2.11} \label{eq:2.11}
\end{equation}

\noindent with $\kappa^2 = 0.02$ as the so-called mixing factor and $\theta_W$ the Weinberg angle, related to the Electroweak force.

From (\ref{eq:2.8}) and (\ref{eq:2.9}), we can see that when $\hat{s} = M_s$ the scattering amplitudes reach a local maximum. This is known as a resonance and occurs when the the center-of-mass energy of the parton collision $\hat{s}$ approaches the string scale $M_s$. The string scale can be probed from LHC collision data by looking for such resonances in invariant mass distributions of scattering events in the ATLAS detector.

\section{STRINGS Validation}

\subsection{Differential Cross-sections}

To validate $\gamma + \textup{jet}$ resonances simulated by STRINGS, we refer to the black distribution using parton distribution function set CTEQ6L1 \cite{Pumplin_2002} from \cite{Anchordoqui_2014}:

\begin{figure}[H]
   \centering
   \includegraphics[width=0.5\textwidth]{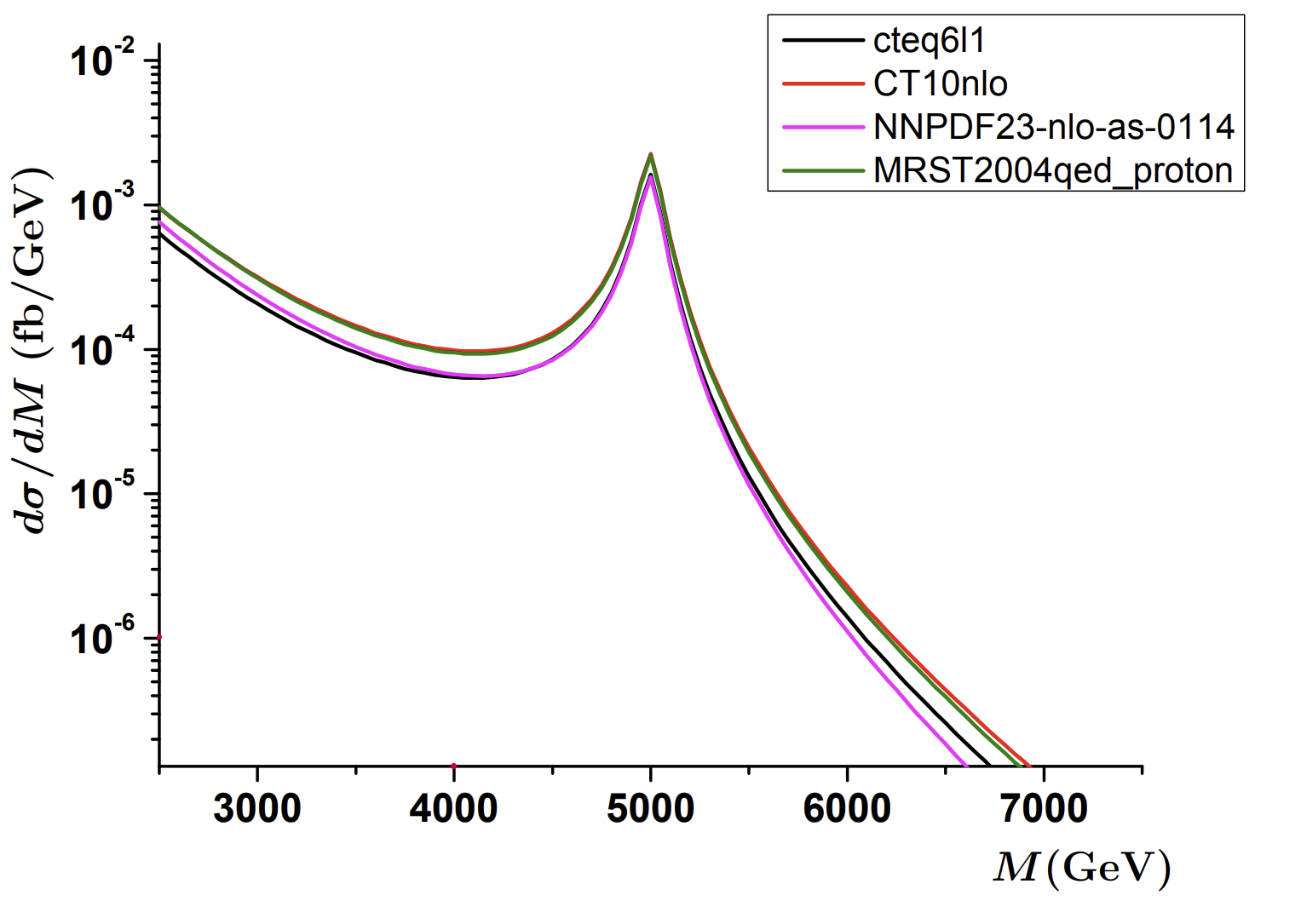}
   \caption{Differential cross-sections of $\gamma + \textup{jet}$ scattering events at $M_{s}$ = 5 TeV}
   \label{fig:3}
\end{figure}

\begin{figure}[H]
   \centering
   \includegraphics[width=0.49\textwidth]{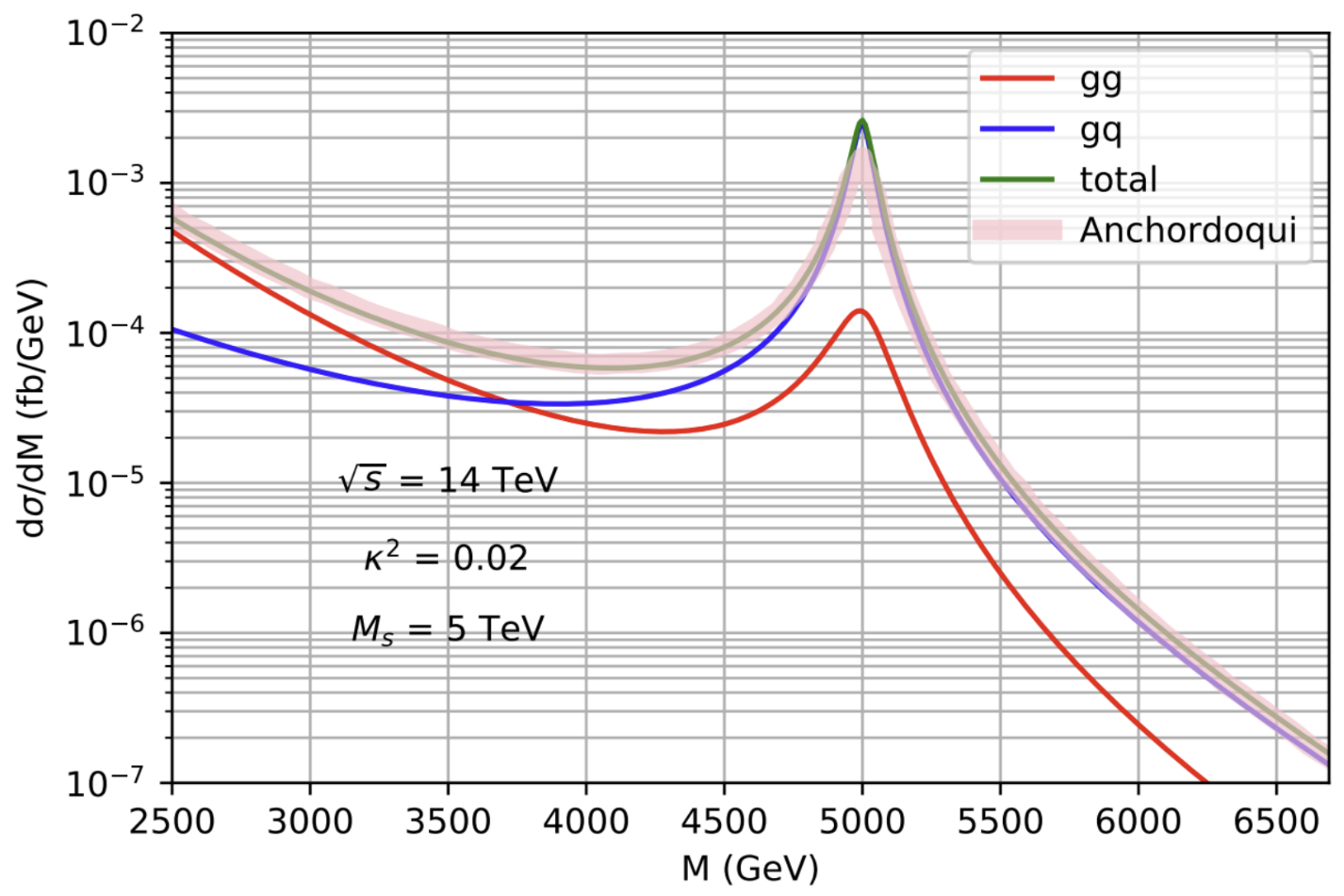}
   \includegraphics[width=0.49\textwidth]{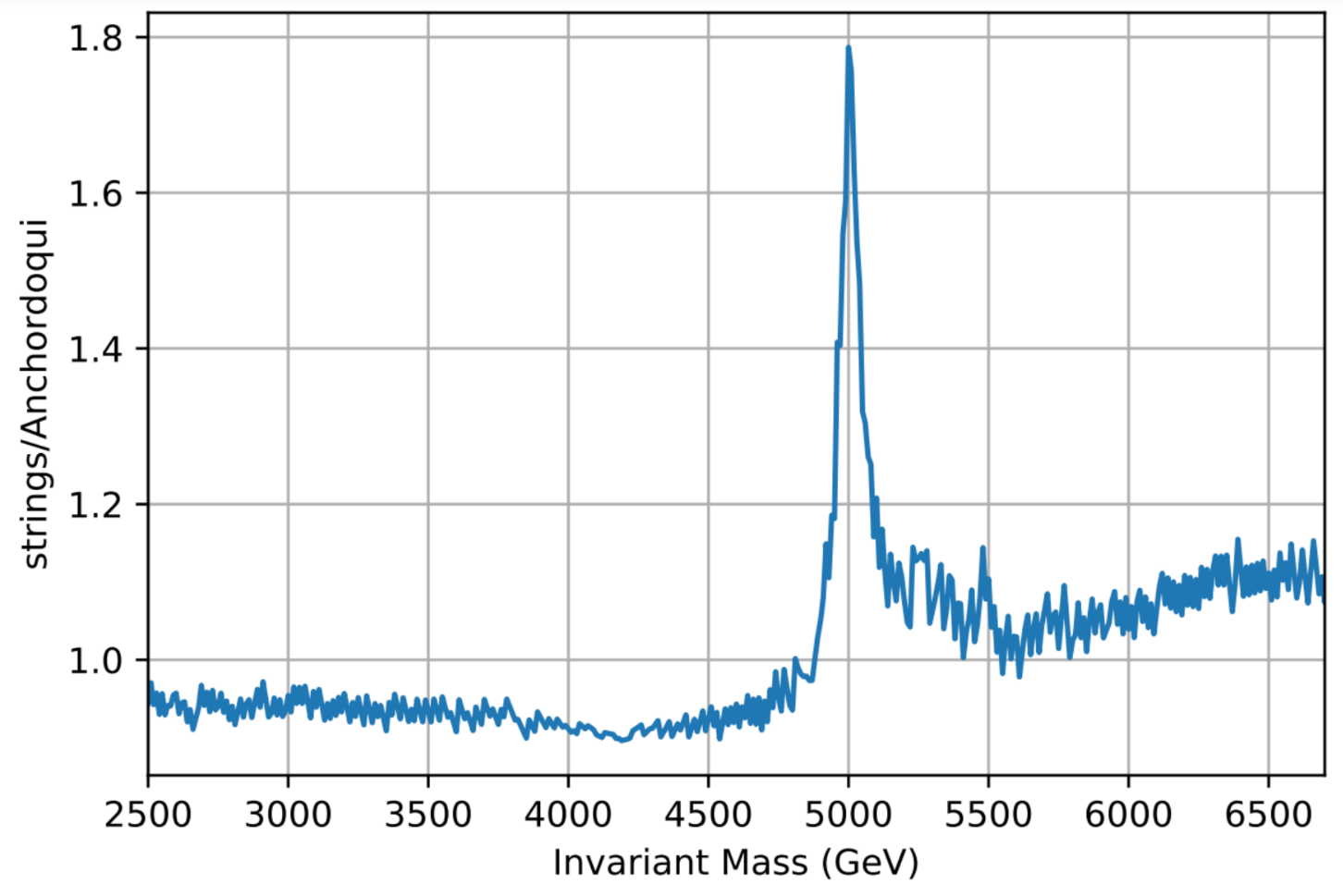}
   \caption{Differential cross-sections of single parton scattering events produced by \textbf{pre-bug-fix} STRINGS compared with \cite{Anchordoqui_2014}, $\alpha_{s}$ = running coupling constant. The right plot shows the relative error.}
   \label{fig:4}
\end{figure}

\begin{figure}
   \centering
   \includegraphics[width=0.49\textwidth]{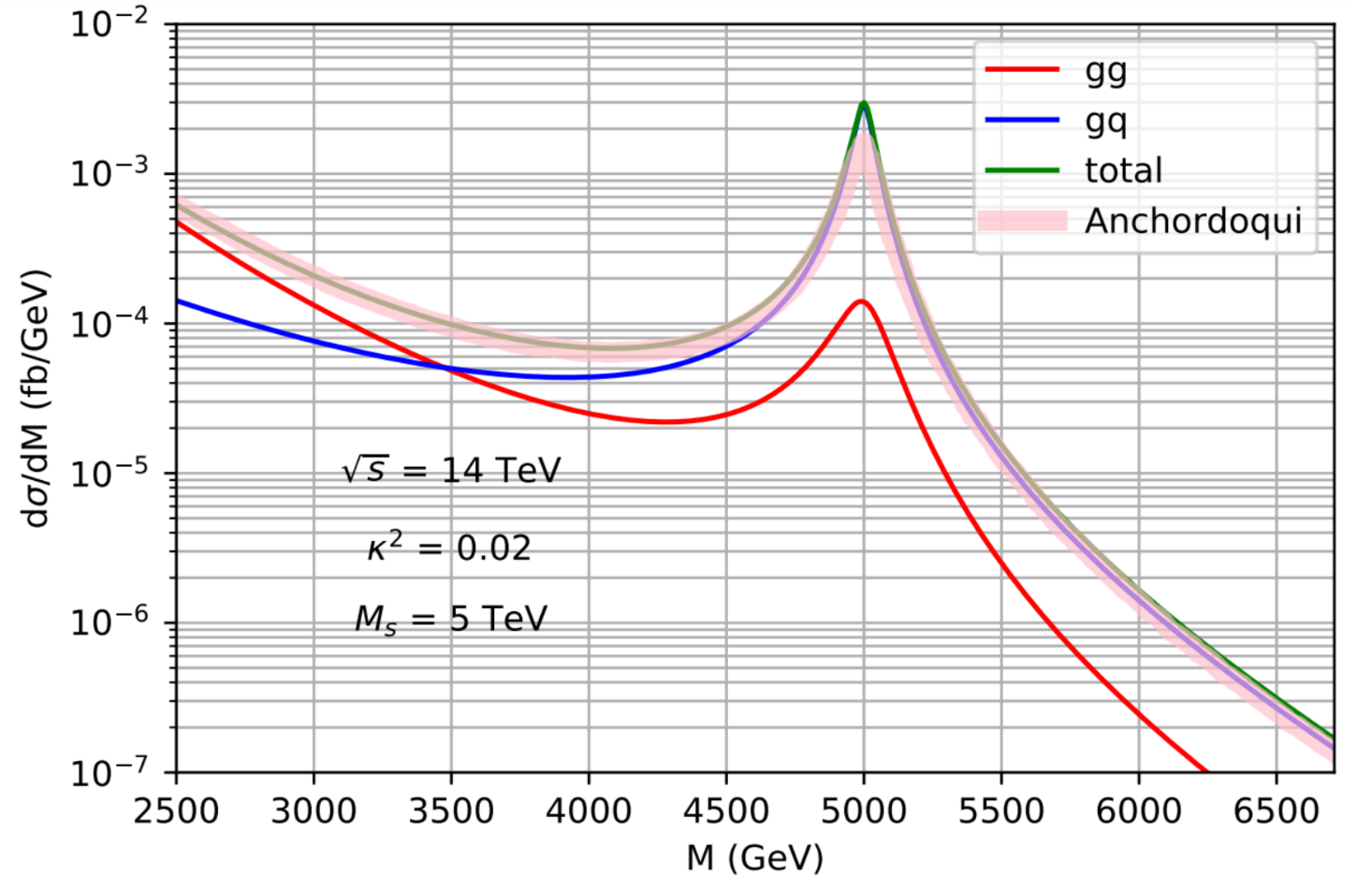}
   \includegraphics[width=0.49\textwidth]{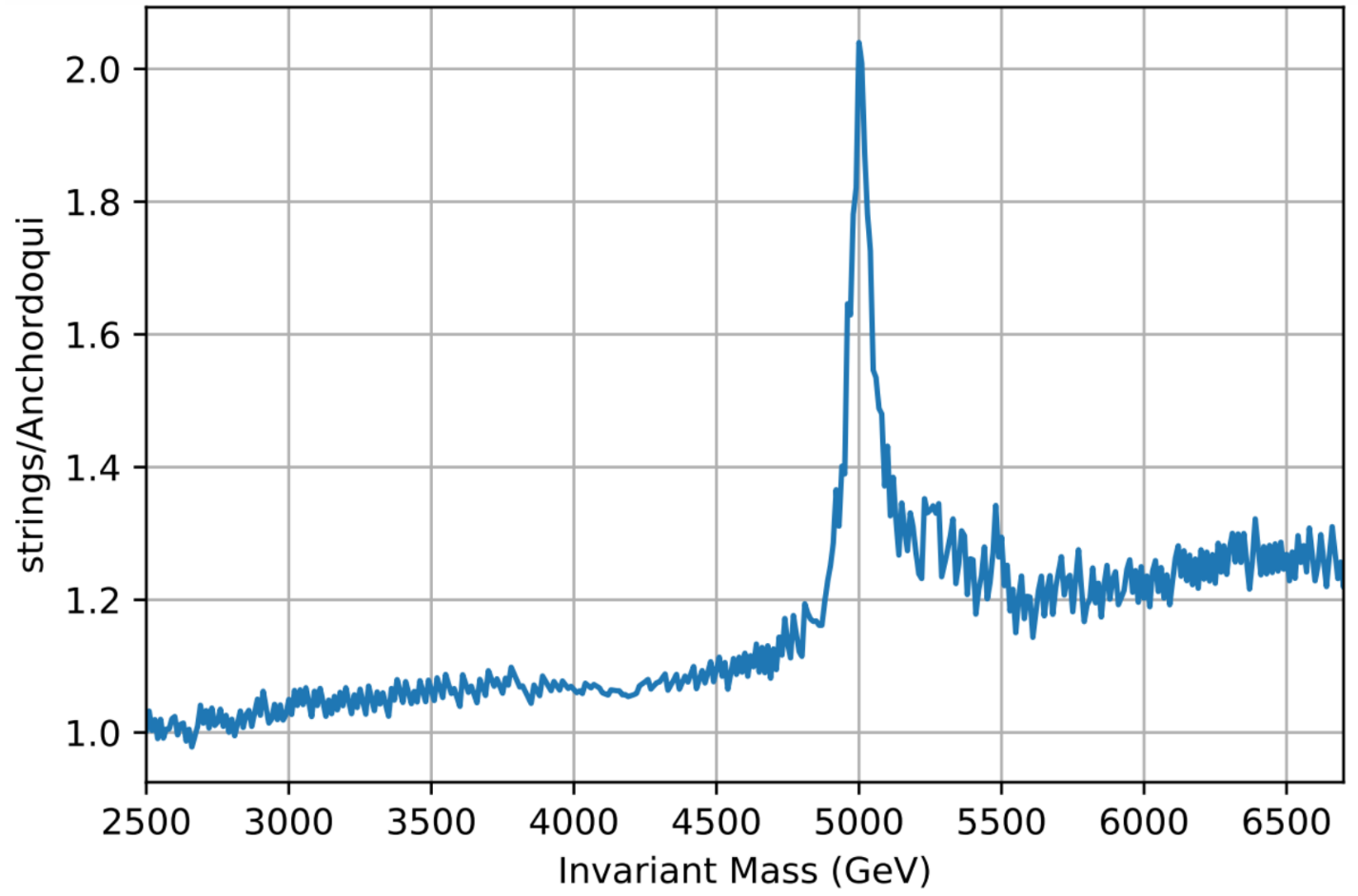}
   \caption{Differential cross-sections of single parton scattering events produced by \textbf{post-bug-fix} STRINGS compared with \cite{Anchordoqui_2014}, $\alpha_{s}$ = running coupling constant.}
   \label{fig:5}
\end{figure}

We can recreate the above plot by running STRINGS for $M = [2500, 6720]$ GeV at $M_{s}$ = 5000 GeV, $\sqrt{s} = 14$ TeV, for $gg$ and $gq$ processes that lead to $\gamma$ + jet. Several attempts were made using varying parameters. Partway through testing, a bug was located in the code that was excluding $t$ and $u$ channel contributions to the calculation of $\mathcal{M}^2$ (see Figures \ref{fig:4} and \ref{fig:5}).

It was also uncovered in the paper that rather using the resonance widths as defined in (\ref{eq:2.10}), \cite{Anchordoqui_2014} used concatenated widths:

\begin{align}
   \Gamma_{g^{*}}^{J=0} = 75 \frac{M_{s}}{\textup{TeV}} \textup{GeV} &&
   \Gamma_{g^{*}}^{J=2} = 45 \frac{M_{s}}{\textup{TeV}} \textup{GeV} &&
   \Gamma_{q^{*}}^{J=\frac{1}{2}}=37 \frac{M_{s}}{\textup{TeV}} \textup{GeV} &&
   \Gamma_{q^{*}}^{J=\frac{3}{2}}=19 \frac{M_{s}}{\textup{TeV}} \textup{GeV} \tag{3.1}
\end{align}

\begin{figure}[H]
   \centering
   \includegraphics[width=0.49\textwidth]{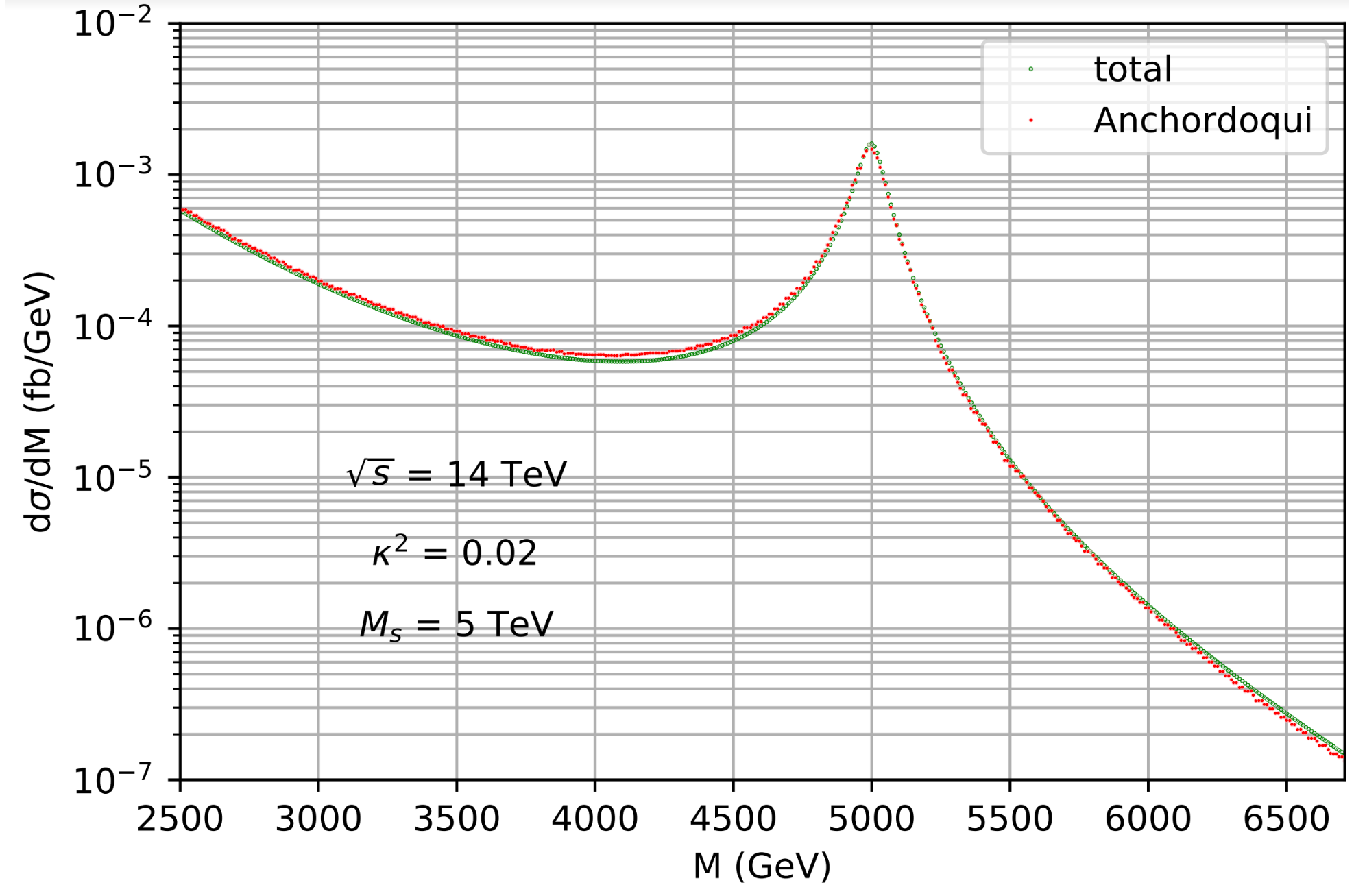}
   \includegraphics[width=0.49\textwidth]{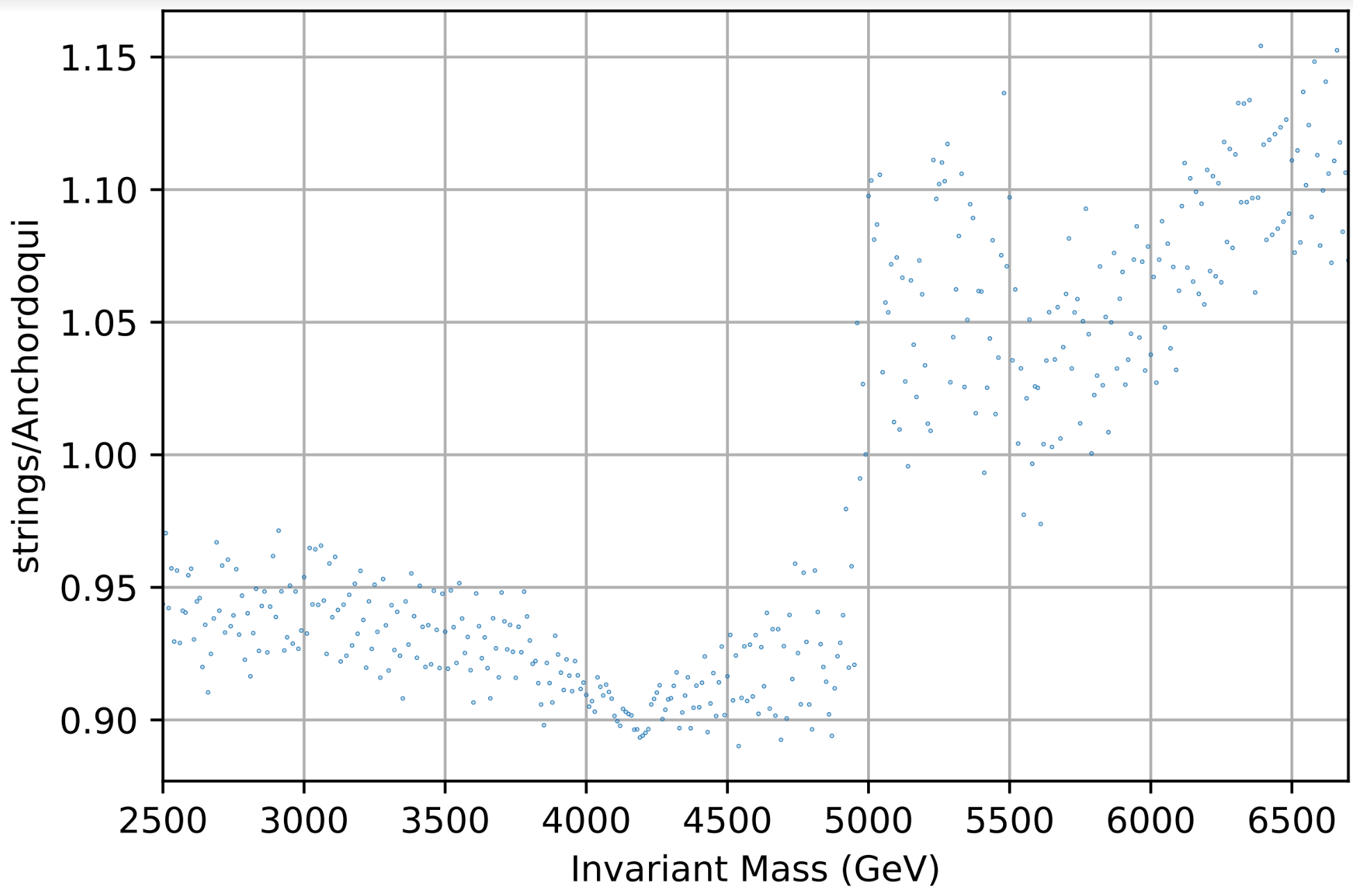}
   \caption{Differential cross-sections of single parton scattering events produced by \textbf{pre-bug-fix} STRINGS compared with \cite{Anchordoqui_2014}, $\alpha_{s}$ = running coupling constant, using concatenated widths.}
   \label{fig:6}
\end{figure}

\begin{figure}[H]
   \centering
   \includegraphics[width=0.49\textwidth]{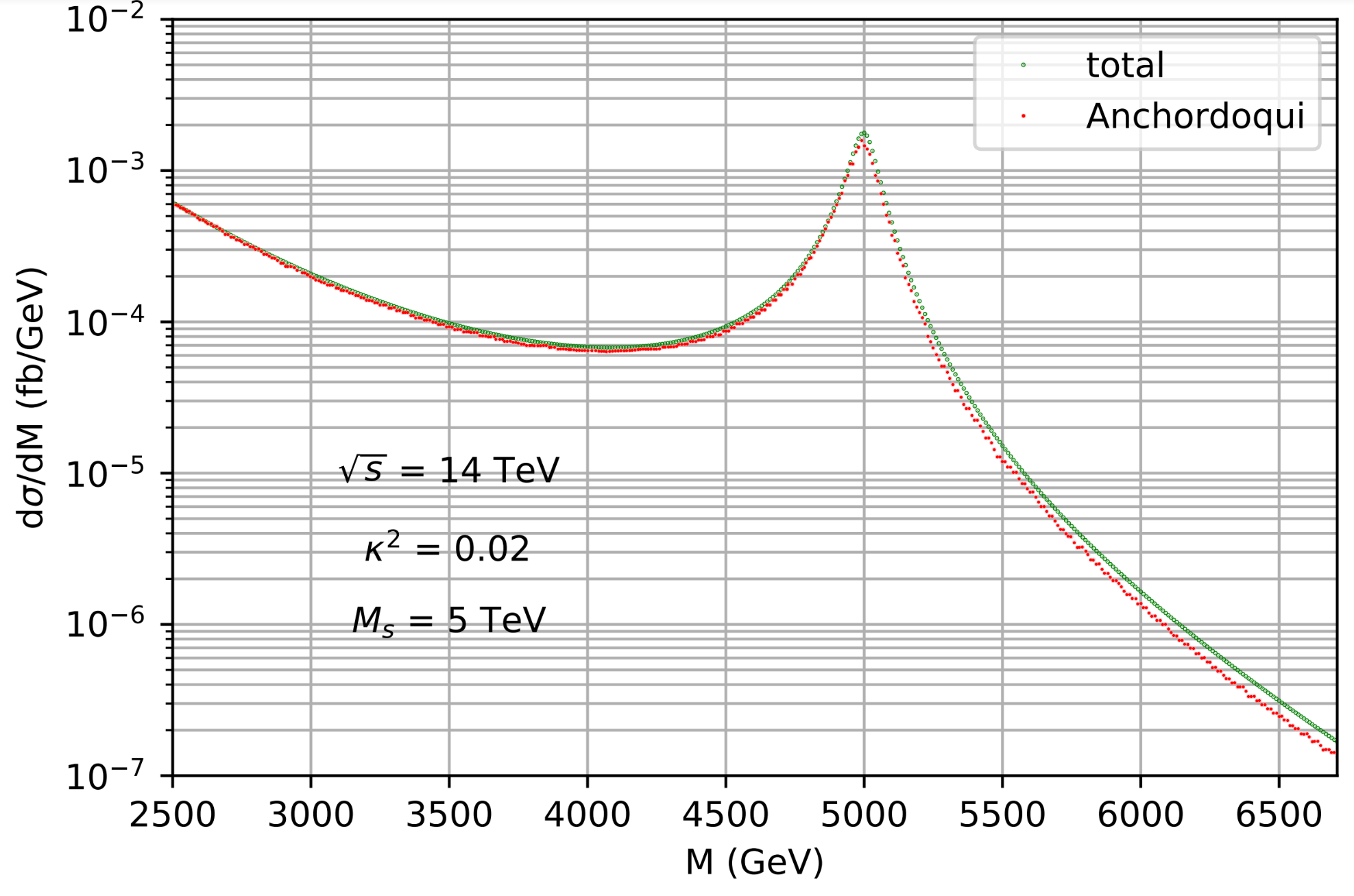}
   \includegraphics[width=0.49\textwidth]{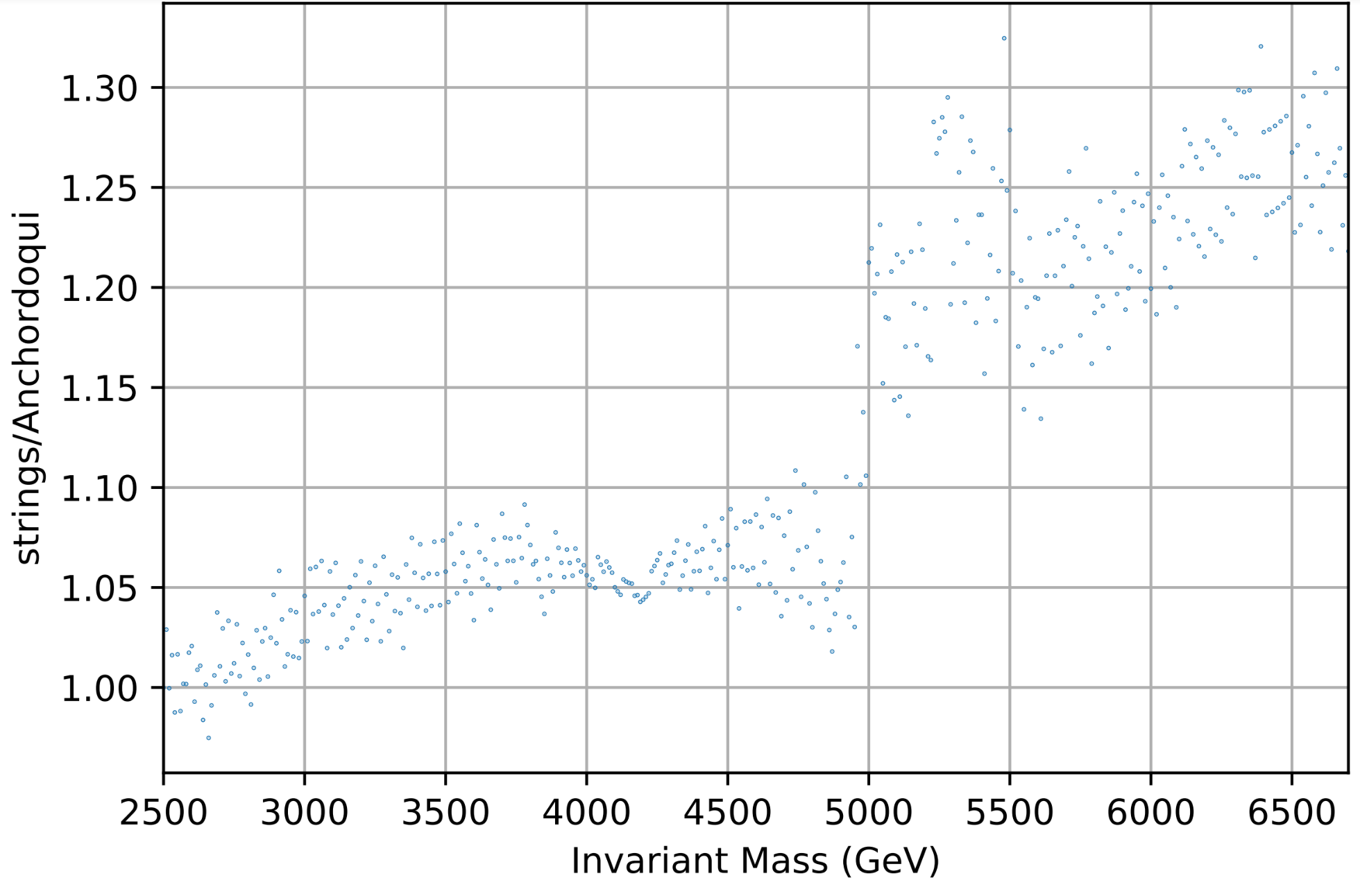}
   \caption{Differential cross-sections of single parton scattering events produced by \textbf{post-bug-fix} STRINGS compared with \cite{Anchordoqui_2014}, $\alpha_{s}$ = running coupling constant, using concatenated widths.}
   \label{fig:7}
\end{figure}

\noindent Using these widths; the plots in Figures \ref{fig:6} and \ref{fig:7} are produced.

By using small markers, we can observe the reasoning for the $\pm \sim$ 0.05 fluctuations on either side of the peak: our data is smooth, but the data collected from \cite{Anchordoqui_2014} (pink) moves in a step-like fashion, as a result of linear interpolation of the original points. We also observe a much closer peak agreement using the concatenated widths.

To see if a better agreement may be reached, we will attempt to recreate the curve using several different $\alpha_{s}$ coupling values, and then scaling by an appropriate factor to get the peaks to match.

The best results were achieved using a coupling of 0.1 and scaling the data by a factor of exactly 0.5278861221857577. On the left side of the peak, the error is about -15\%, and on the right, +15\%; an even spread. Going lower or above 0.1 results in the ratio plot getting 'tilted' to one side and moving up or down, depending if the coupling is increasing or decreasing. Using $\alpha_s=0.1$ also demands a scale factor closer to 1/2 the other attempted values of $\alpha_s$. Due to this scale factor, in STRINGS we divide $\mathcal{M}^2$ by $\sim 2$ to adhere as closely as possible to \cite{Anchordoqui_2014}.

\begin{figure}[H]
   \centering
   \includegraphics[width=0.32\textwidth]{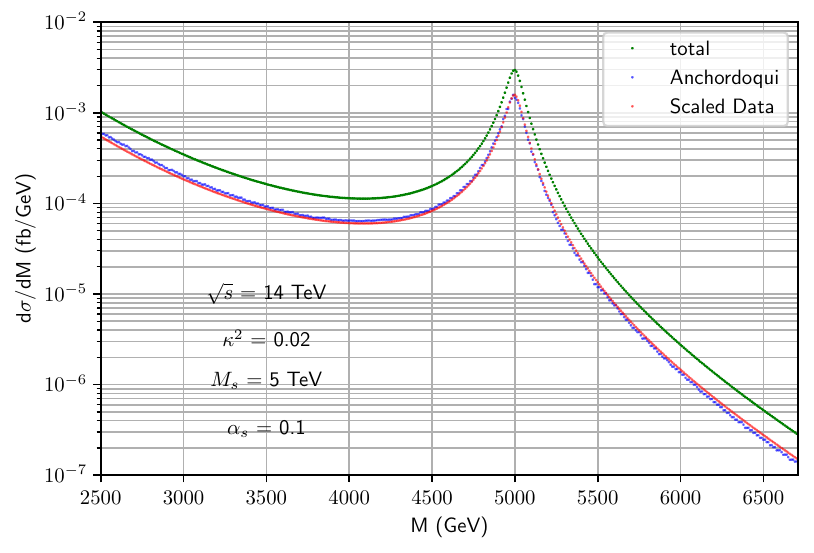}
   \includegraphics[width=0.32\textwidth]{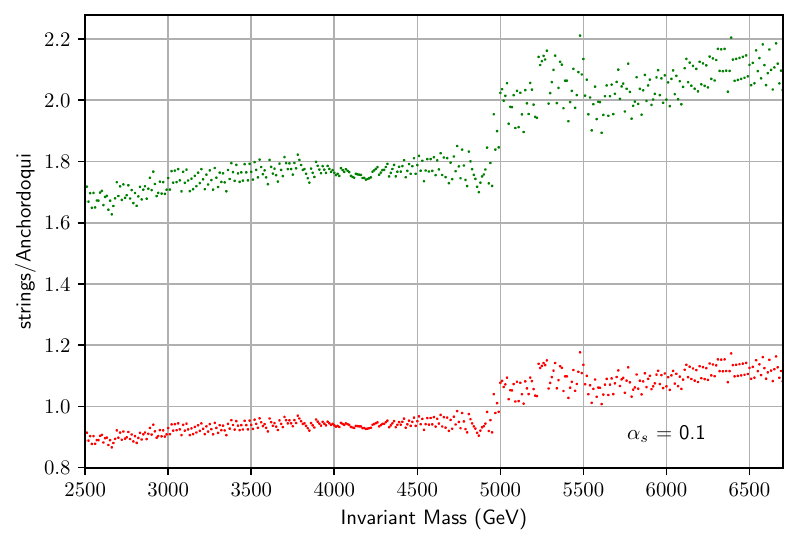}
   \includegraphics[width=0.32\textwidth]{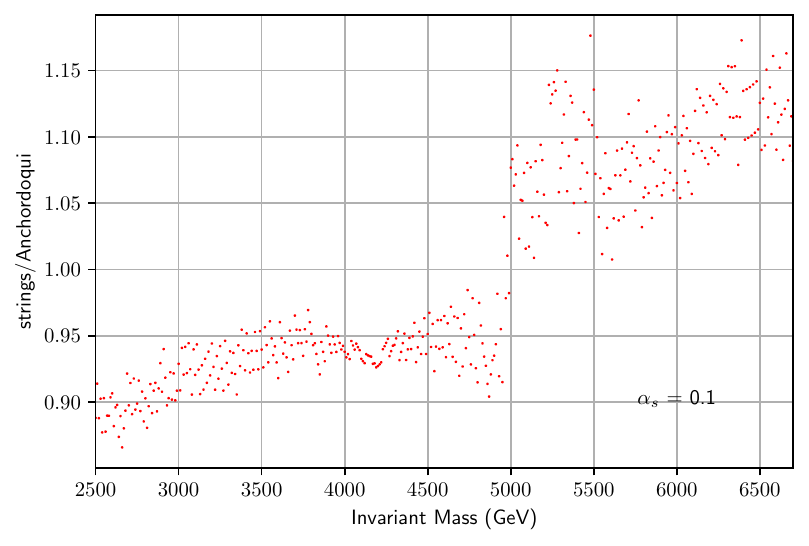}
   \caption{$\alpha_{s} = 0.1$}
   \label{fig:8}
\end{figure}

\subsection{Cross-sections}

We also aim to reproduce the following $\sigma$ vs $M_s$ plot from \cite{Anchordoqui_2008}:

\begin{figure}[H]
   \centering
   \includegraphics[width=0.5\textwidth]{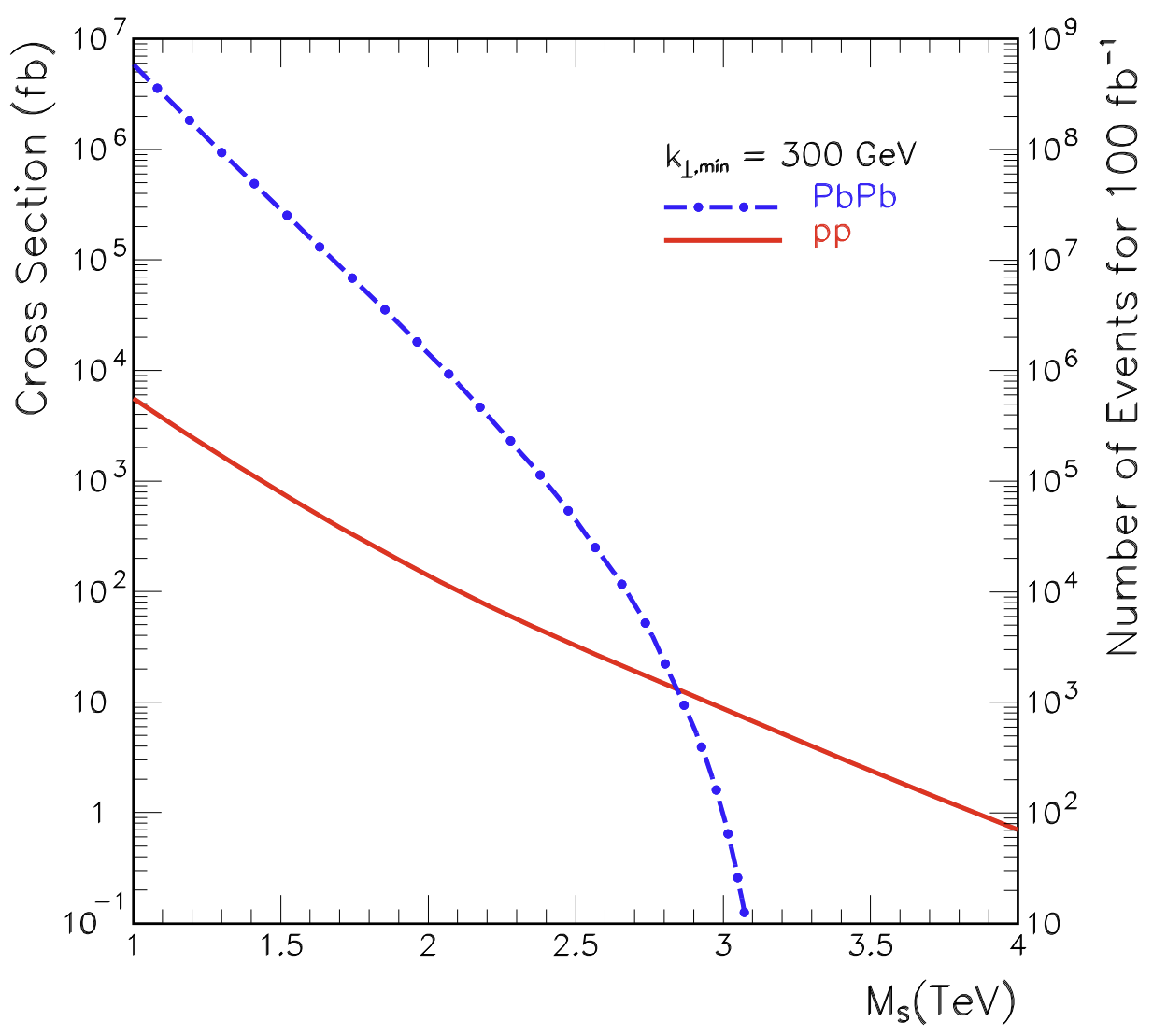}
   \caption{$\alpha_{s} = 0.100$}
   \label{fig:9}
\end{figure}

We can most accurately recreate this plot by integrating the differential cross-section over the invariant mass window $[M_{\text{cut}}, \sqrt{s}]$ where $\sqrt{s}$ = 14 TeV and each $M_{\text{cut}}$ is shown in Table \ref{tab:1}.

\begin{table}[H]
   \centering
   \begin{tabular}{c|c}
       String Scale $M_{s}$ [TeV] & $M_{\text{cut}}$ [GeV] \\
       \hline
       1 & NA \\
       1.5 & 1170 \\
       2 & 1590 \\
       2.5 & 2010 \\
       3 & 2450 \\
       3.5 & 2890 \\
       4 & 3350\\
   \end{tabular}
   \caption{String scale vs $M_{\text{cut}}$, which is taken to be the local minimum to the left of the resonance.}
   \label{tab:1}
\end{table}

\begin{figure}[H]
   \centering
   \includegraphics[width=0.49\textwidth]{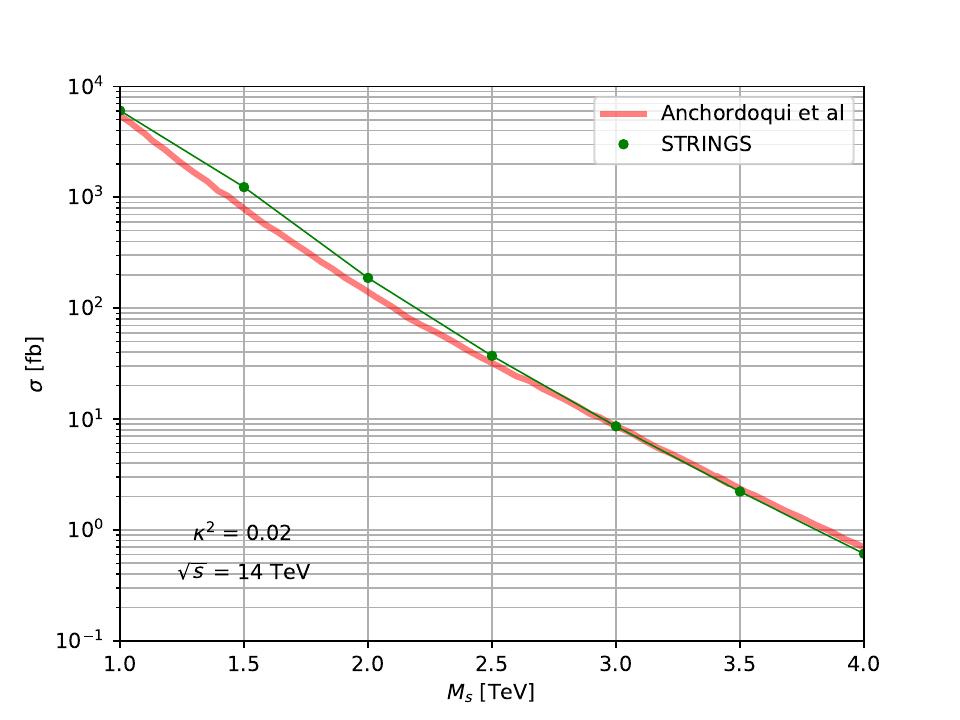}
   \includegraphics[width=0.49\textwidth]{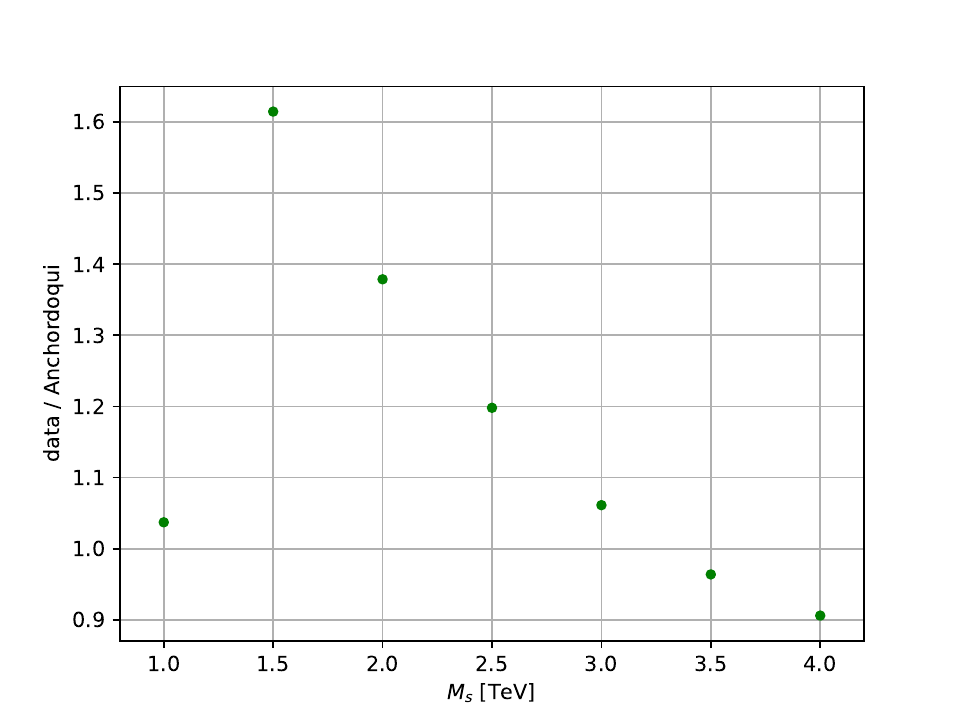}
   \caption{Total cross-section as a function of $M_{s}$; STRINGS vs. \cite{Anchordoqui_2008} (left) with ratio plot (right).}
   \label{fig:10}
\end{figure}

\section{Sample Generation}

\subsection{String Scale Selection}

On the order of a few TeV, it is important to ensure that our choice of string scales provides enough events for us to analyze. We begin by looking at the differential cross-sections for $M_{s}$ = [7, 9] TeV (Appendix 8.3.1). For $\sqrt{s}$ = 13 TeV, and another for $\sqrt{s}$ = 13.6 TeV, separate plots will be made. The region of integration is $M$ = [$M_{\text{cut}}$, $\sqrt{s}$], where $M_{\text{cut}}$ denotes the local minimum to the left of the resonance. The total number of events as a function of the string scale with the following formula:

\begin{align}
   N = 140 \sigma_{13} + 115 , \sigma_{13.6} , \tag{4.1}
\end{align}

\noindent where $\sigma_{i}$ represents the summed cross-section using $\sqrt{s} = i \,\, \textup{TeV}$.

\begin{figure}[H]
   \centering
   \includegraphics[width=0.32\textwidth]{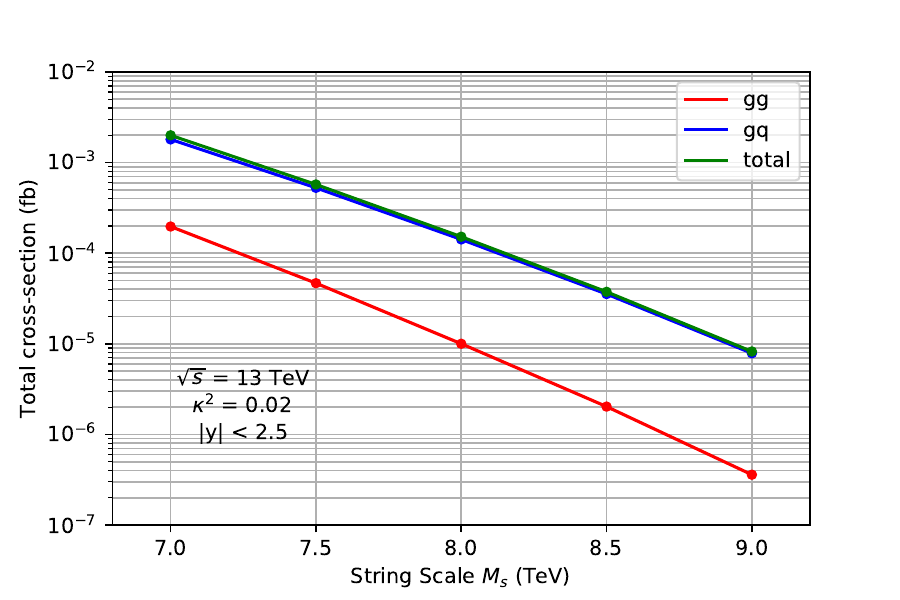}
   \includegraphics[width=0.32\textwidth]{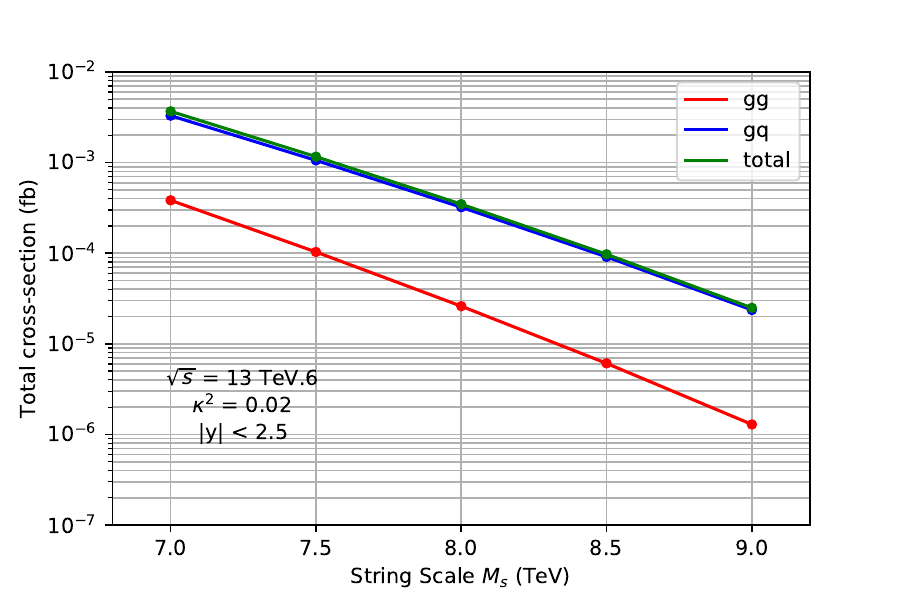}
   \includegraphics[width=0.32\textwidth]{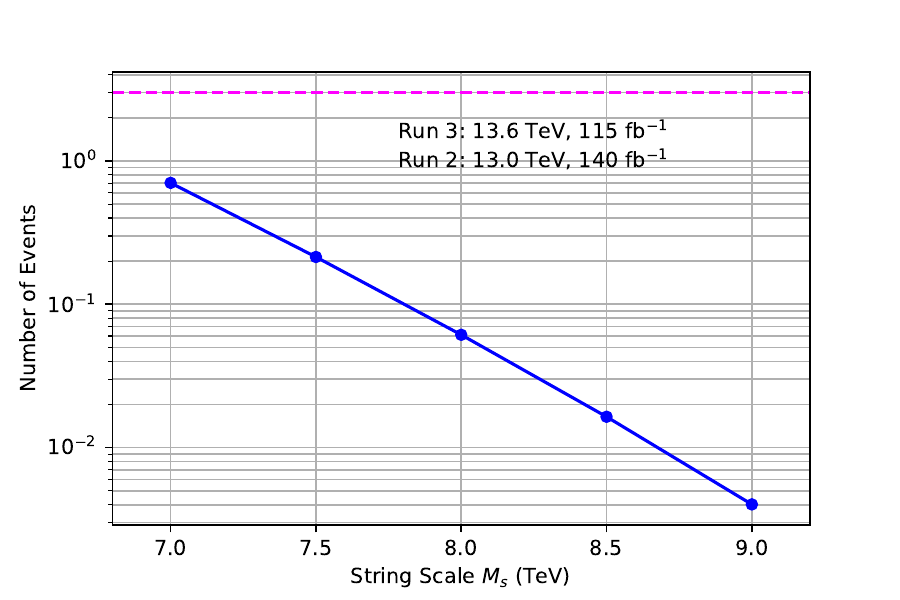}
   \caption{$\sigma$ vs $M_{s}$, $M$ = [$M_{\text{cut}}$, $\sqrt{s}$], along with the number of events as a function of $M_s$.}
   \label{fig:11}
\end{figure}

\noindent The curve in Figure \ref{fig:11} is far too low on the y-axis, so more events will need to be produced. Repeating the same process for $M_{s}$ = [5, 7] TeV yields the following plots:

\begin{figure}[H]
   \centering
   \includegraphics[width=0.32\textwidth]{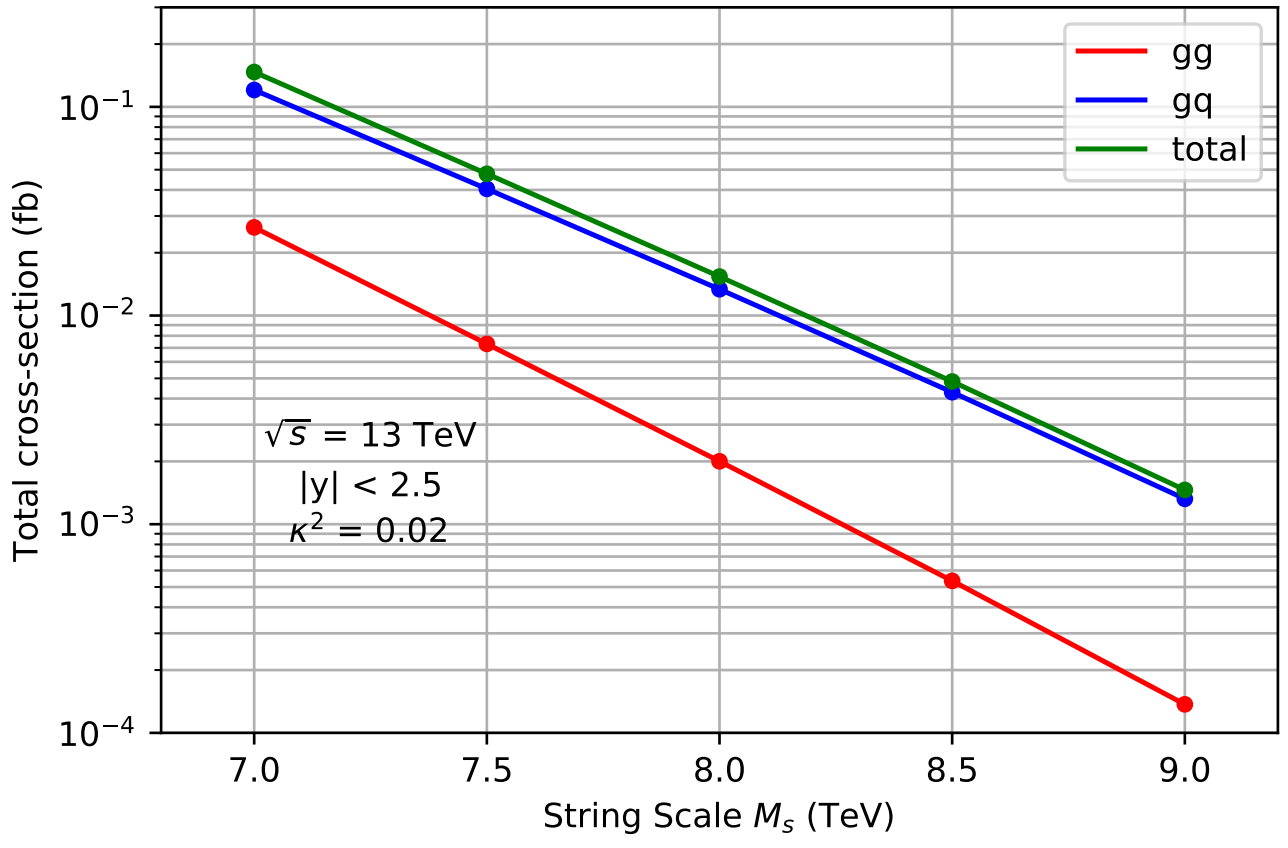}
   \includegraphics[width=0.32\textwidth]{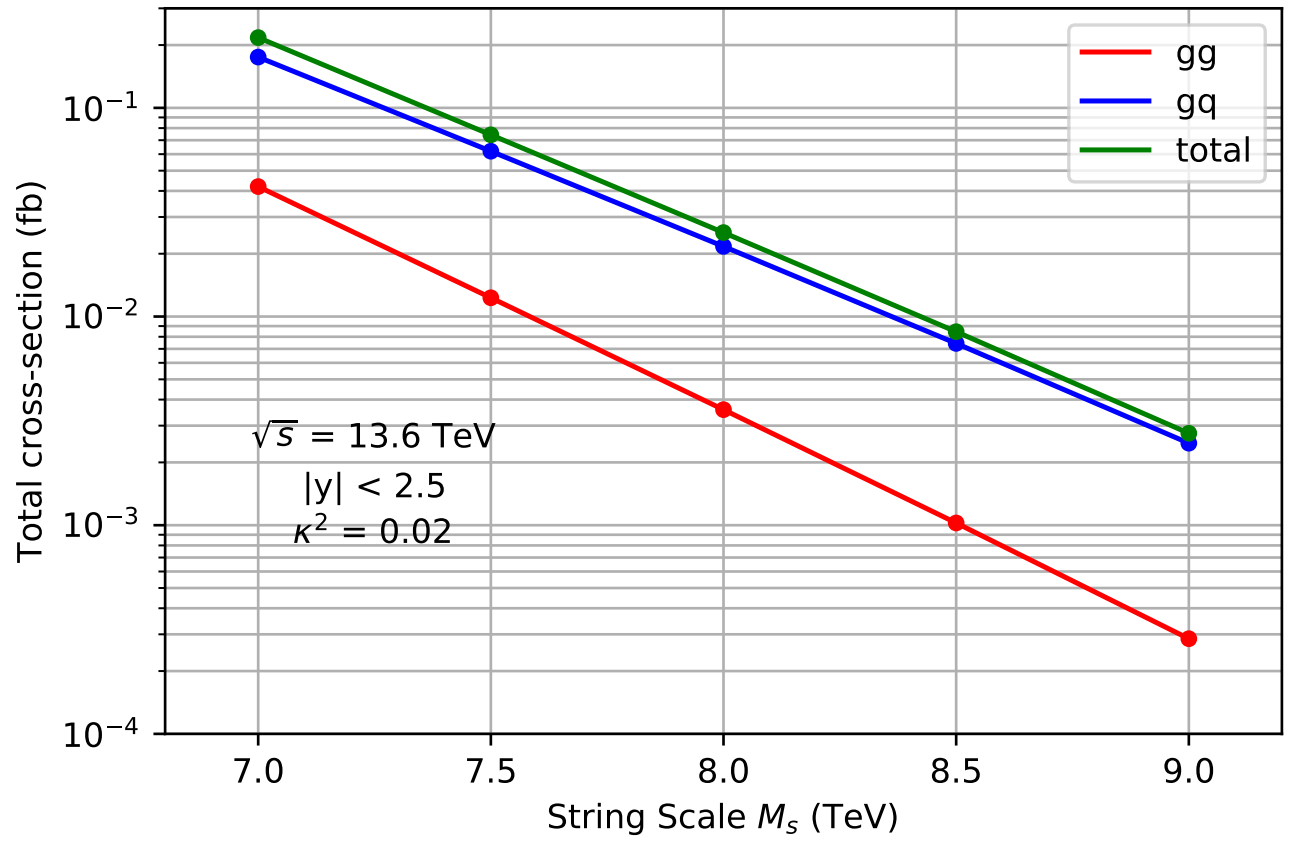}
   \includegraphics[width=.32\textwidth]{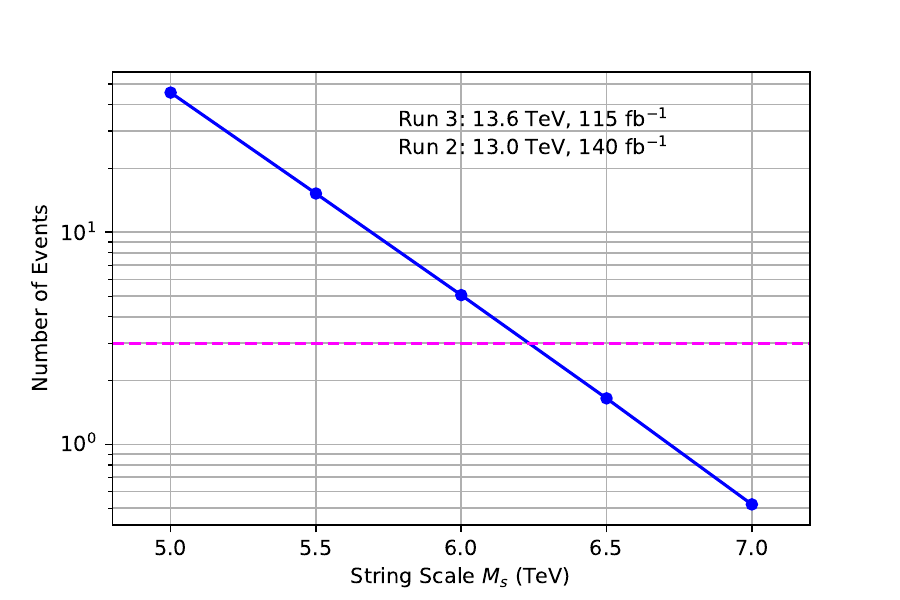}
   \caption{$\sigma$ vs $M_{s}$, $M$ = [$M_{\text{cut}}$, $\sqrt{s}$], along with the number of events as a function of $M_s$.}
   \label{fig:12}
\end{figure}

\noindent The $M_{s}$ values are optimal, as 3 data points are above the $N = 3$ line, which indicates that 95\% of events will be observed.

\subsection{Validity of Samples}

The events generated in these MC samples are either the $gg \rightarrow g\gamma$ subprocess or the $gq \rightarrow q\gamma$ subprocess. At different string scales and $\sqrt{s}$, these portions change as indicated in Tables \ref{tab:2} and \ref{tab:3}.

\begin{table}[H]
   \centering
   \begin{tabular}{c|c|c|c|c|c}
        & \textbf{$M_{s}$ = 5 TeV} & \textbf{$M_{s}$ = 5.5 TeV} & \textbf{$M_{s}$ = 6 TeV} & \textbf{$M_{s}$ = 6.5 TeV} & \textbf{$M_{s}$ = 7 TeV}\\
        \hline
       $gg \rightarrow g\gamma$ & 18.010\% & 15.310\% & 13.028\% & 11.737\% & 9.3811\% \\
       $gq \rightarrow q\gamma$ & 81.990\% & 84.680\% & 86.972\% & 88.263\% & 90.618\% \\
   \end{tabular}
   \caption{Event fractions for $\sqrt{s}$ = 13 TeV}
   \label{tab:2}
\end{table}

\begin{table}[H]
   \centering
   \begin{tabular}{c|c|c|c|c|c}
        & \textbf{$M_{s}$ = 5 TeV} & \textbf{$M_{s}$ = 5.5 TeV} & \textbf{$M_{s}$ = 6 TeV} & \textbf{$M_{s}$ = 6.5 TeV} & \textbf{$M_{s}$ = 7 TeV}\\
        \hline
       $gg \rightarrow g\gamma$ & 19.319\% & 16.564\% & 14.172\% & 12.128\% & 10.367\% \\
       $gq \rightarrow q\gamma$ & 80.681\% & 83.436\% & 85.828\% & 87.872\% & 89.633\% \\
   \end{tabular}
   \caption{Event fractions for $\sqrt{s}$ = 13.6 TeV}
   \label{tab:3}
\end{table}

Since two processes are being studied, both types of interactions will be generated by STRINGS. It is important to be confident that the number of $gg$ and $gq$ events being generated are proportional to the cross-sections of the two processes. For example, if the $gg$ cross-section constitutes 30\% of the total cross-section for a certain invariant mass window, then 30\% of the events generated should be $gg$. In this Table \ref{tab:4}, this is investigated.

\begin{table}[H]
   \centering
   \begin{tabular}{c|c|c|c}
       Events Generated & Quadrature ($gg, gq$) & Monte Carlo Integration ($gg, gq$) & Event Fraction ($gg, gq$) \\
       \hline
       11,000 & 17.999\%, 82.001\% & 17.931\%, 82.069\% & 18.010\%, 81.990\% \\
       150,000 & || & 17.981\%, 82.019\% & 18.006\%, 81.994\%
   \end{tabular}
   \caption{Proportion of cross-section and generated events for the $gg$ and $gq$ processes. Cross-section integrated over $M$ = [$M_{\text{cut}}$, $\sqrt{s}$], $\sqrt{s}$ = 13 TeV, $M_{s}$ = 5 TeV}
   \label{tab:4}
\end{table}

\section{Analysis of Samples}

In the produced $\gamma$ + jet events, there are several kinematic quantities the outgoing partons possess that are of interest. In no particular order, they are; the 4-vectors of each outgoing particle, the radial and azimuthal coordinates $\theta$ and $\phi$, the transverse momentum $p_{T}$, the energy, and the invariant mass. When the parton and photon quantities are summed, the characteristics of the resonance may be deduced. Histograms that convey this data for $M_s = 5$ with $\sqrt{s} = 13$ TeV are in Appendix A.

\subsection{Invariant Mass Distributions from STRINGS}

\begin{figure}[H]
   \centering
   \includegraphics[width=0.49\textwidth]{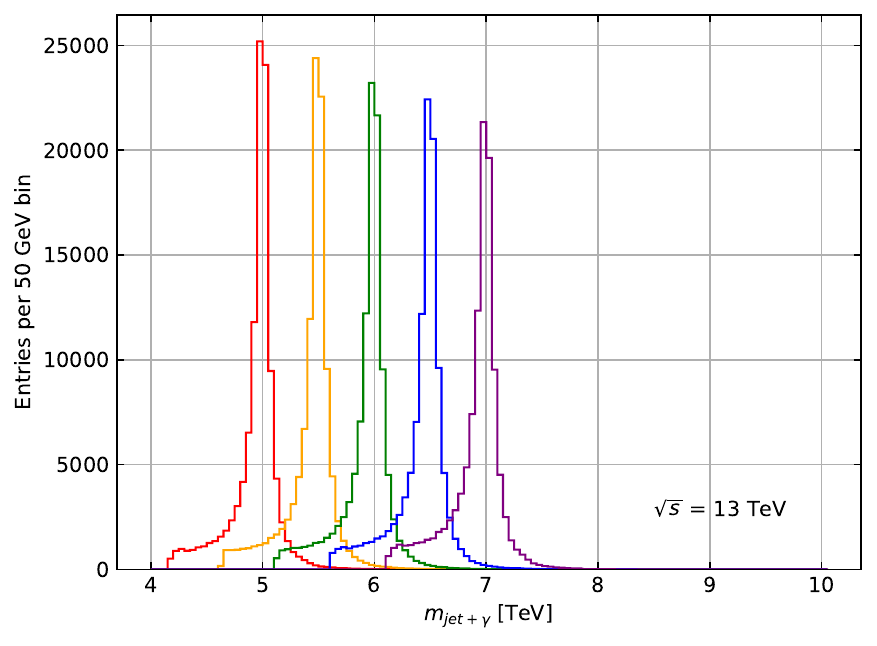}
   \includegraphics[width=0.49\textwidth]{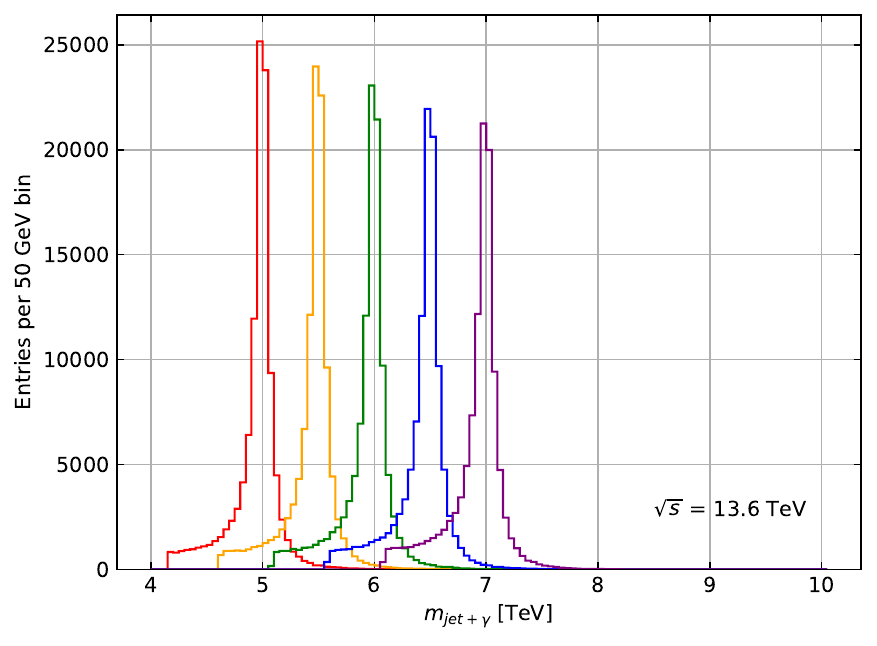}
   \caption{(Linear Axis)}
   \label{fig:13}
\end{figure}

\begin{figure}[H]
   \centering
   \includegraphics[width=0.49\textwidth]{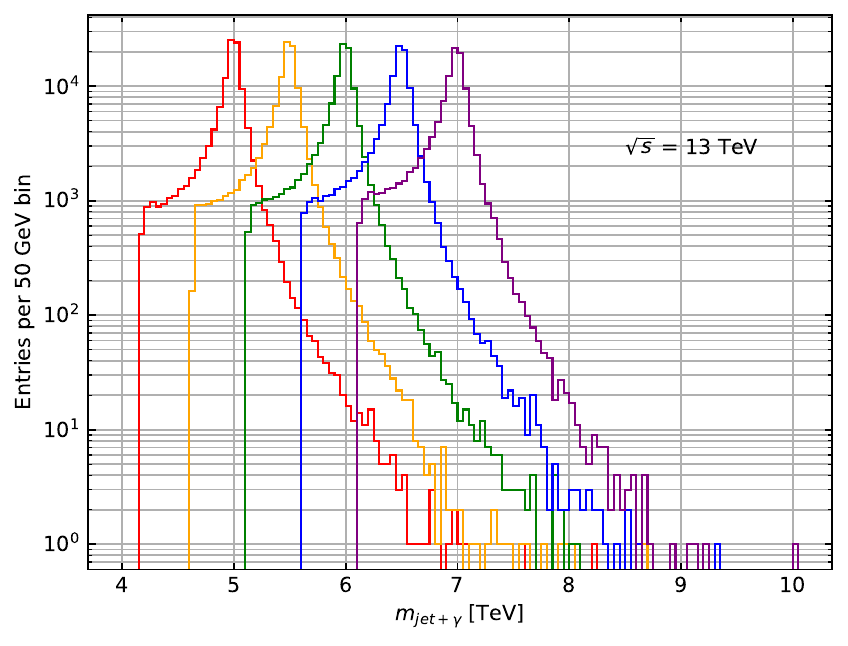}
   \includegraphics[width=0.49\textwidth]{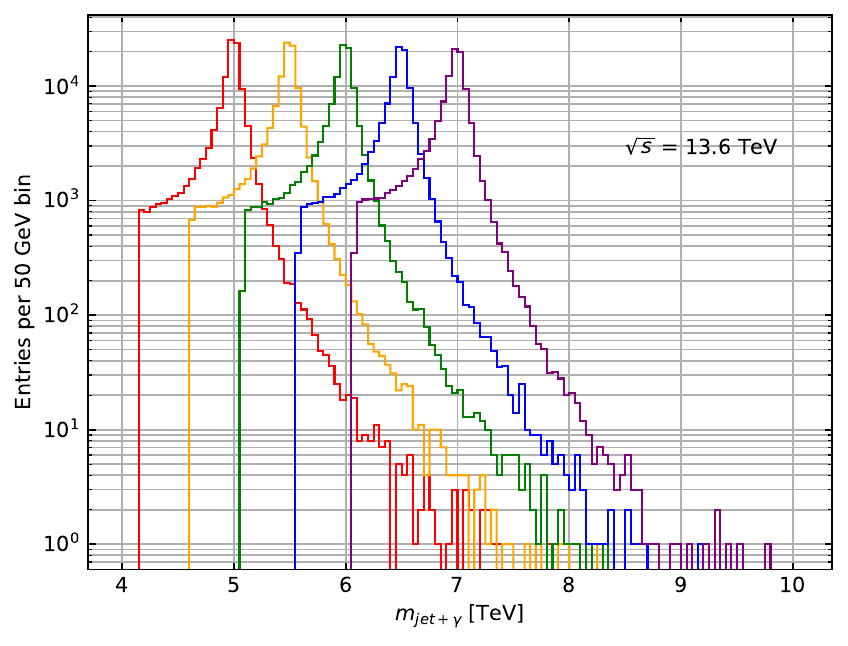}
   \caption{(Logarithmic Axis)}
   \label{fig:14}
\end{figure}

The histograms in Figures \ref{fig:13} and \ref{fig:14} were created by superimposing each $\gamma$ + jet invariant mass histogram using each string scale $M_s$ for $\sqrt{s}$ = 13 TeV and 13.6 TeV. More kinematic data can be found in Appendix A. 

\subsection{Angular Distributions}

The radial angle $\theta$ changes depending on whether is being measured in the lab frame or the resonance frame, where the radial angle is defined as $\theta^{*}$. In the resonance frame, the $z$ momentum cancels out and the radial and azimuthal angles are opposite to each other for each outgoing parton. The cosine of the radial angle is given by $\theta = \frac{p_{z}}{p}$. To convert $p_{z}$ into the resonance frame, a Lorentz boost is applied along the $z$-axis:

\begin{equation}
   p_{z}^{*} = \gamma (p_{z} - E \beta) , \tag{5.1} \label{eq:5.1}
\end{equation}

\noindent where

\begin{align}
   \gamma = \frac{E^{s}}{M^{s}} , \qquad \beta = \frac{p_{z}^{s}}{E^{s}} . \tag{5.2} \label{eq:5.2}
\end{align}

The superscript $s$ stands for 'string.' The distribution of the radial cosines is described by the curve $2+3x^2$ in the $gq$ case, and $\frac{94}{25}+6x^2+x^4$, where $x = \cos{\theta^{*}}$.

\begin{figure}[H]
   \centering
   \includegraphics[width=0.5\textwidth]{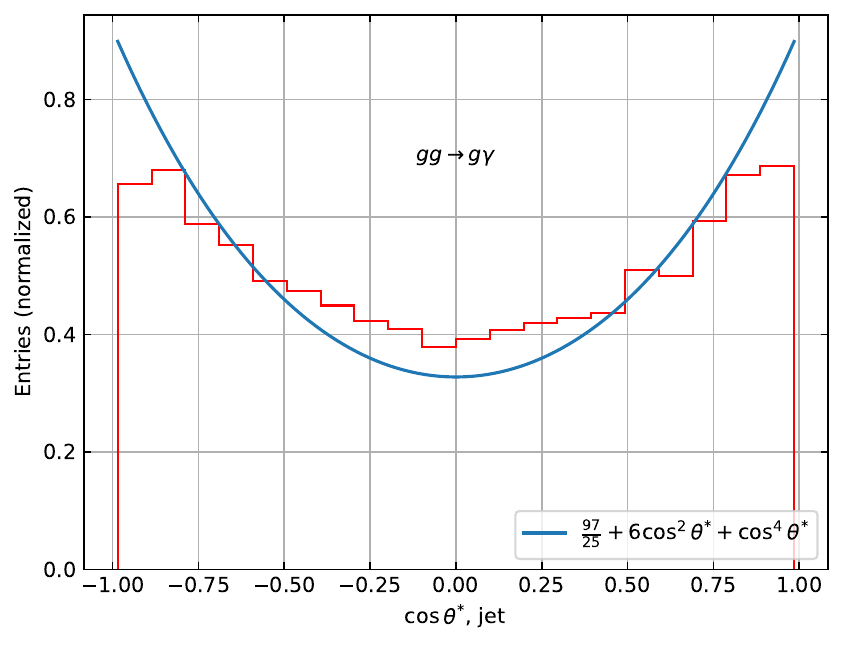}
   \caption{Angular distribution for $M_{s}$ = 5 TeV}
   \label{fig:15}
\end{figure}
  
One can observe that toward the +1 and -1, the histogram does not match the predicted distribution. The STRINGS generator has changed since the last ATLAS note. The theory approximated being at the resonance, but there is a mass distribution in practice. In past trials, this approximation seemed to work well \cite{gingrich2022}.

\begin{figure}[H]
   \centering
   \includegraphics[width=0.49\textwidth]{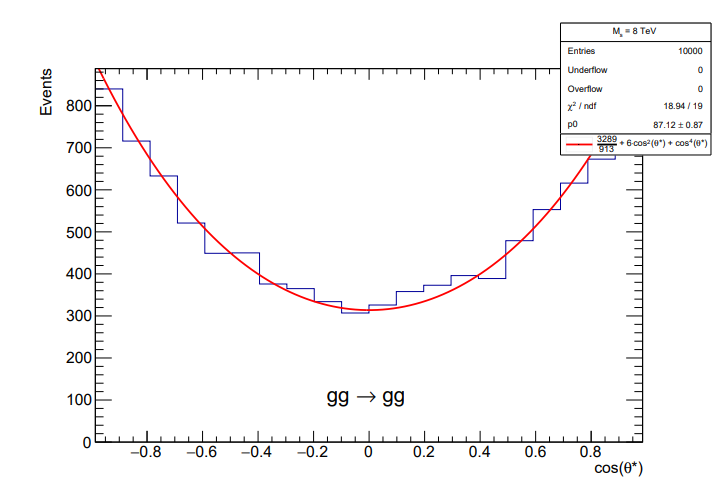}
   \includegraphics[width=0.49\textwidth]{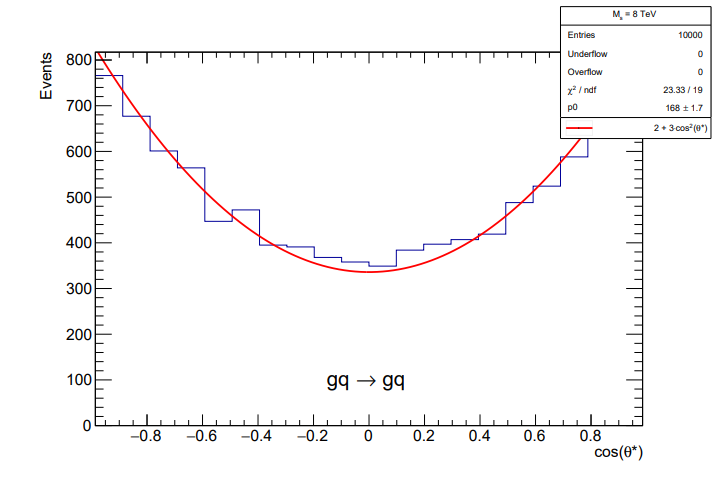}
   \caption{Angular distributions}
   \label{fig:16}
\end{figure}

It is possible the old version of STRINGS was more accurate than it should have been. It is noted that for higher-energy $M_s$, the distribution may be more accurate.

\begin{figure}[H]
   \centering
   \includegraphics[width=0.49\textwidth]{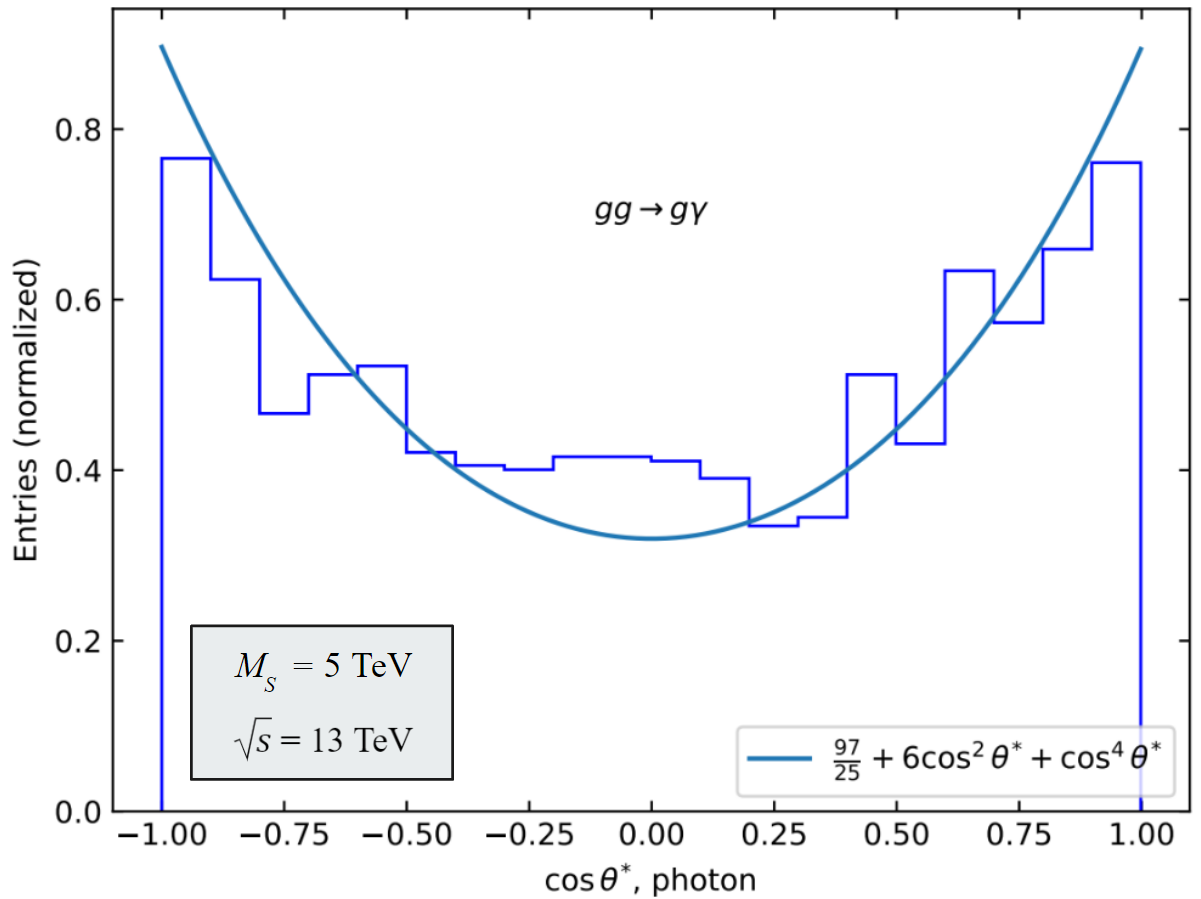}
   \includegraphics[width=0.49\textwidth]{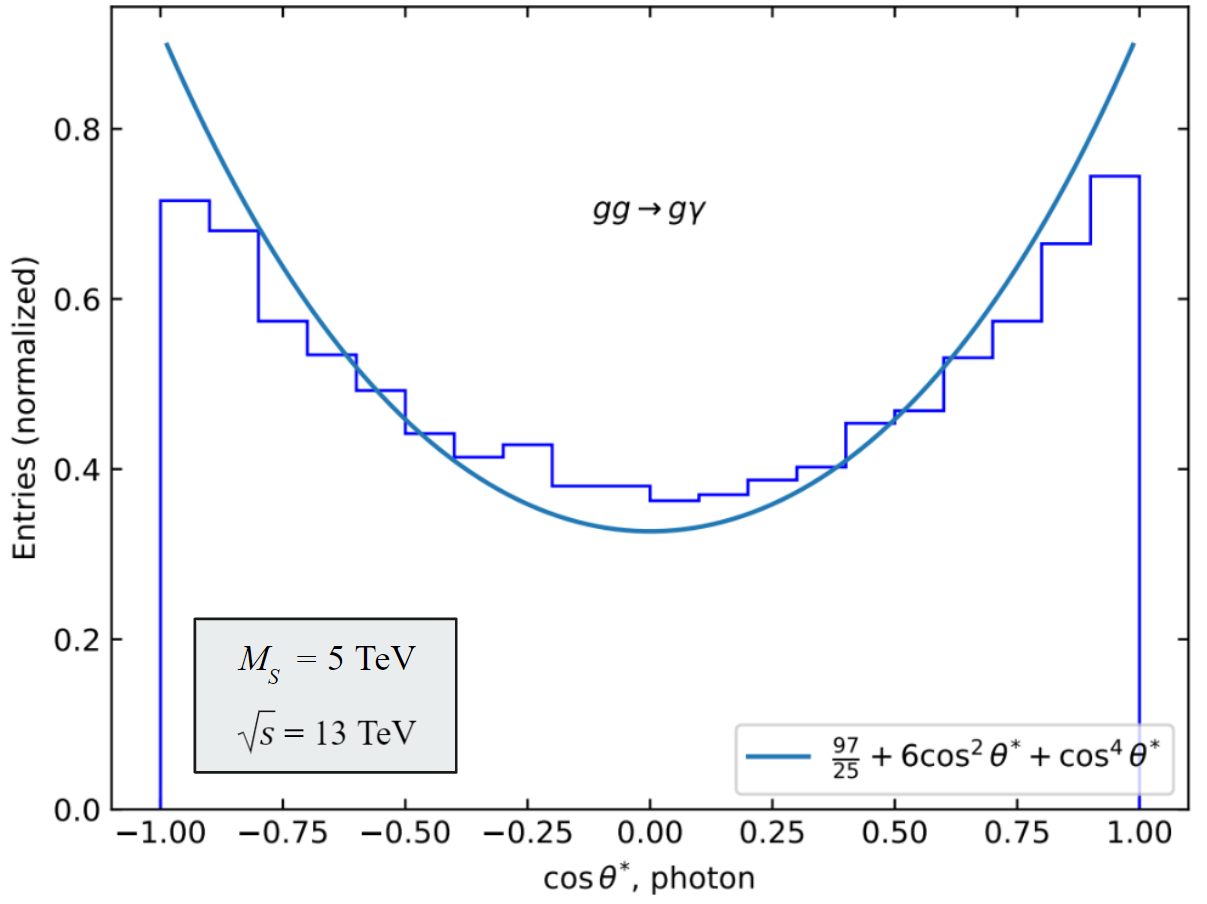}
   \caption{11,000 events, $y_{cut}$ = 10; 110,000 events, $y_{cut}$ = 6}
   \label{fig:17}
\end{figure}

Furthermore, the maximum rapidity cut on the data generated was 2.5. This cut minimizes how close an outgoing parton's trajectory can be to the beam axis. With a higher rapidity cut, the distribution may be more closely matched by allowing more events to scatter partons close to the beam axis (such events would have a $\cos{\theta^{*}} \approx \pm$ 1). Given the data in Figure \ref{fig:17}, this seems to be the case; a cut of 6 gives a shape closer to the predicted distribution. A cut of 10 would likely be even better, but because imposing greater cuts vastly increases the time it takes for STRINGS to run, not many events were generated. The distribution would likely be smoother with more events.

\subsection{Invariant Mass Distributions from Pythia}

Following the STRINGS simulations, LHE files are fed into Pythia \cite{Sj_strand_2015, Sj_strand_2006}, a more sophisticated and realistic generator for simulating scattering events. The invariant mass histograms on the next page are generated using ROOT, and follow the same color scheme as the corresponding previous STRINGS histograms. Other kinematic data for $M_s = 5$ TeV with $\sqrt{s} = 13$ TeV can be found in Appendix B. 

\begin{figure}[H]
   \centering
   \includegraphics[width=.49\textwidth]{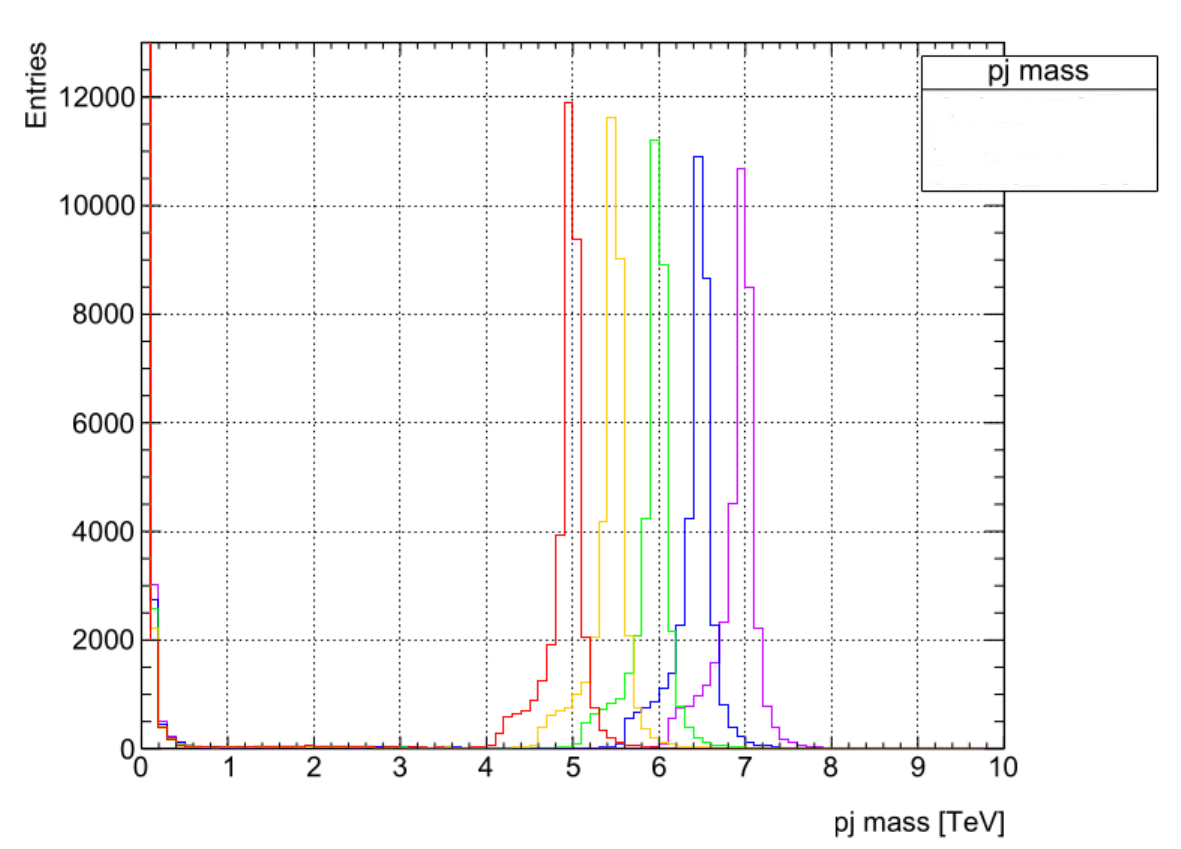}
   \includegraphics[width=.49\textwidth]{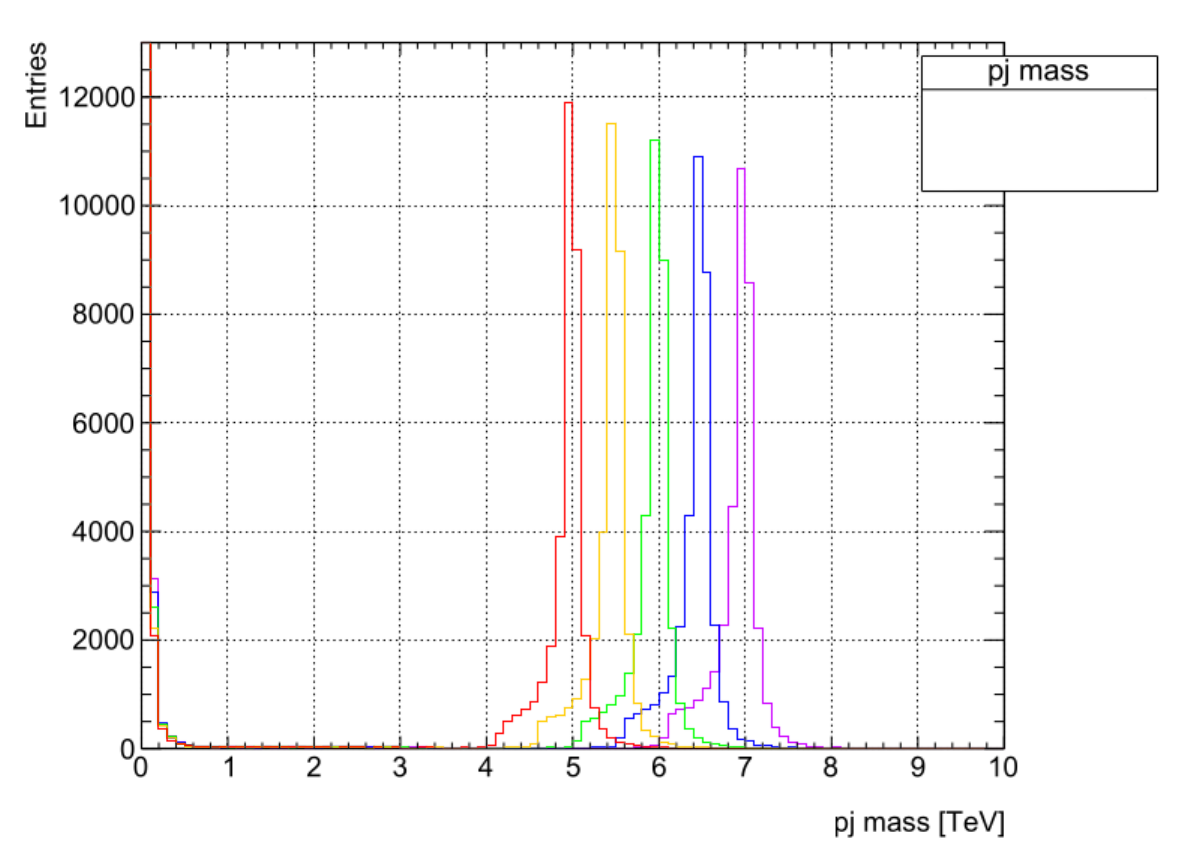}
   \caption{Linear Axis; $\sqrt{s}$ = 13 TeV (top) and 13.6 TeV (bottom)}
   \label{fig:18}
\end{figure}

\begin{figure}[H]
   \centering
   \includegraphics[width=.49\textwidth]{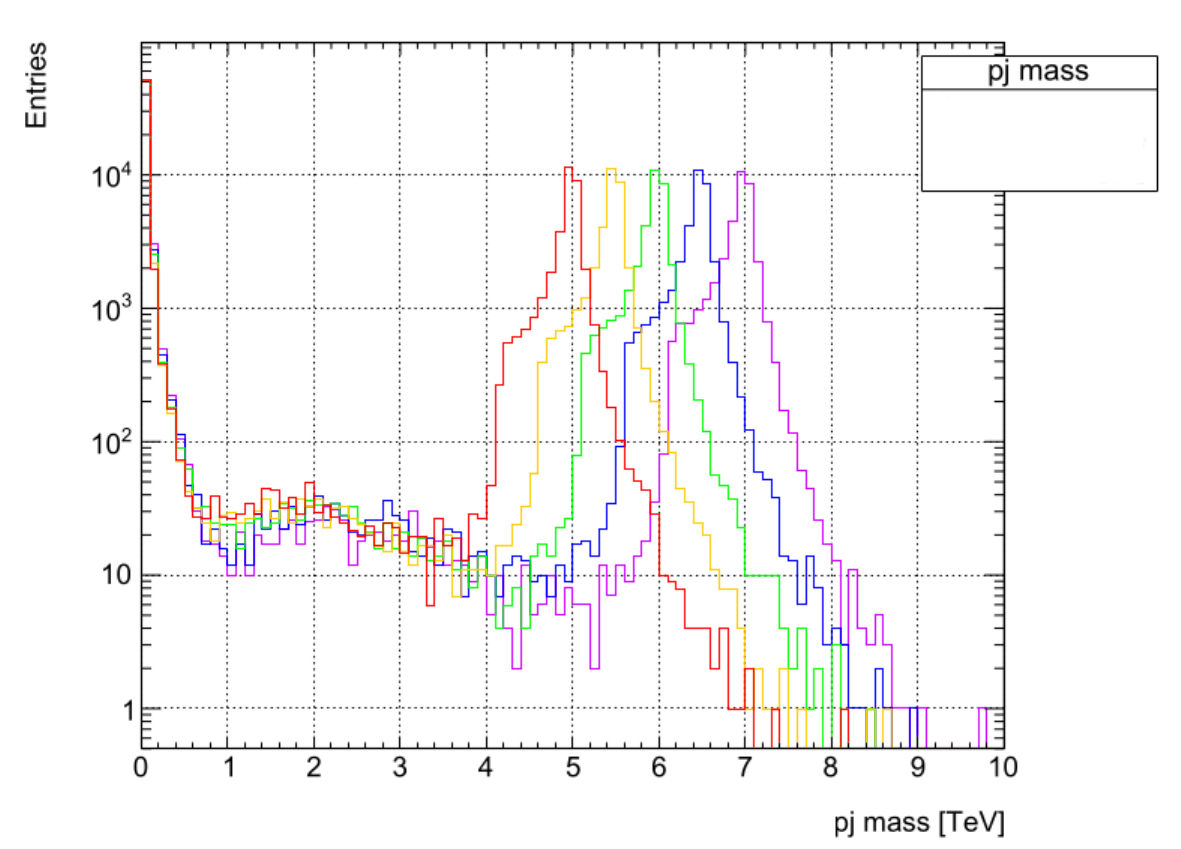}
   \includegraphics[width=.49\textwidth]{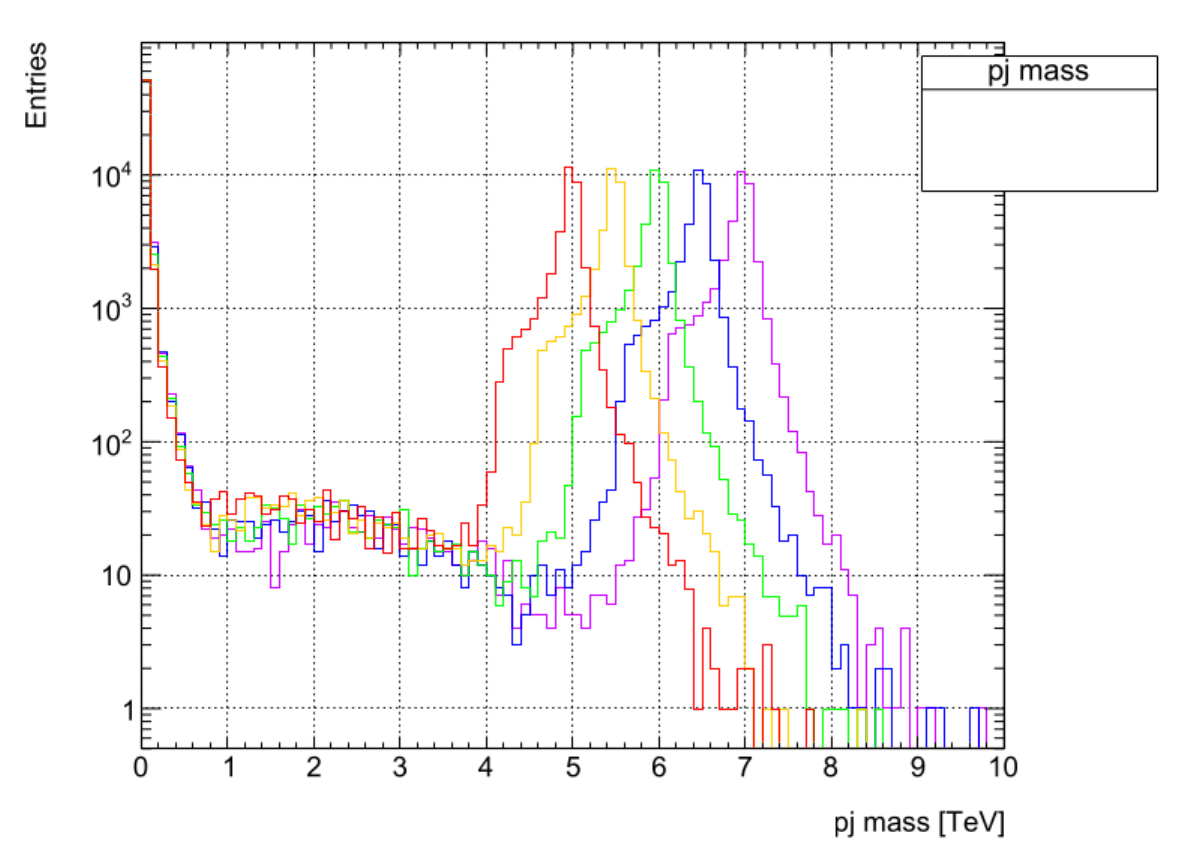}
   \caption{Logarithmic Axis; $\sqrt{s}$ = 13 TeV (top) and 13.6 TeV (bottom)}
   \label{fig:19}
\end{figure}

\section{Discussion}

The Pythia events generated demonstrate the resonance peaks are remarkably similar for $\sqrt{s}$ = 13 and 13.6 TeV, meaning that discovery potential at either $\sqrt{s}$ is relatively consistent. The evolution of the resonance peak shape as $M_{s}$ is consistent with our expectation; as $M_{s}$ increases, the peak gets shorter and thicker at the base.

As evidenced in the initial attempt, the discovery potential for string scales on the interval [7,9] TeV is problematic because there is a significant decrease in the number of events as compared to [5,7] TeV. For studying photon jet scattering processes, discovery potential is much higher at scales lower than 7 TeV. We also observe a drastic low-mass tail in all of the invariant mass distributions generated by Pythia. These low-mass events are of little interest and could obscure events we are interested in studying at the LHC.

\bibliographystyle{plain}
\bibliography{reference}

\appendix

\section{STRINGS Kinematic Data}

\begin{figure}[H]
  \centering
  \includegraphics[width=0.49\textwidth]{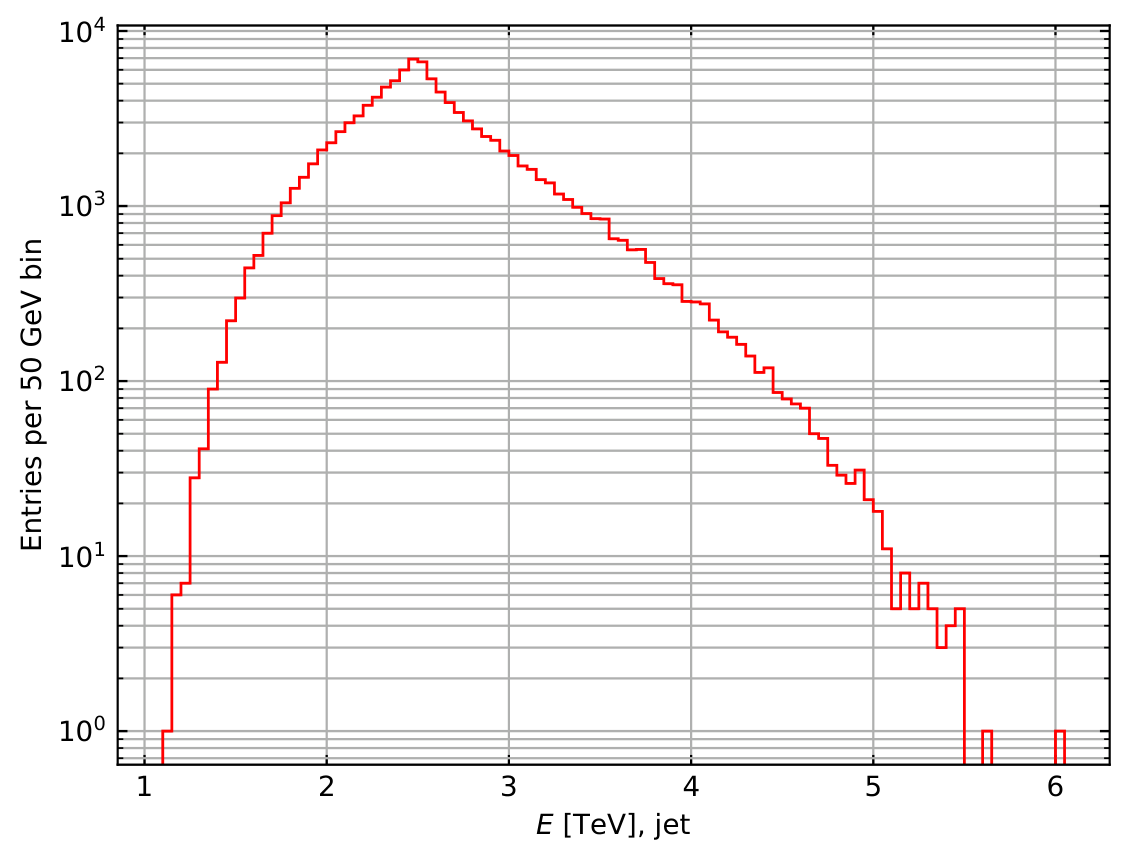}
  \includegraphics[width=0.49\textwidth]{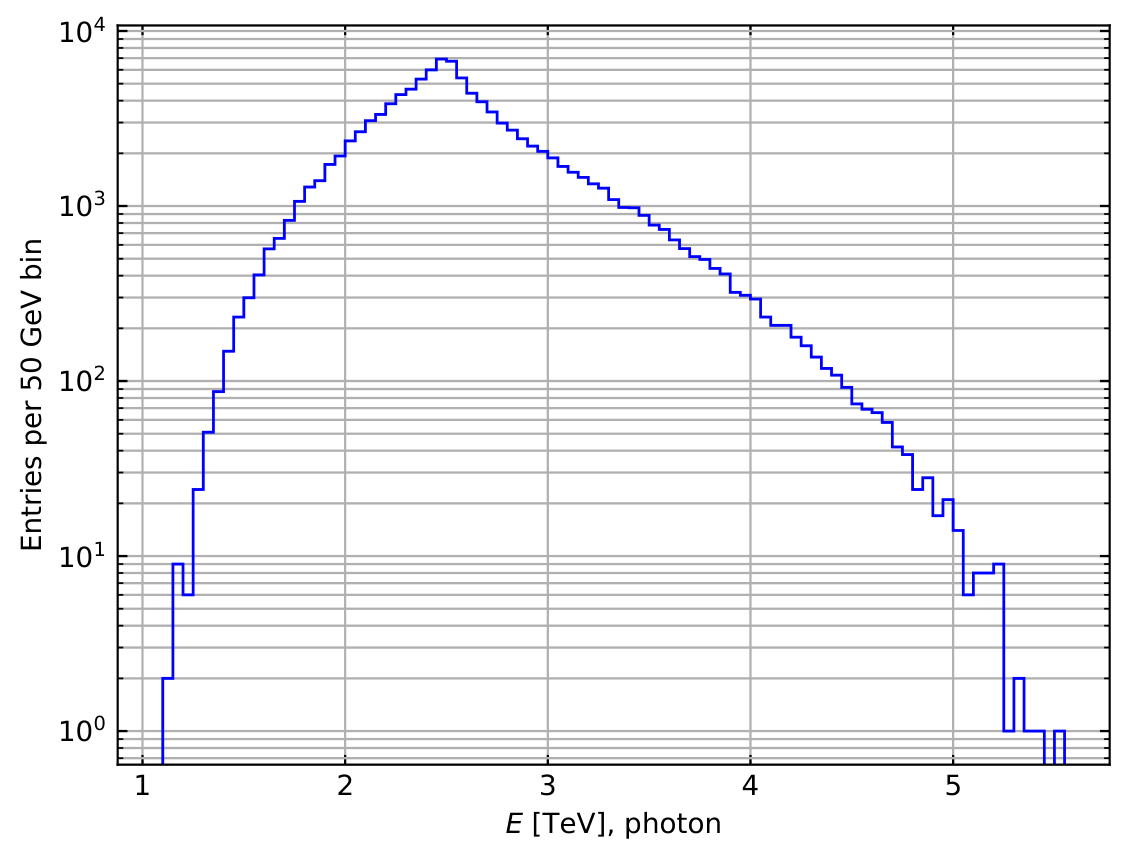}
  \includegraphics[width=0.49\textwidth]{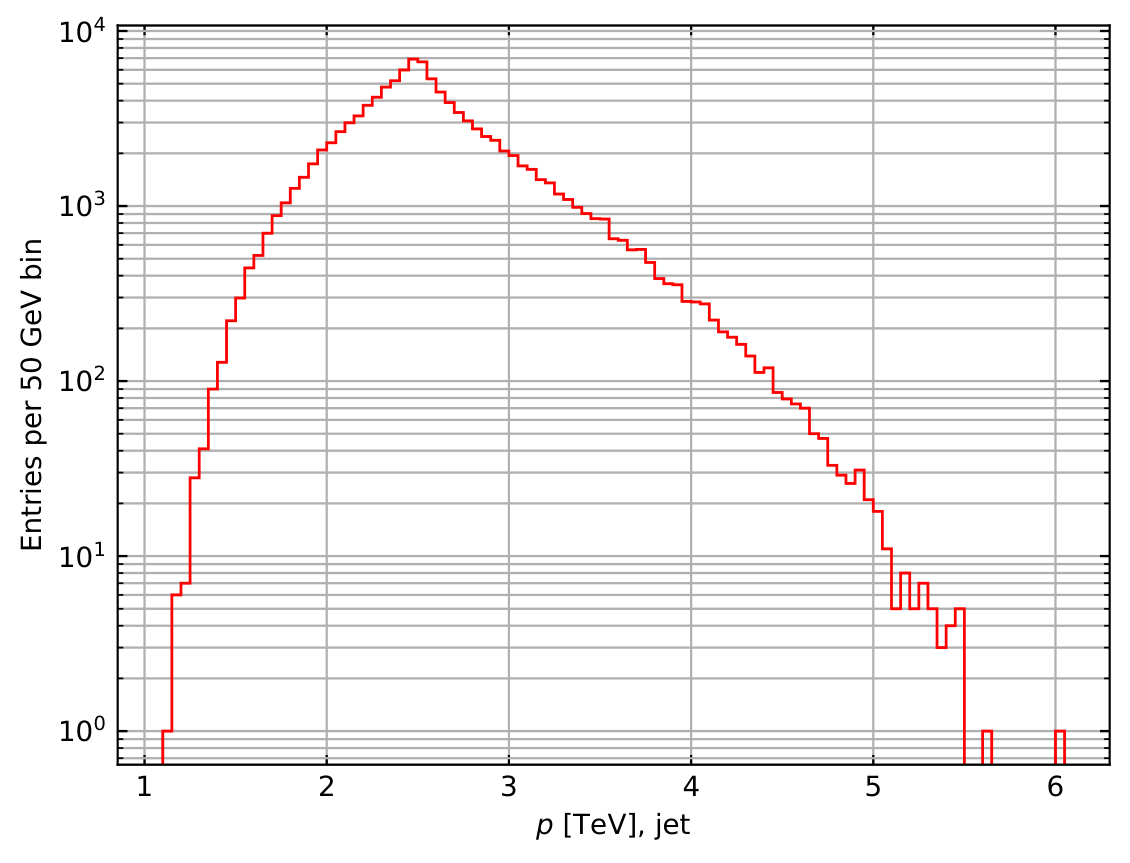}
  \includegraphics[width=0.49\textwidth]{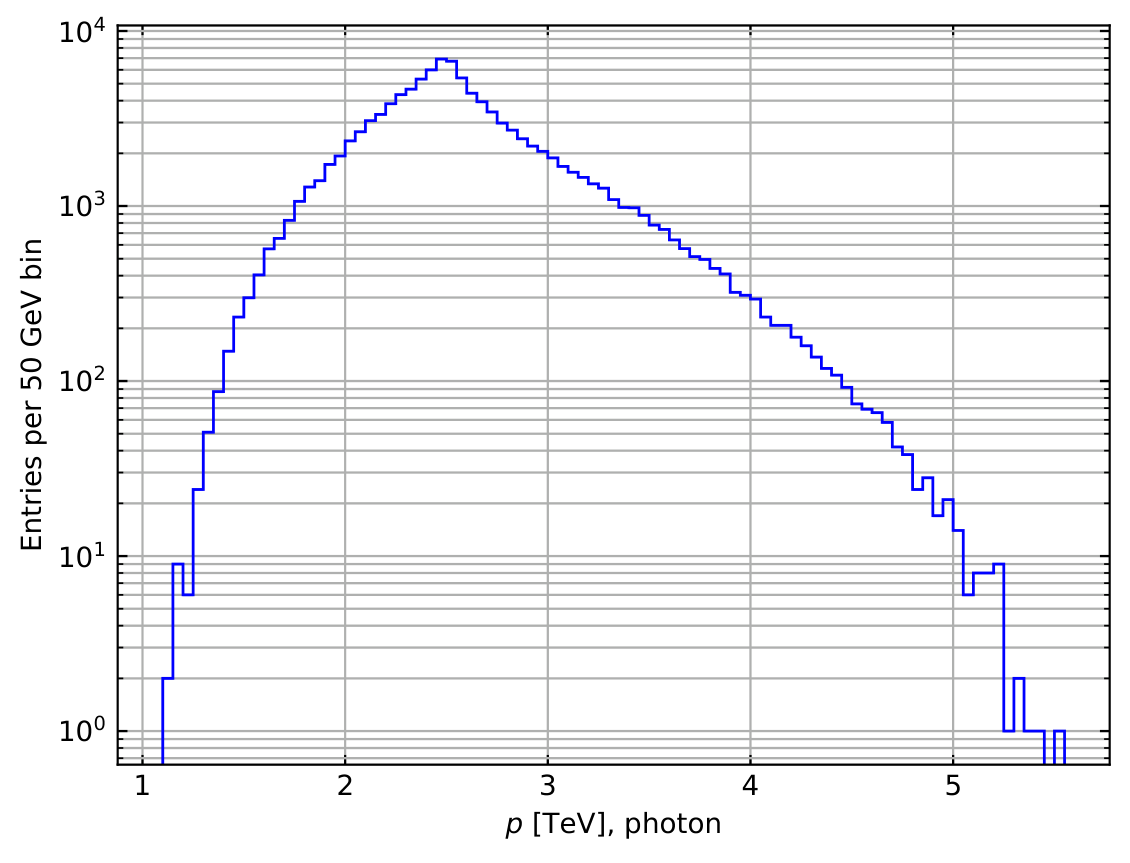}
  \includegraphics[width=0.49\textwidth]{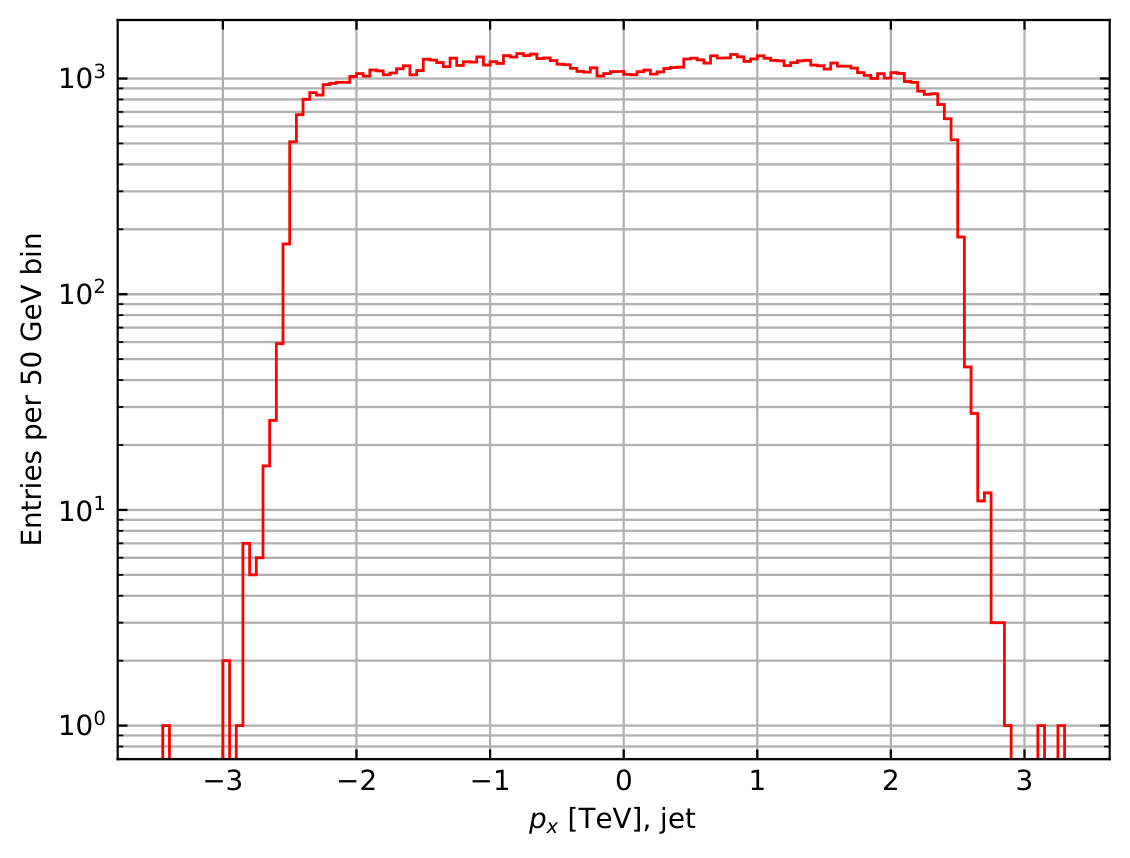}
  \includegraphics[width=0.49\textwidth]{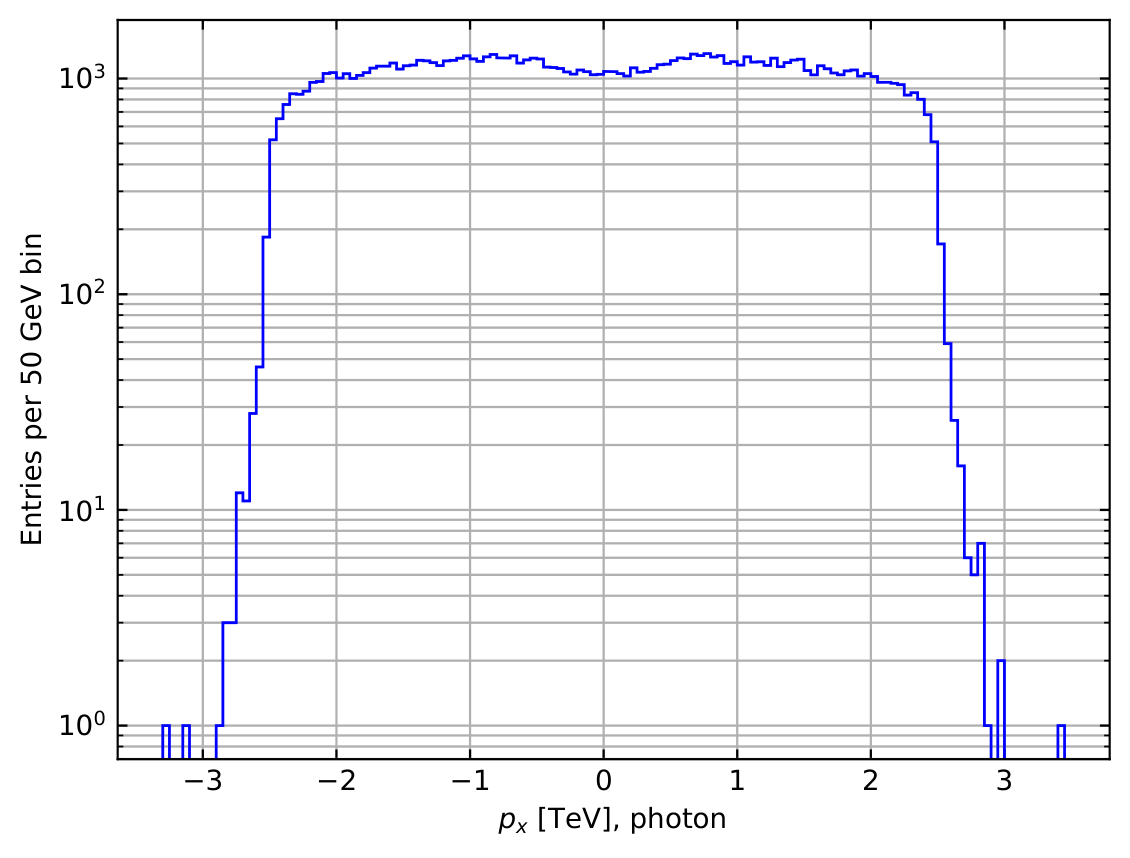}
  \caption{Kinematic data for $M_s = 5$ TeV and $\sqrt{s} = 13$ TeV from STRINGS.}
  \label{fig:20}
\end{figure}

\begin{figure}[H]
  \centering
  \includegraphics[width=0.49\textwidth]{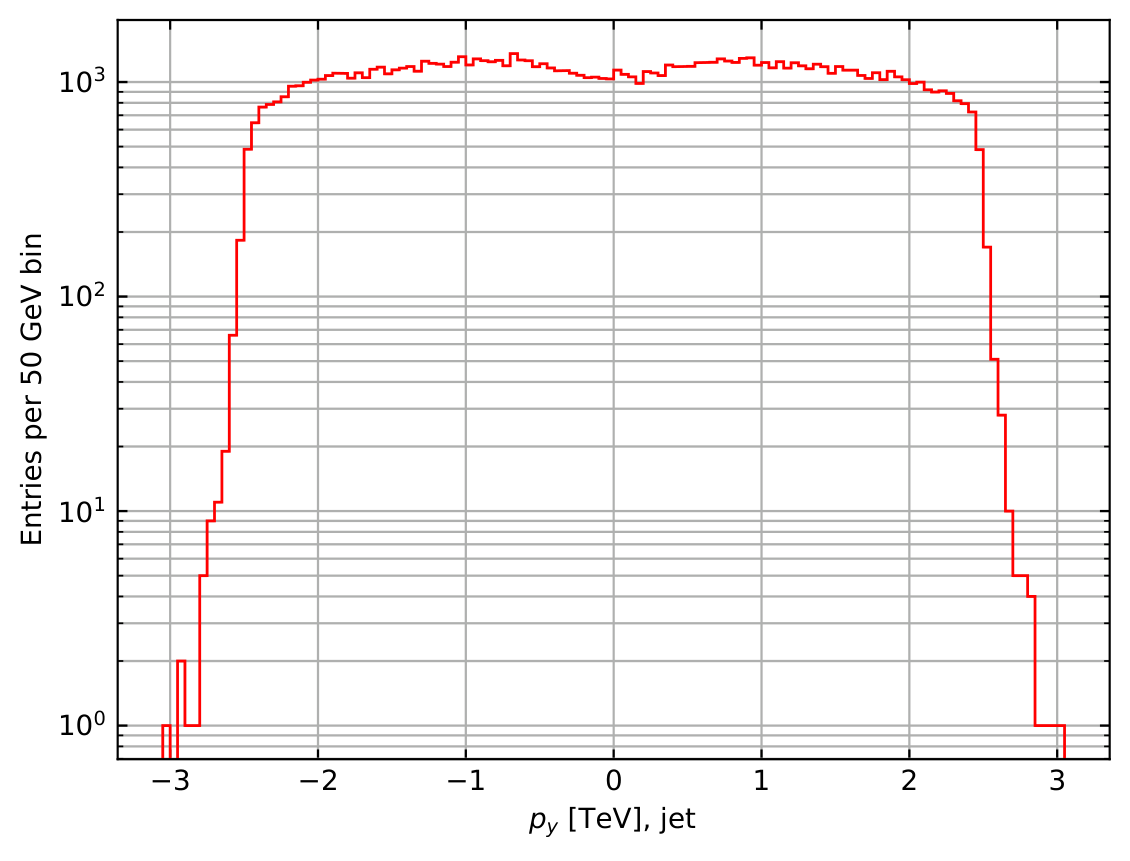}
  \includegraphics[width=0.49\textwidth]{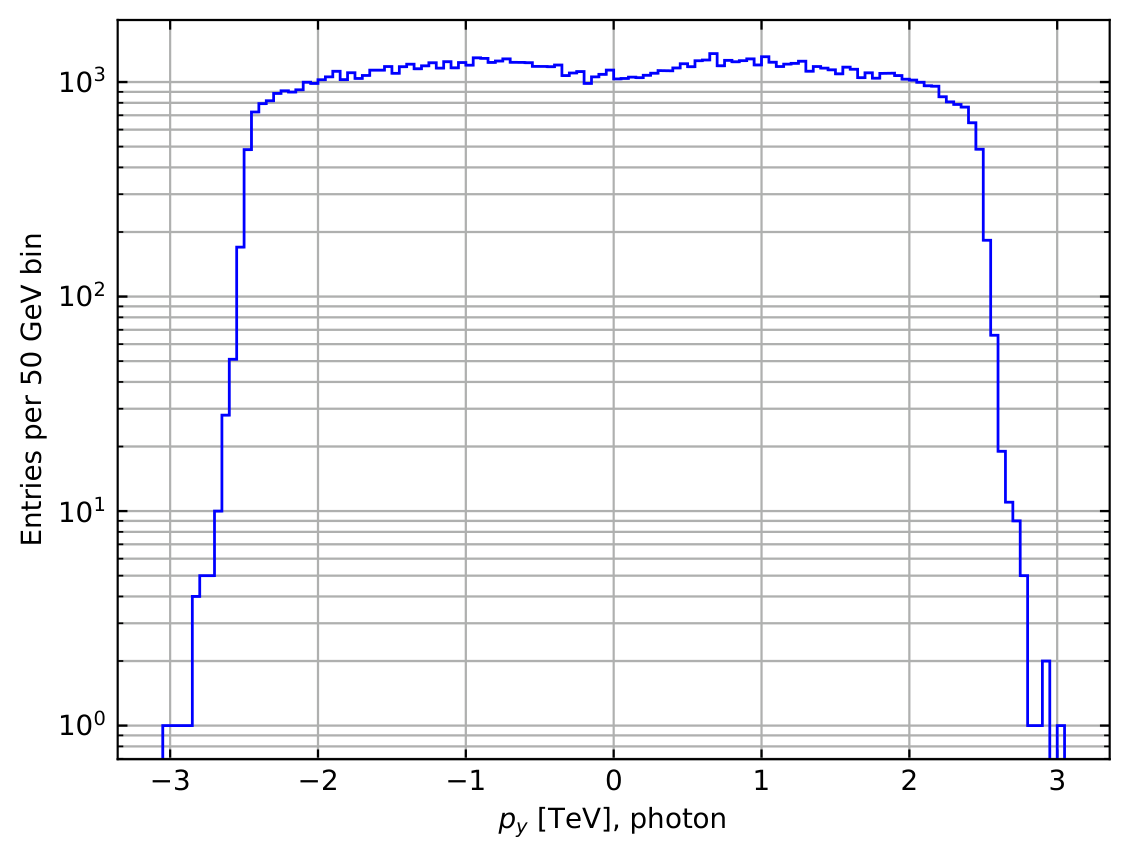}
  \includegraphics[width=0.49\textwidth]{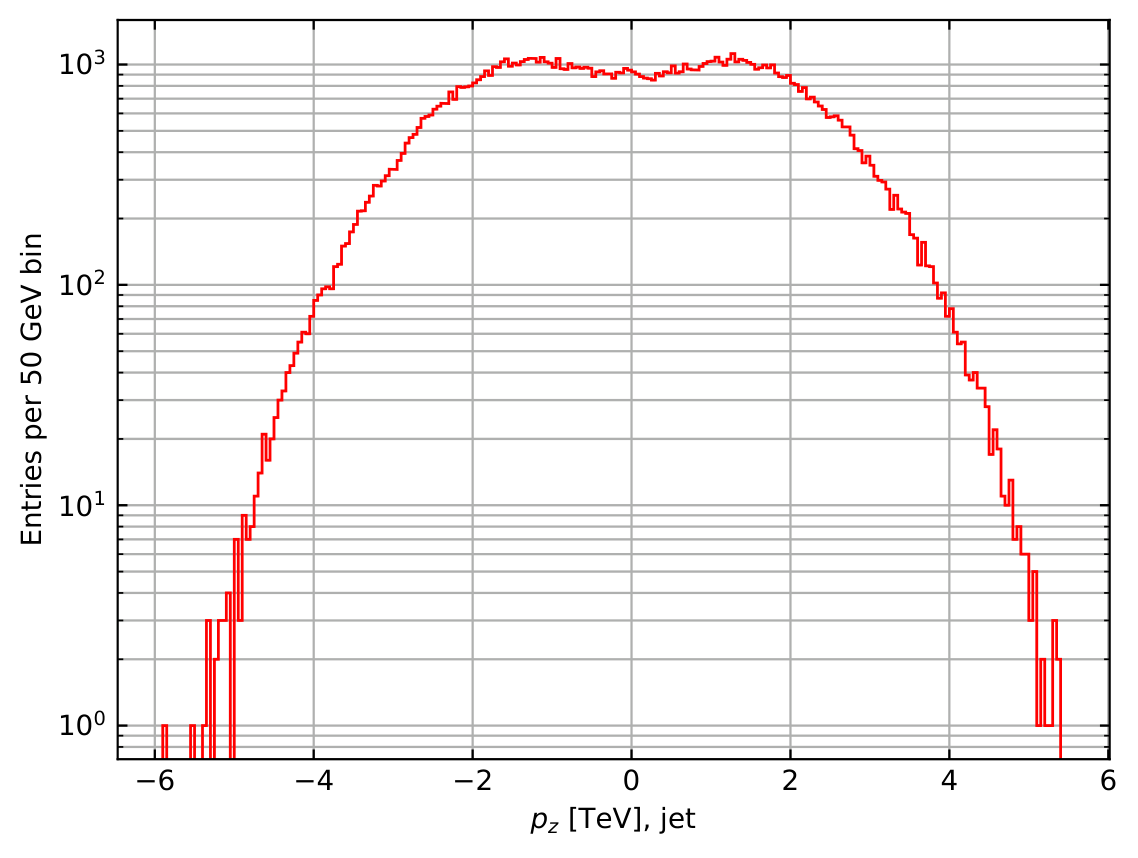}
  \includegraphics[width=0.49\textwidth]{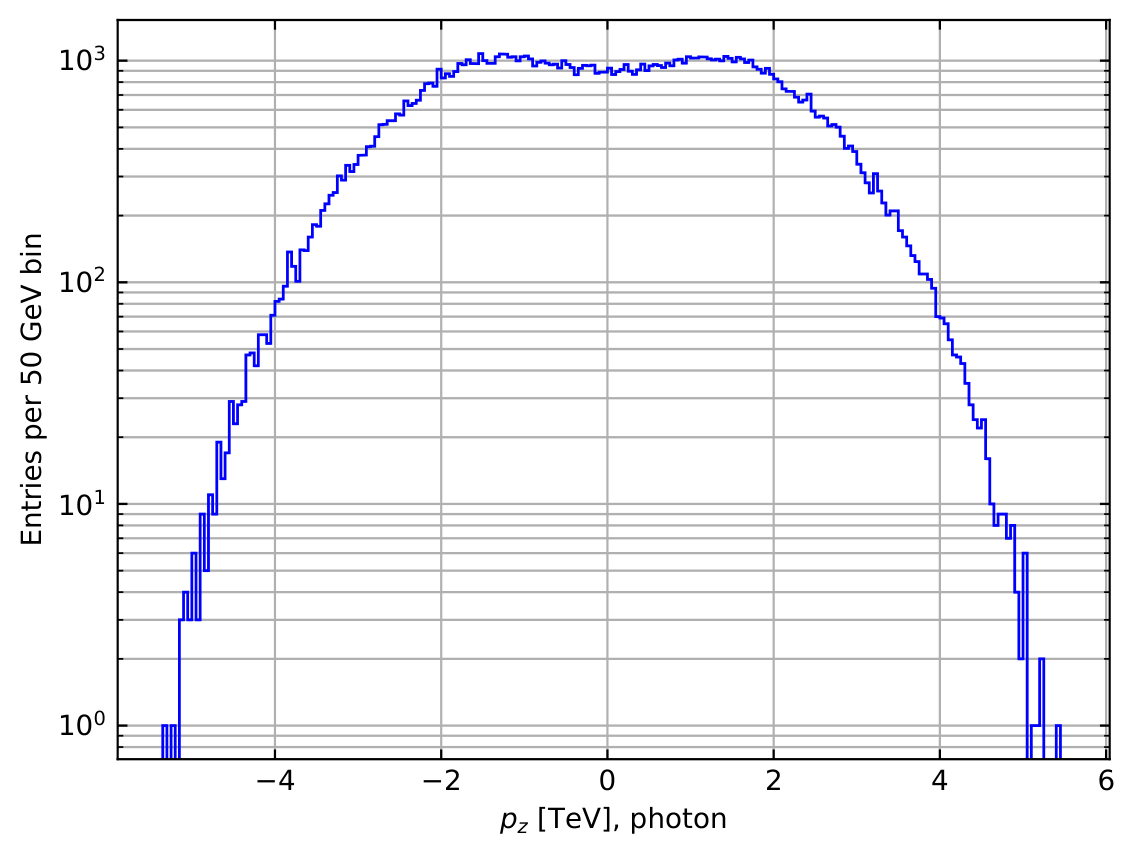}
  \includegraphics[width=0.49\textwidth]{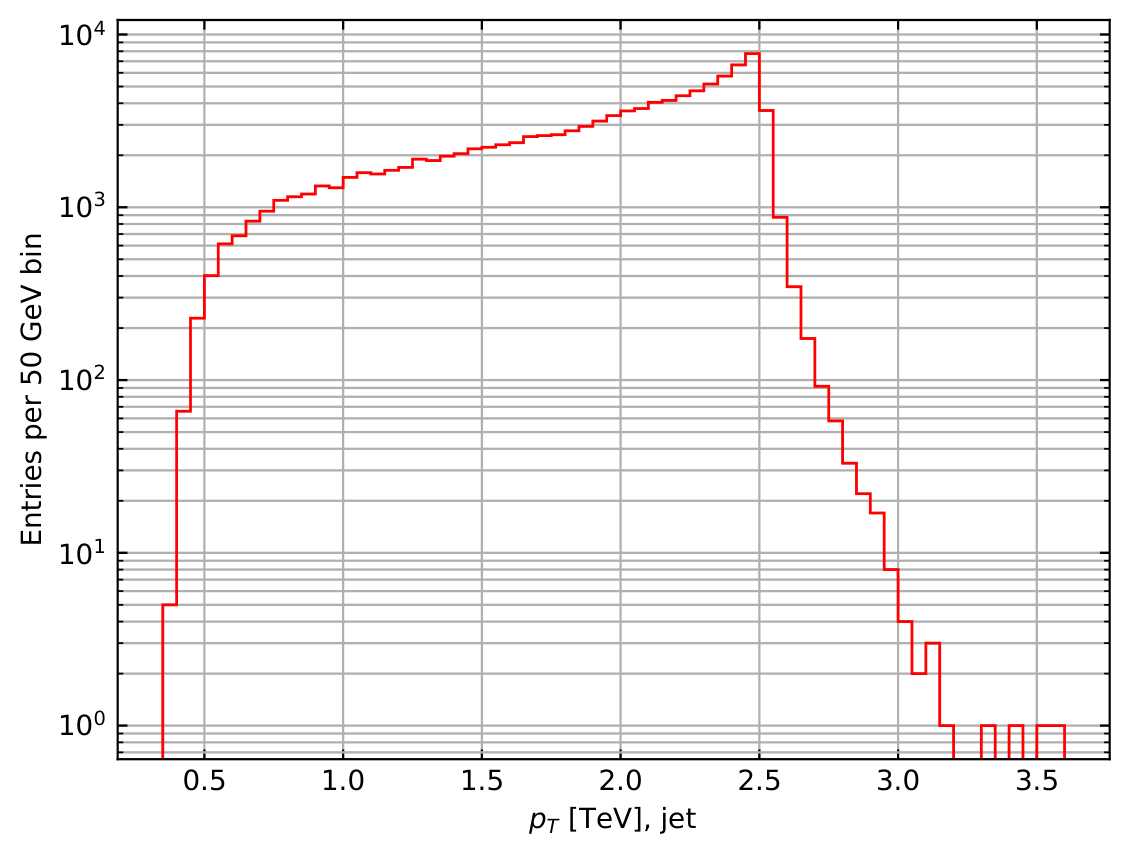}
  \includegraphics[width=0.49\textwidth]{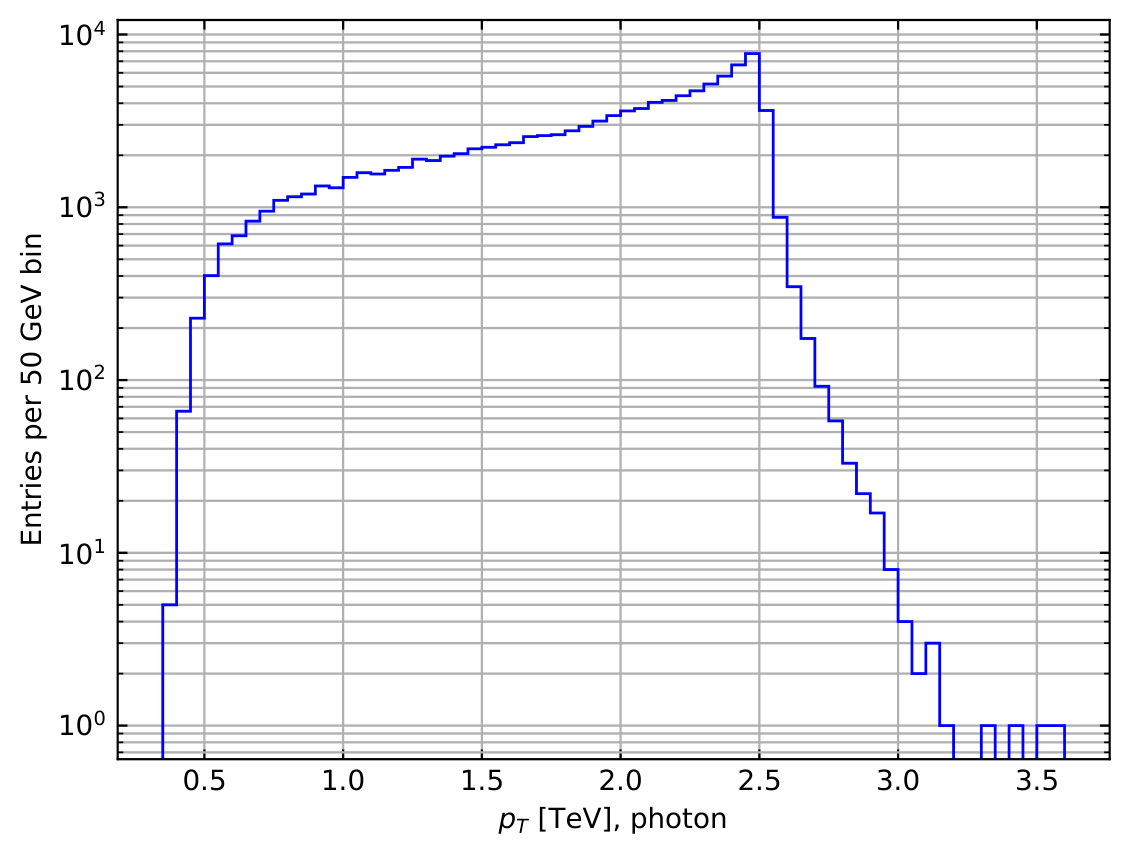}
  \caption{Kinematic data for $M_s = 5$ TeV and $\sqrt{s} = 13$ TeV from STRINGS.}
  \label{fig:21}
\end{figure}

\begin{figure}[H]
  \centering
  \includegraphics[width=0.49\textwidth]{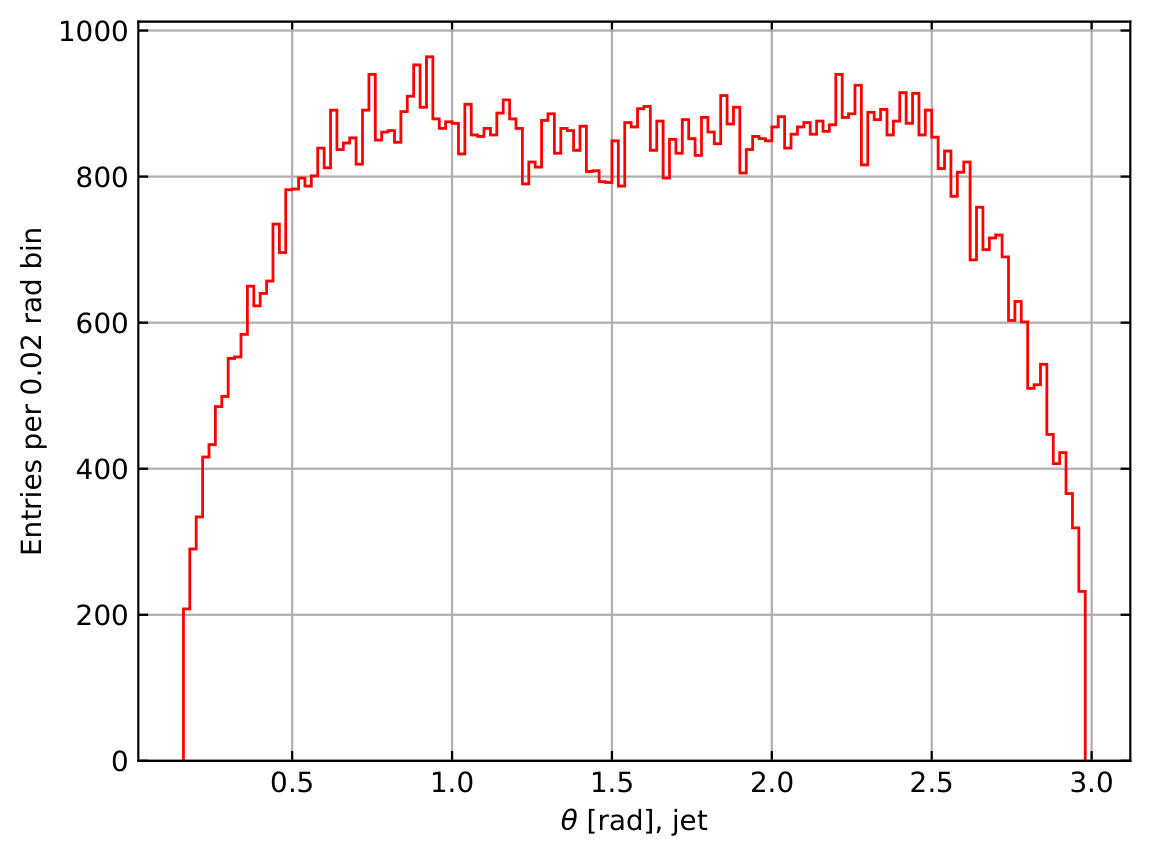}
  \includegraphics[width=0.49\textwidth]{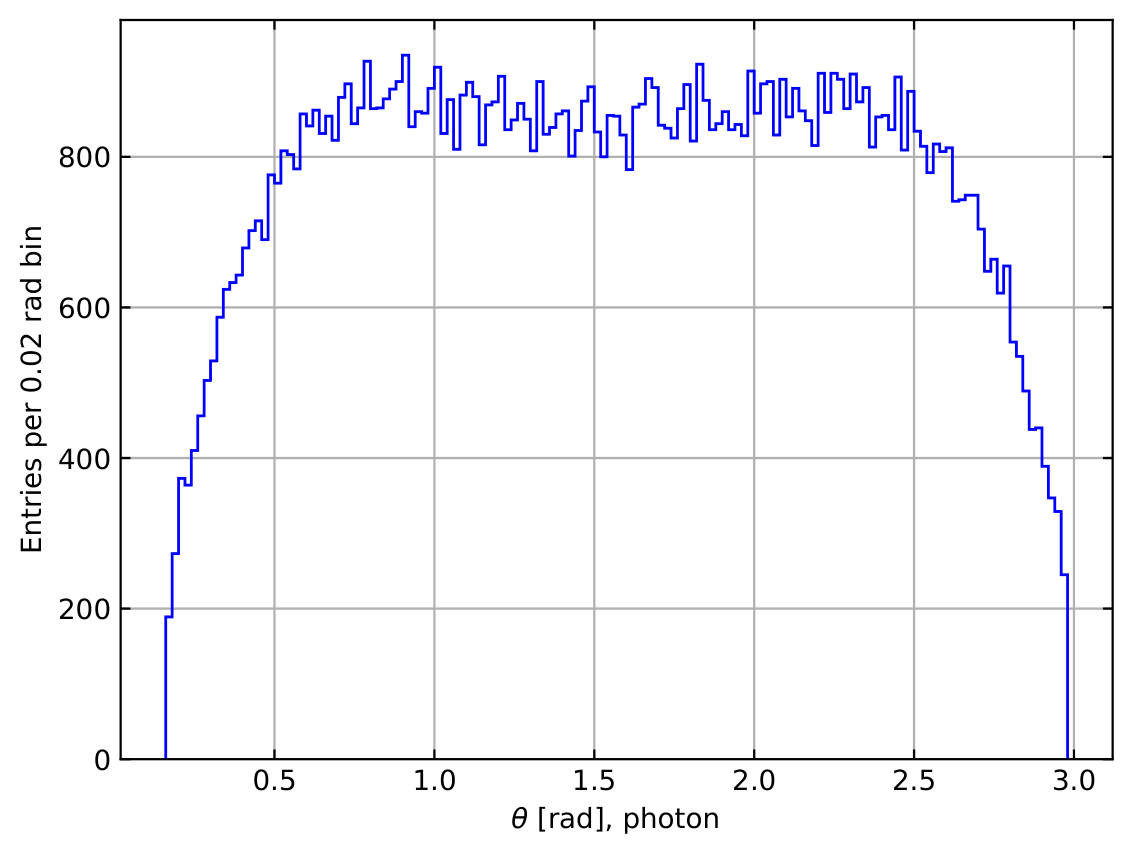}
  \includegraphics[width=0.49\textwidth]{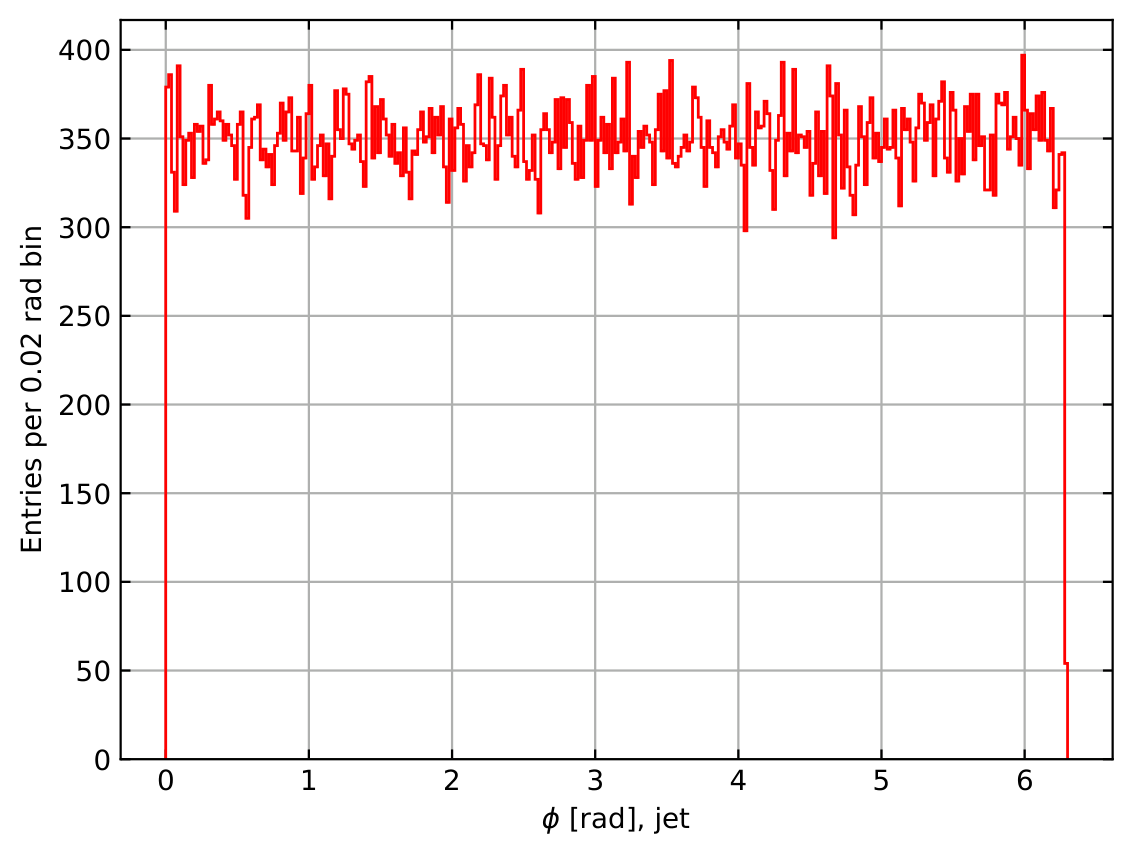}
  \includegraphics[width=0.49\textwidth]{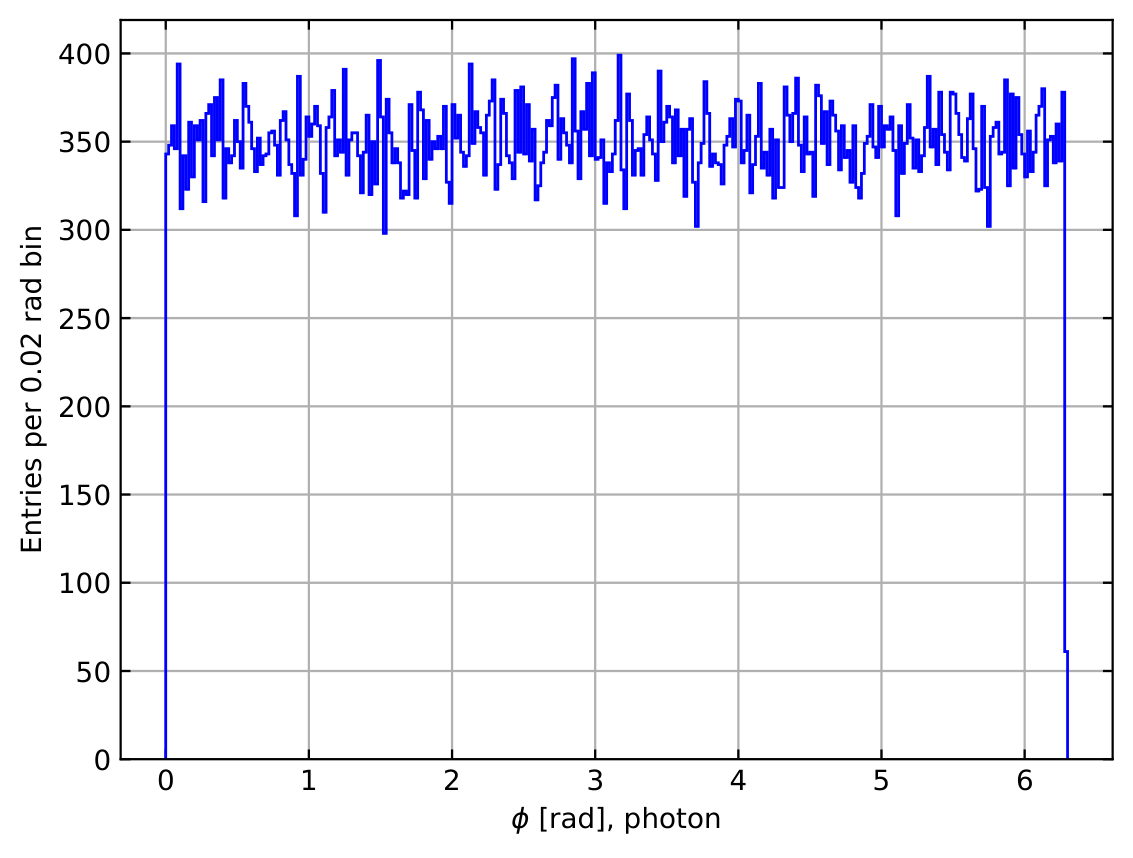}
  \includegraphics[width=0.49\textwidth]{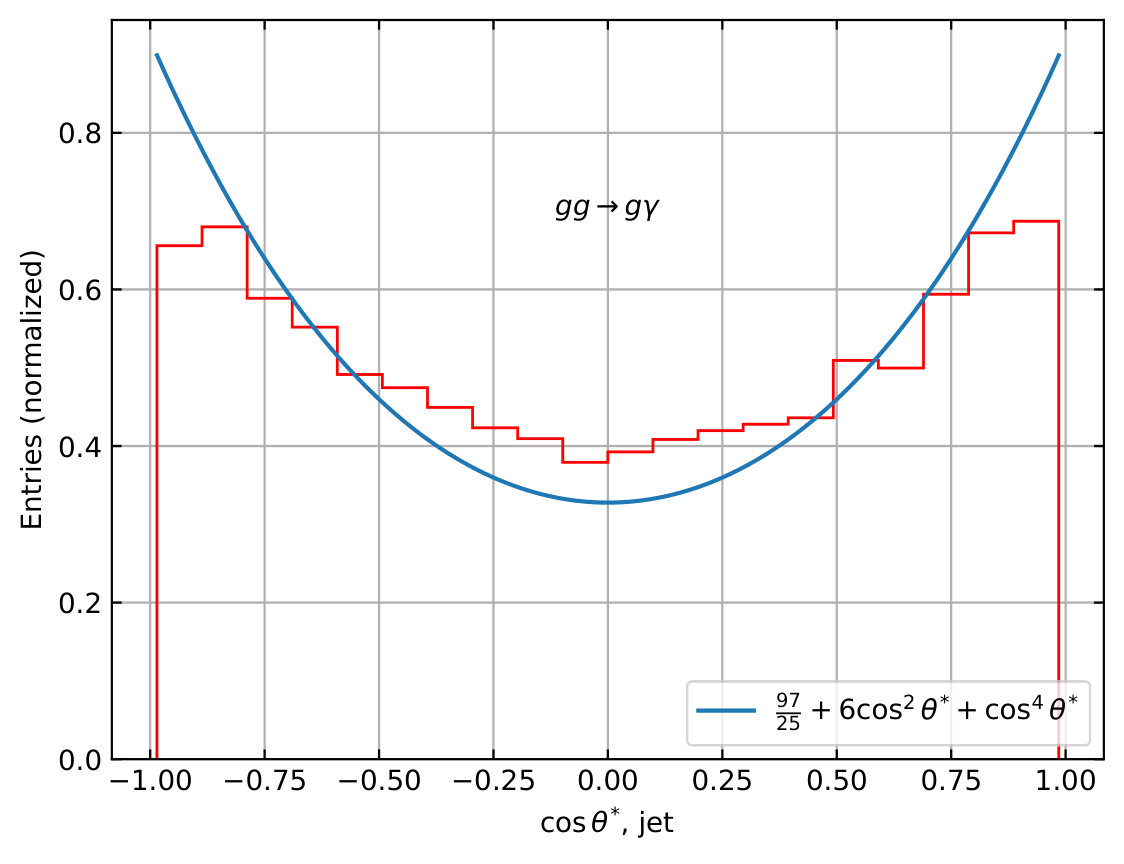}
  \includegraphics[width=0.49\textwidth]{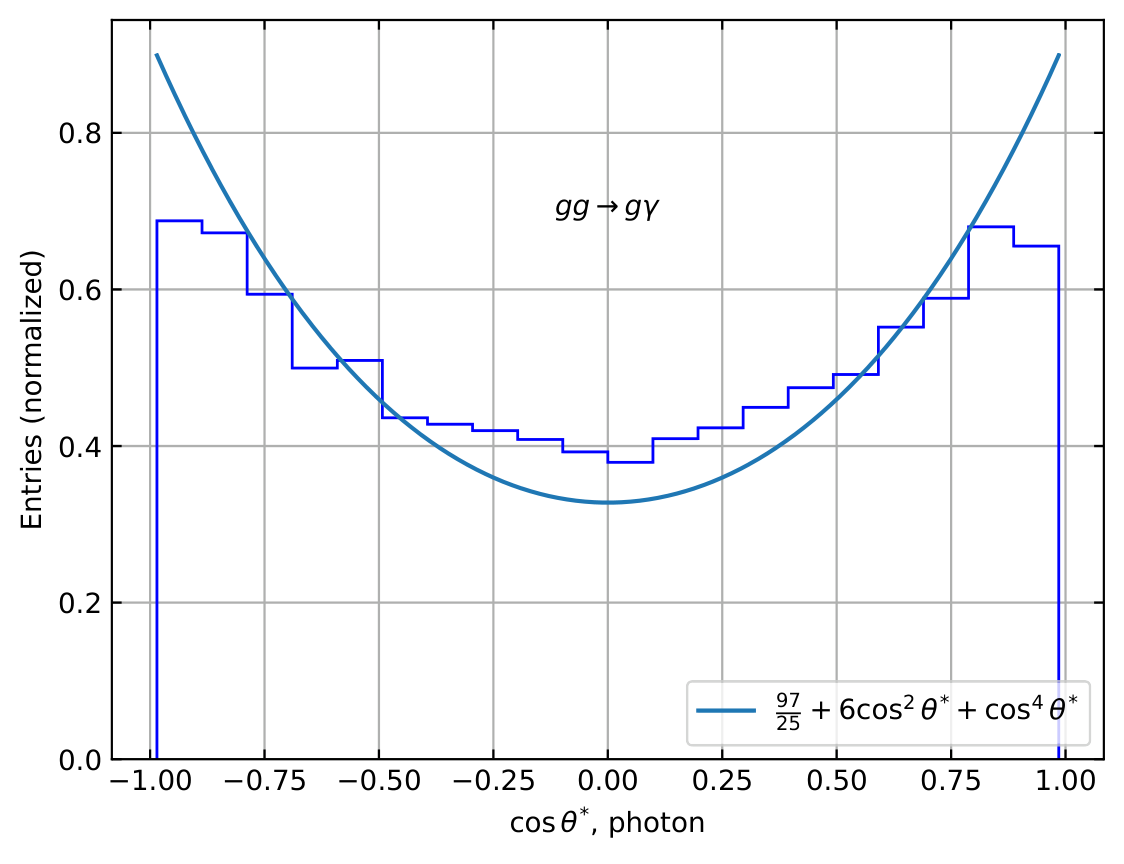}
    \caption{Kinematic data for $M_s = 5$ TeV and $\sqrt{s} = 13$ TeV from STRINGS.}
  \label{fig:22}
\end{figure}

\begin{figure}
  \centering
  \includegraphics[width=0.49\textwidth]{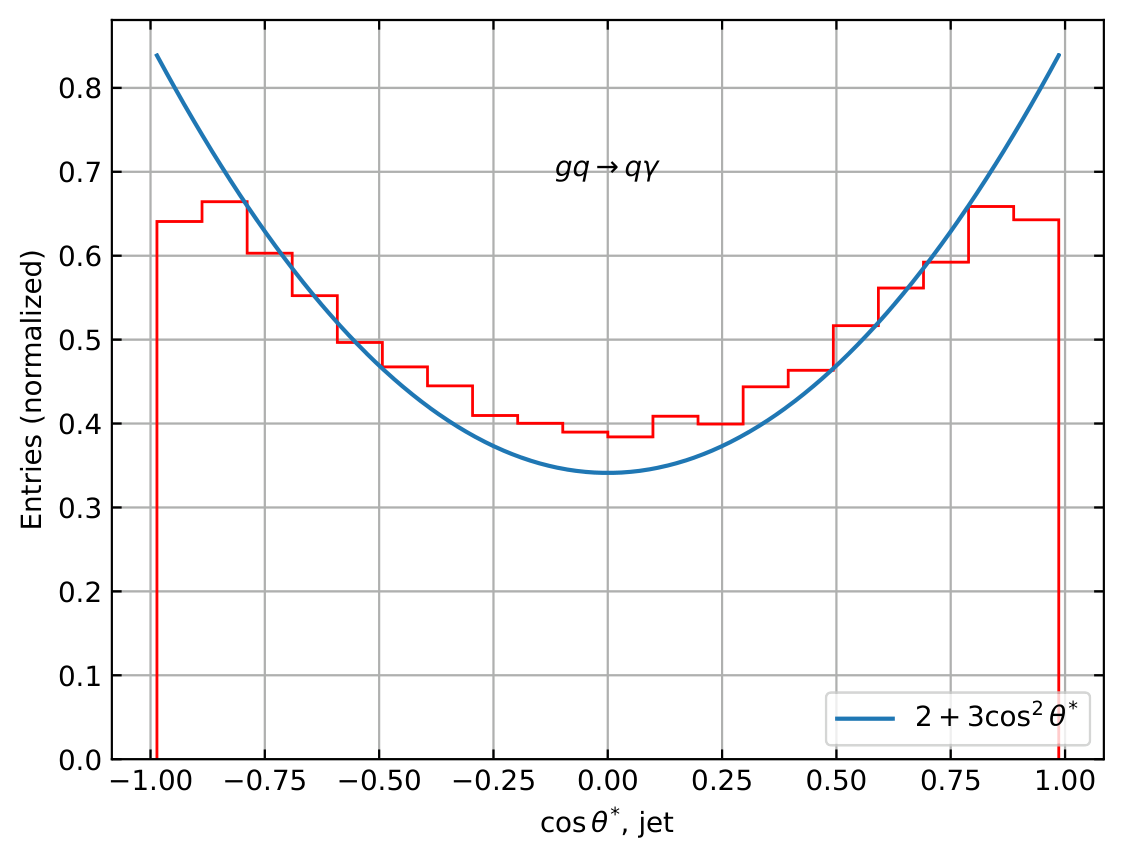}
  \includegraphics[width=0.49\textwidth]{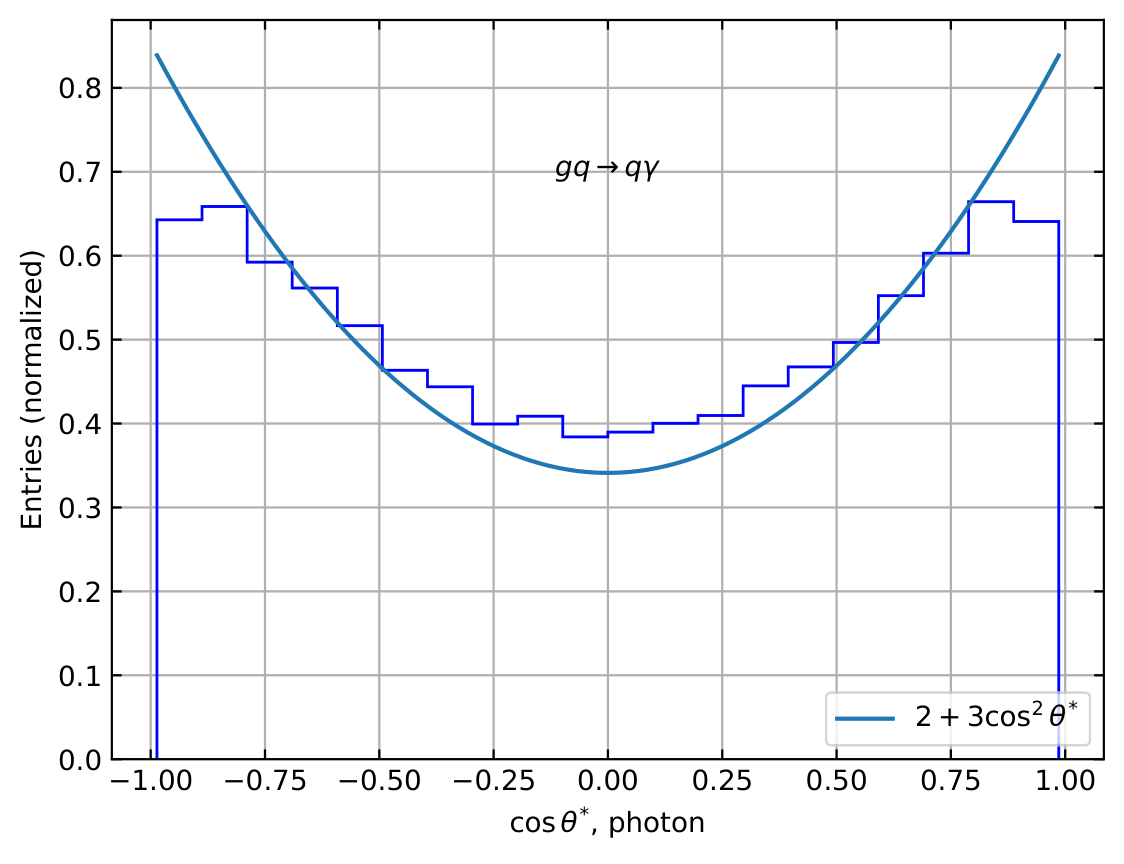}
  \includegraphics[width=0.49\textwidth]{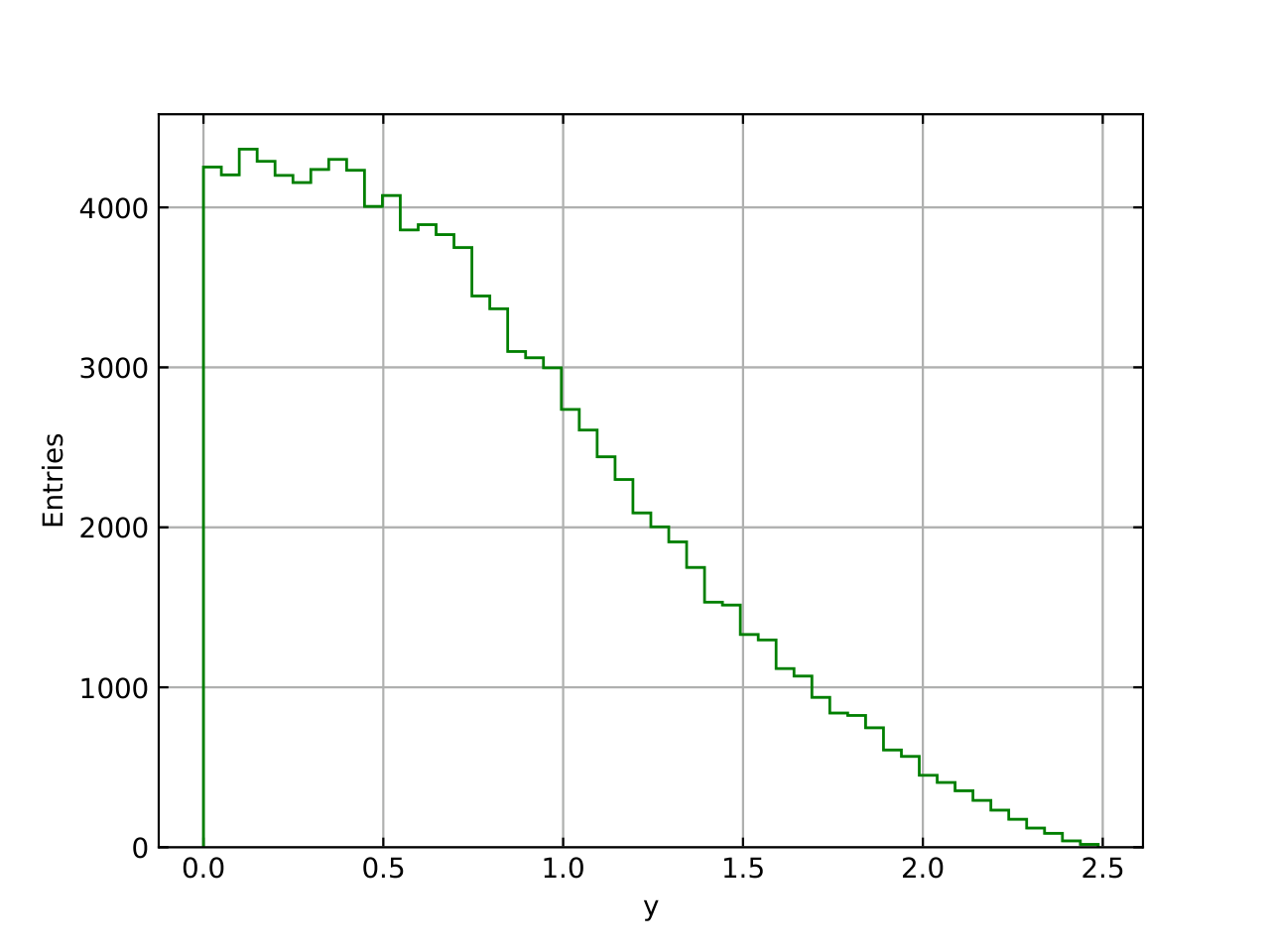}
  \includegraphics[width=0.49\textwidth]{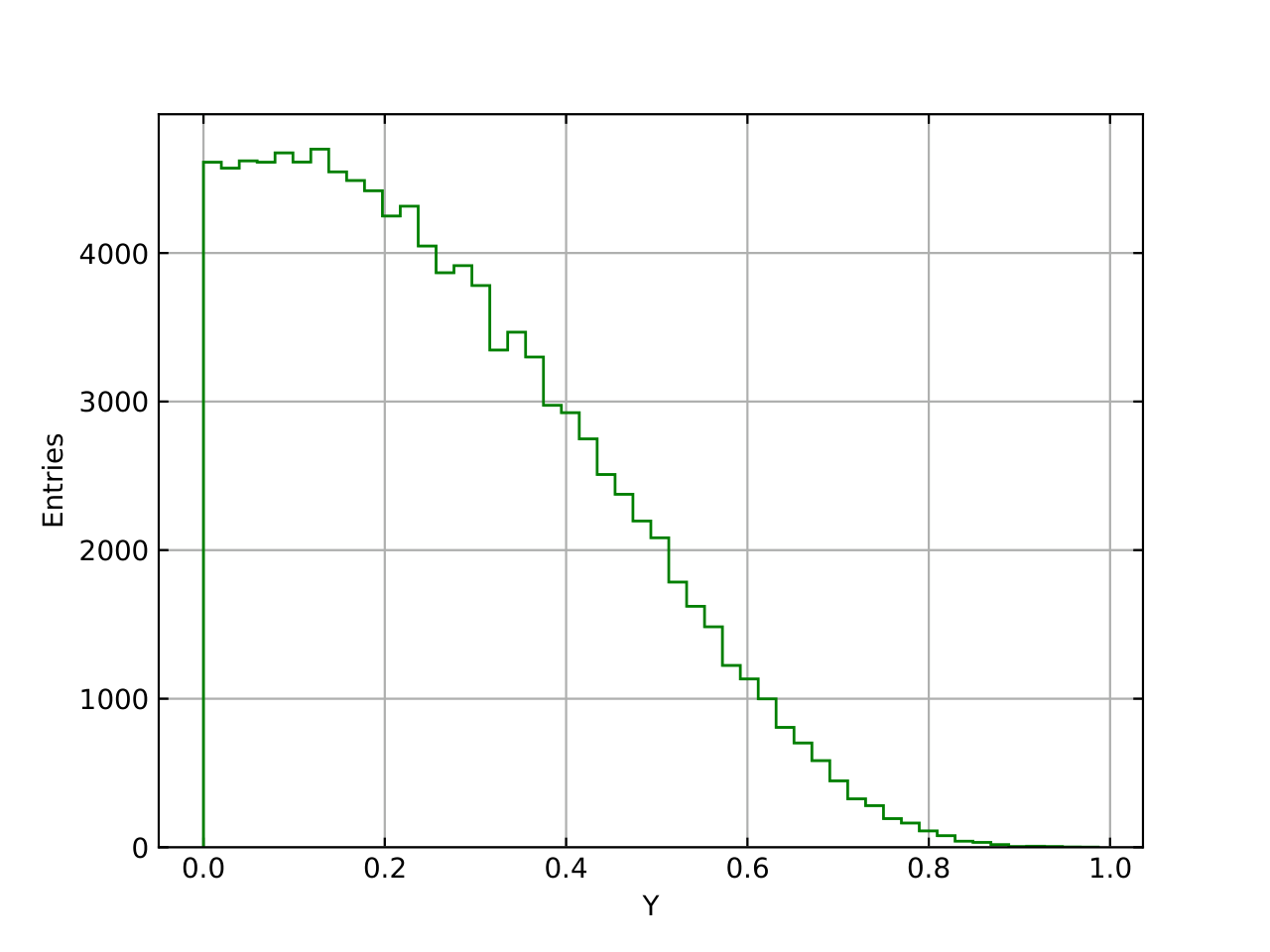}
  \includegraphics[width=0.49\textwidth]{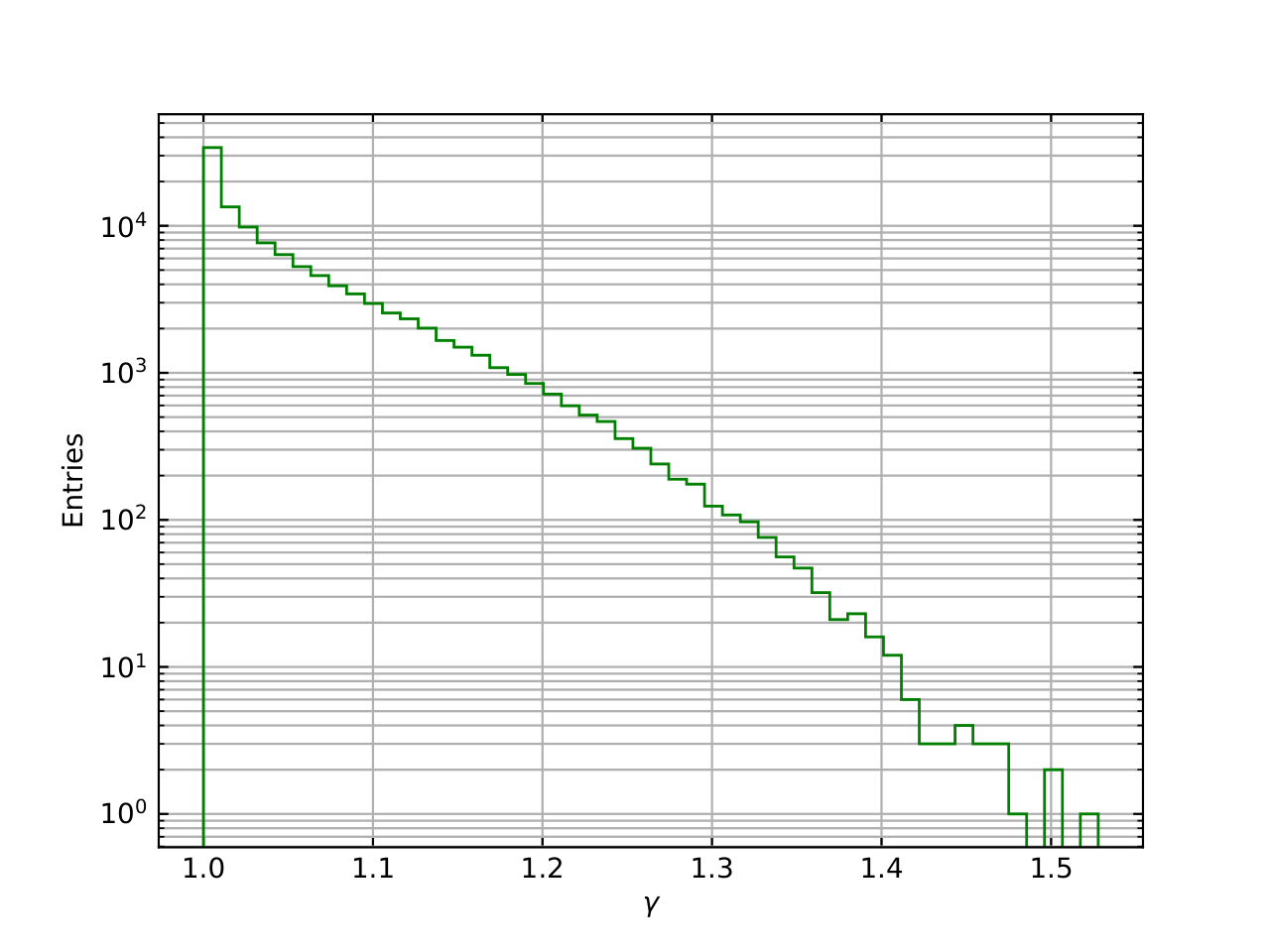}
  \includegraphics[width=0.49\textwidth]{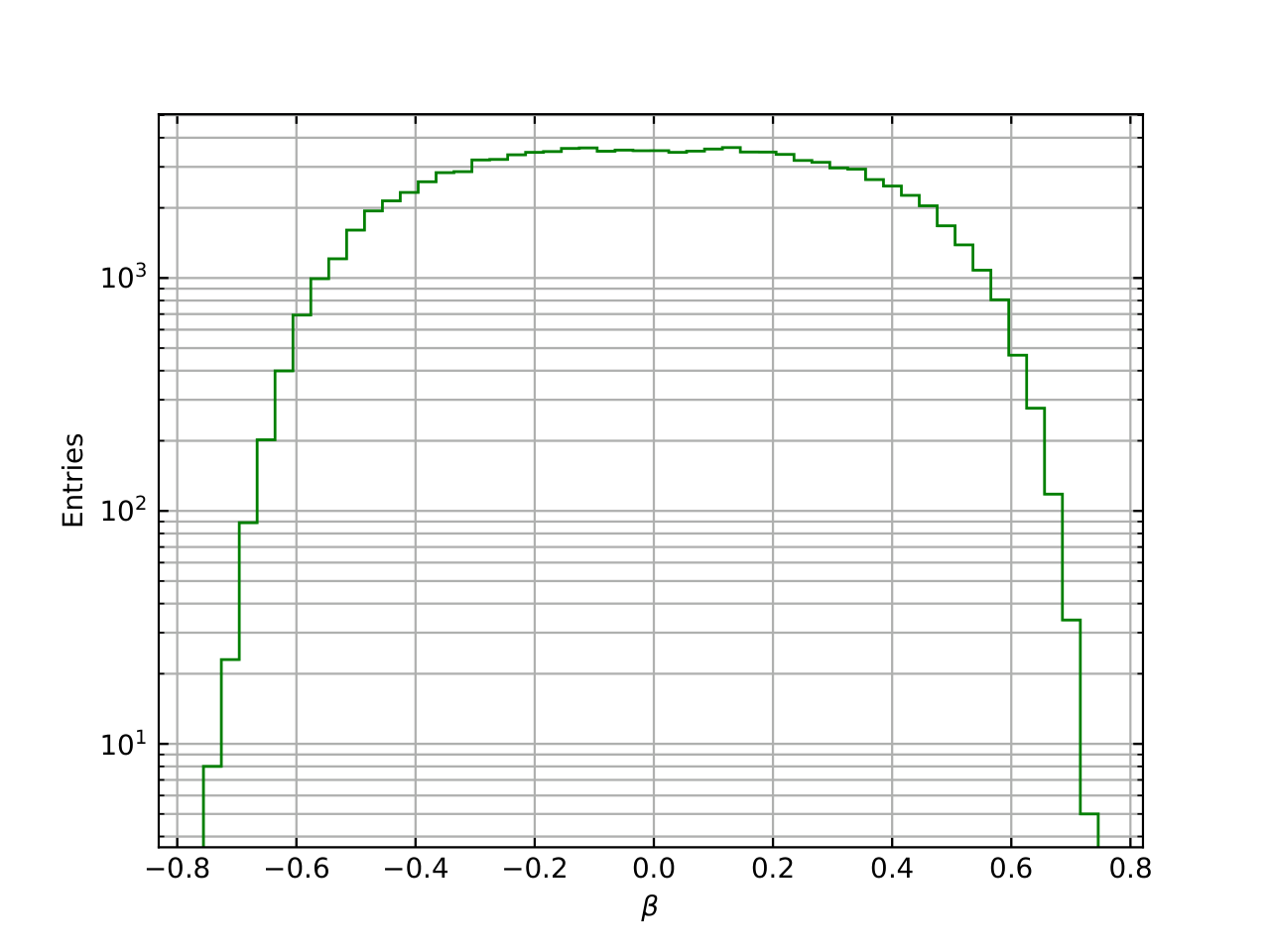}
    \caption{Kinematic data for $M_s = 5$ TeV and $\sqrt{s} = 13$ TeV from STRINGS.}
  \label{fig:23}
\end{figure}

\begin{figure}[H]
  \centering
  \includegraphics[width=0.49\textwidth]{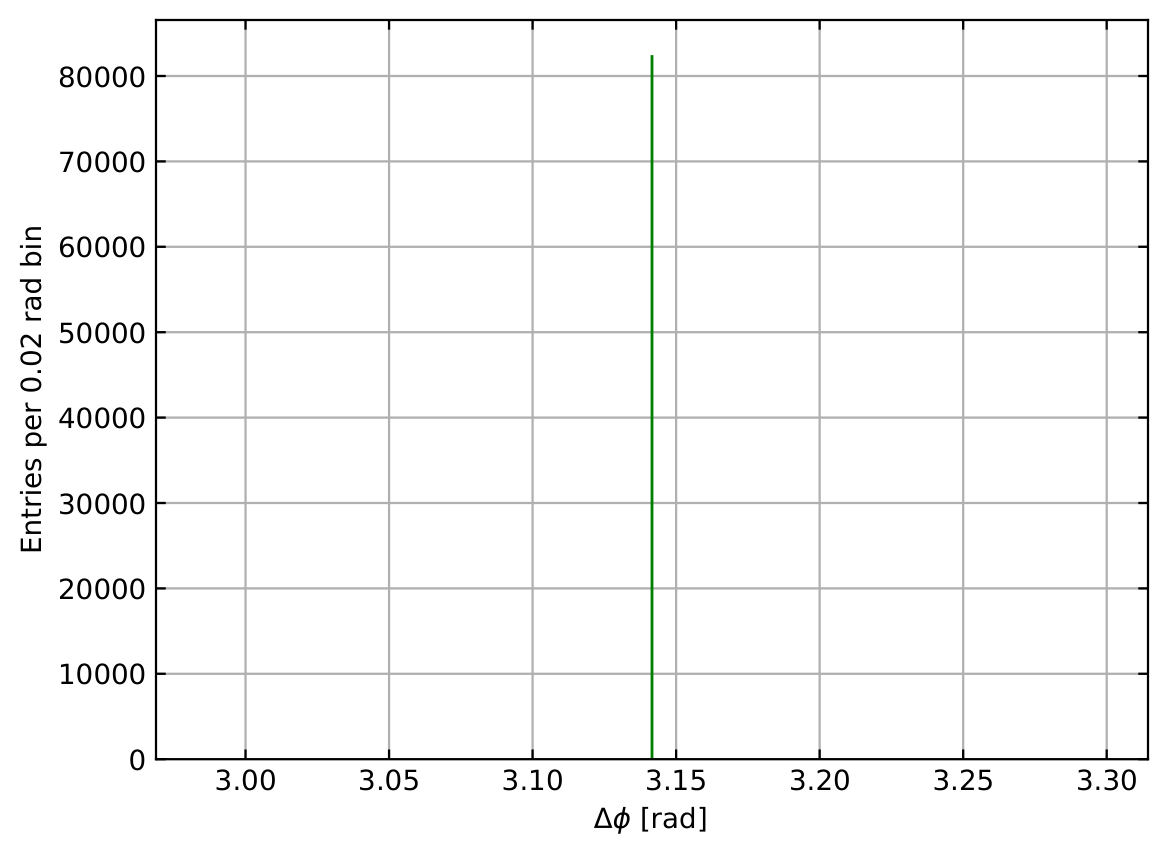}
  \includegraphics[width=0.49\textwidth]{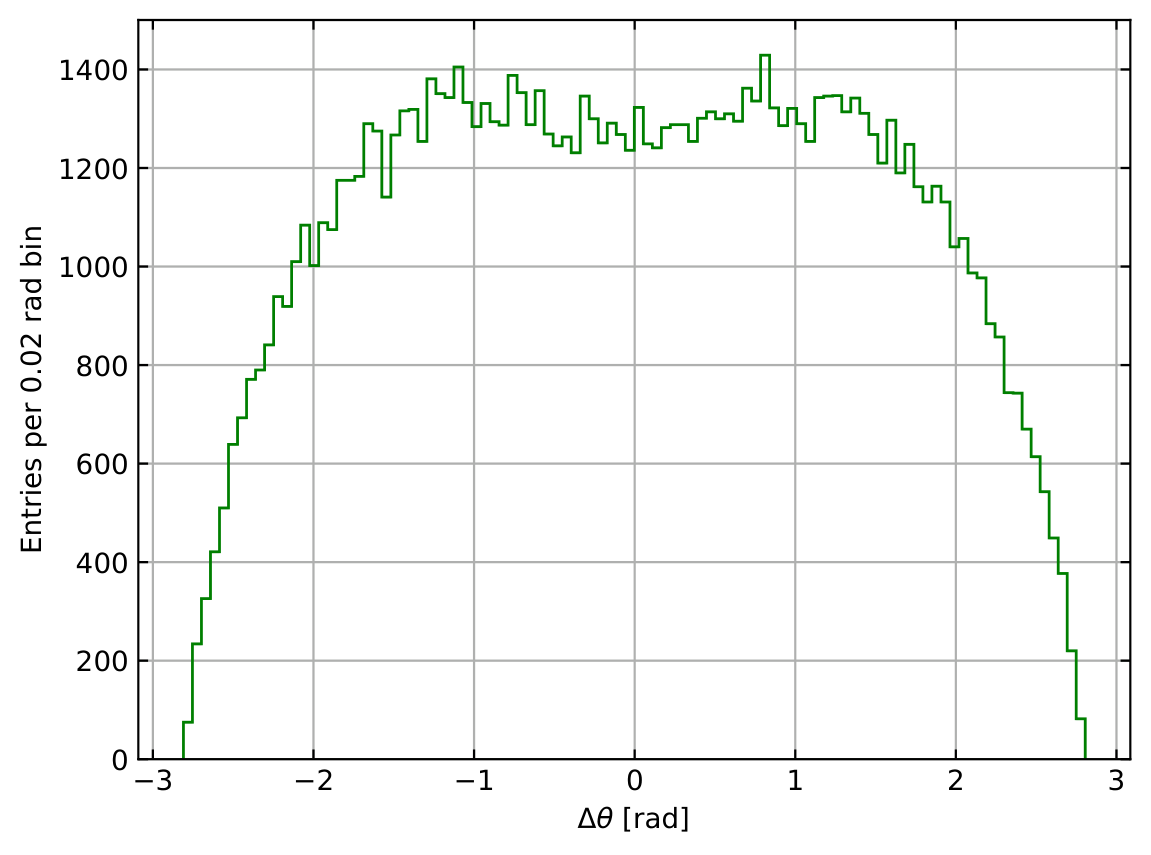}
  \includegraphics[width=0.49\textwidth]{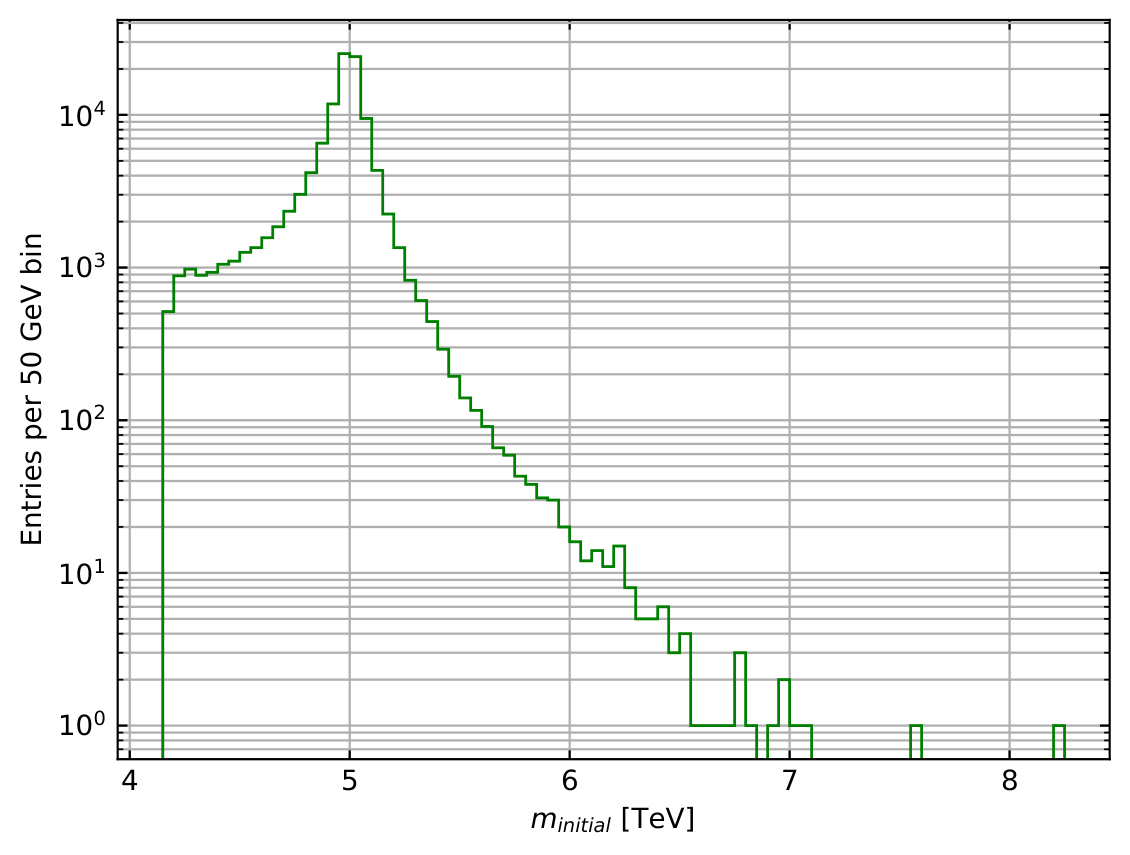}
  \includegraphics[width=0.49\textwidth]{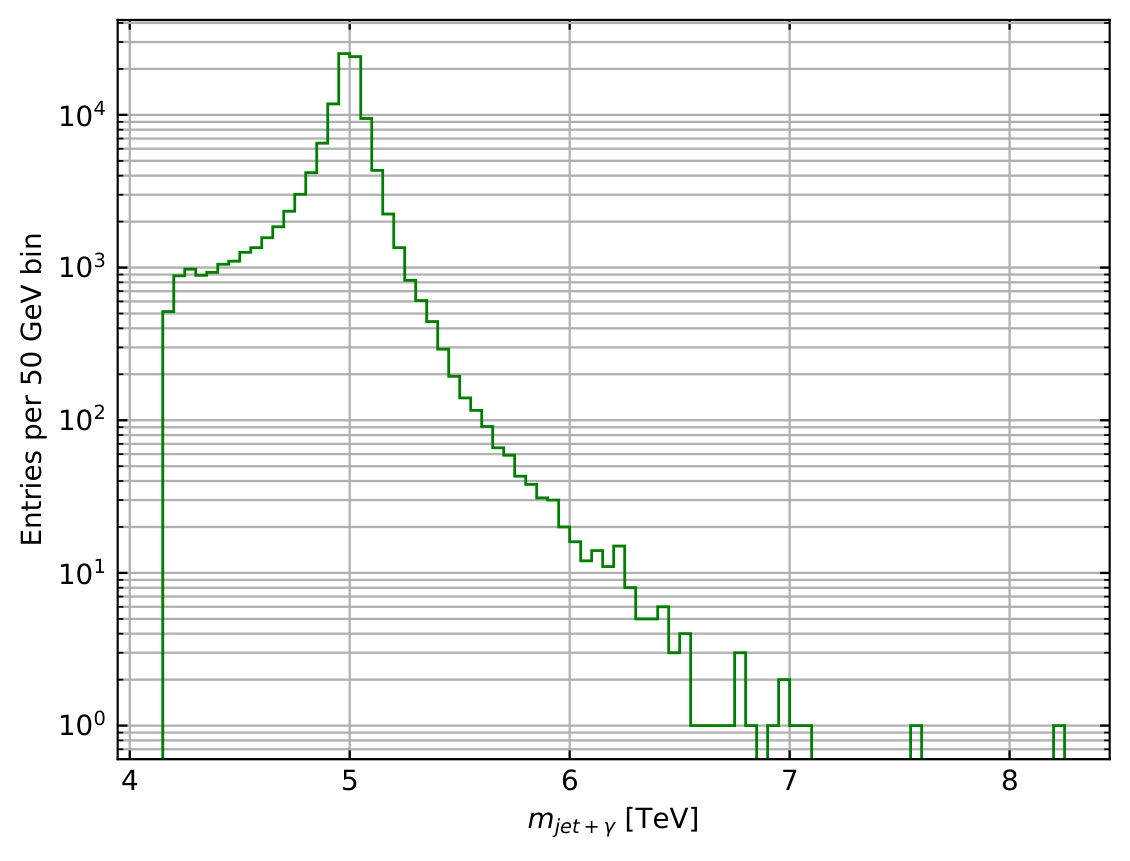}
  \includegraphics[width=0.49\textwidth]{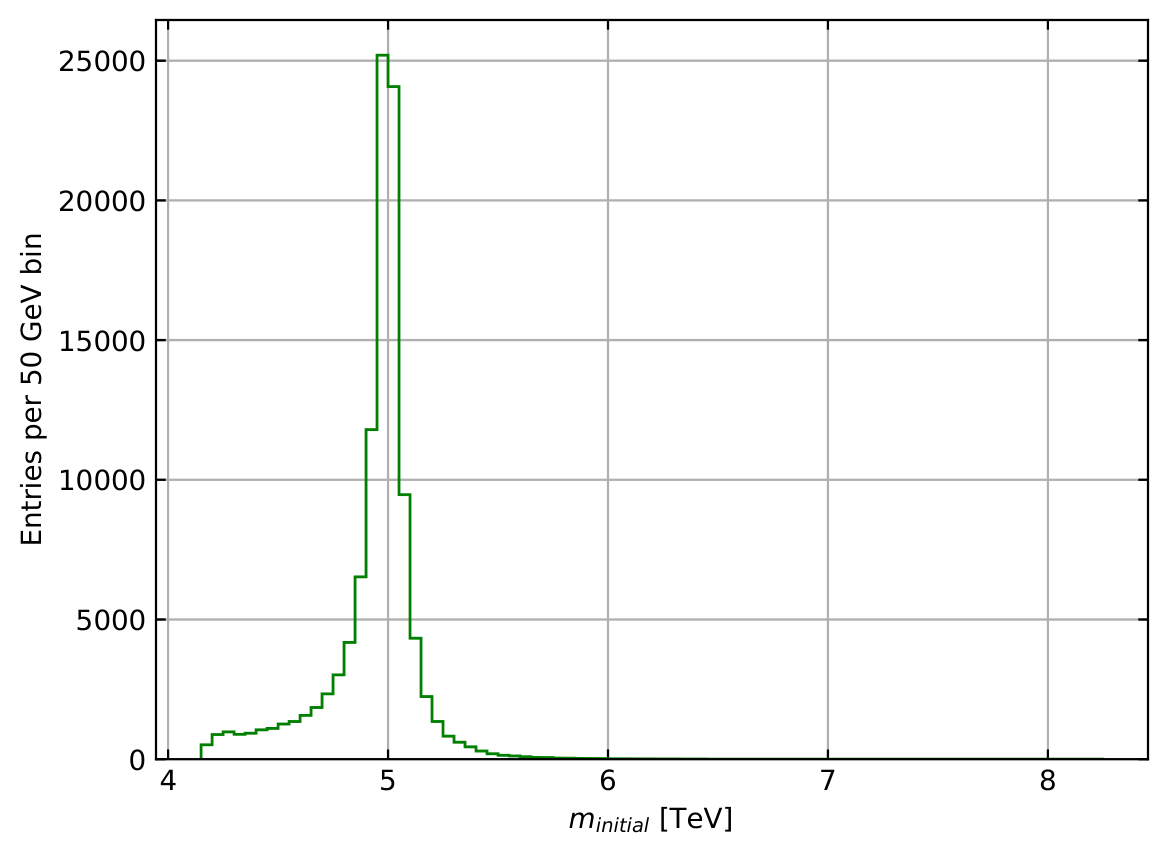}
  \includegraphics[width=0.49\textwidth]{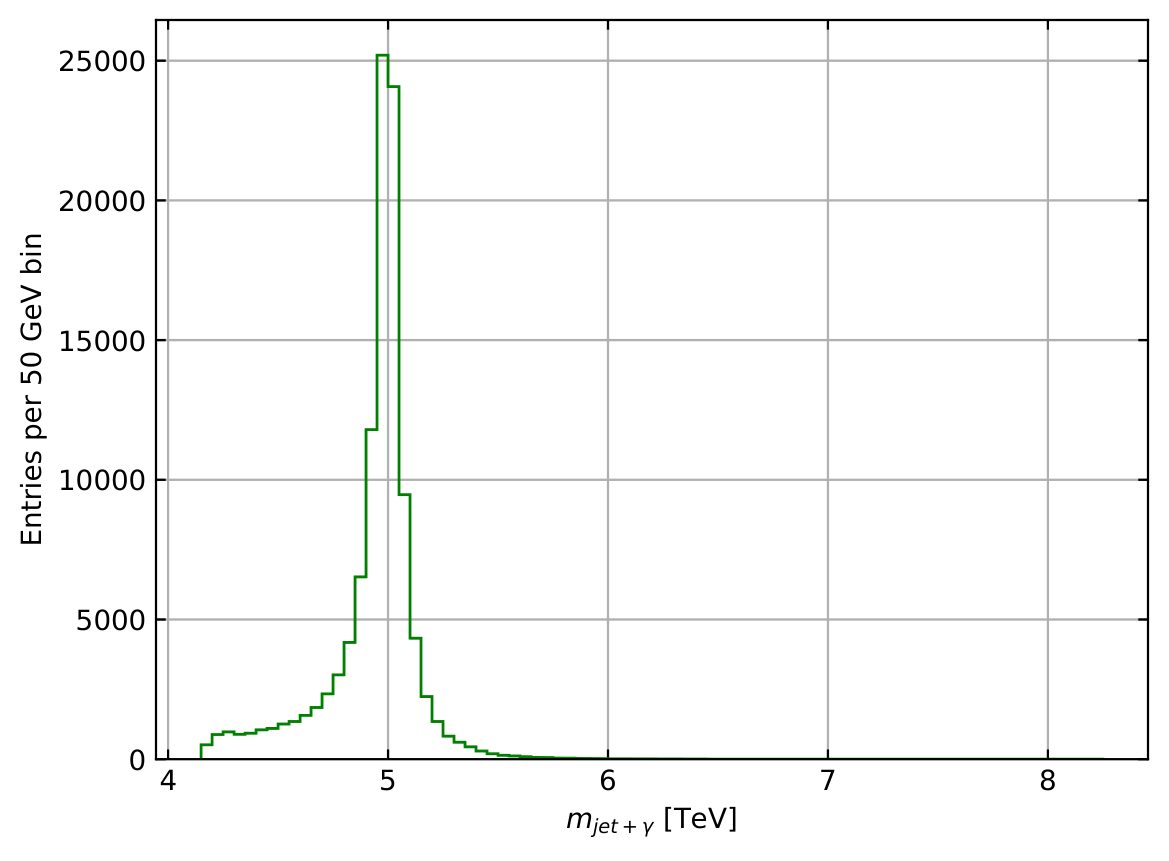}
    \caption{Kinematic data for $M_s = 5$ TeV and $\sqrt{s} = 13$ TeV from STRINGS.}
  \label{fig:24}
\end{figure}

\section{Pythia Kinematic Data}

\begin{figure}[H]
   \centering
   \includegraphics[width=0.49\textwidth]{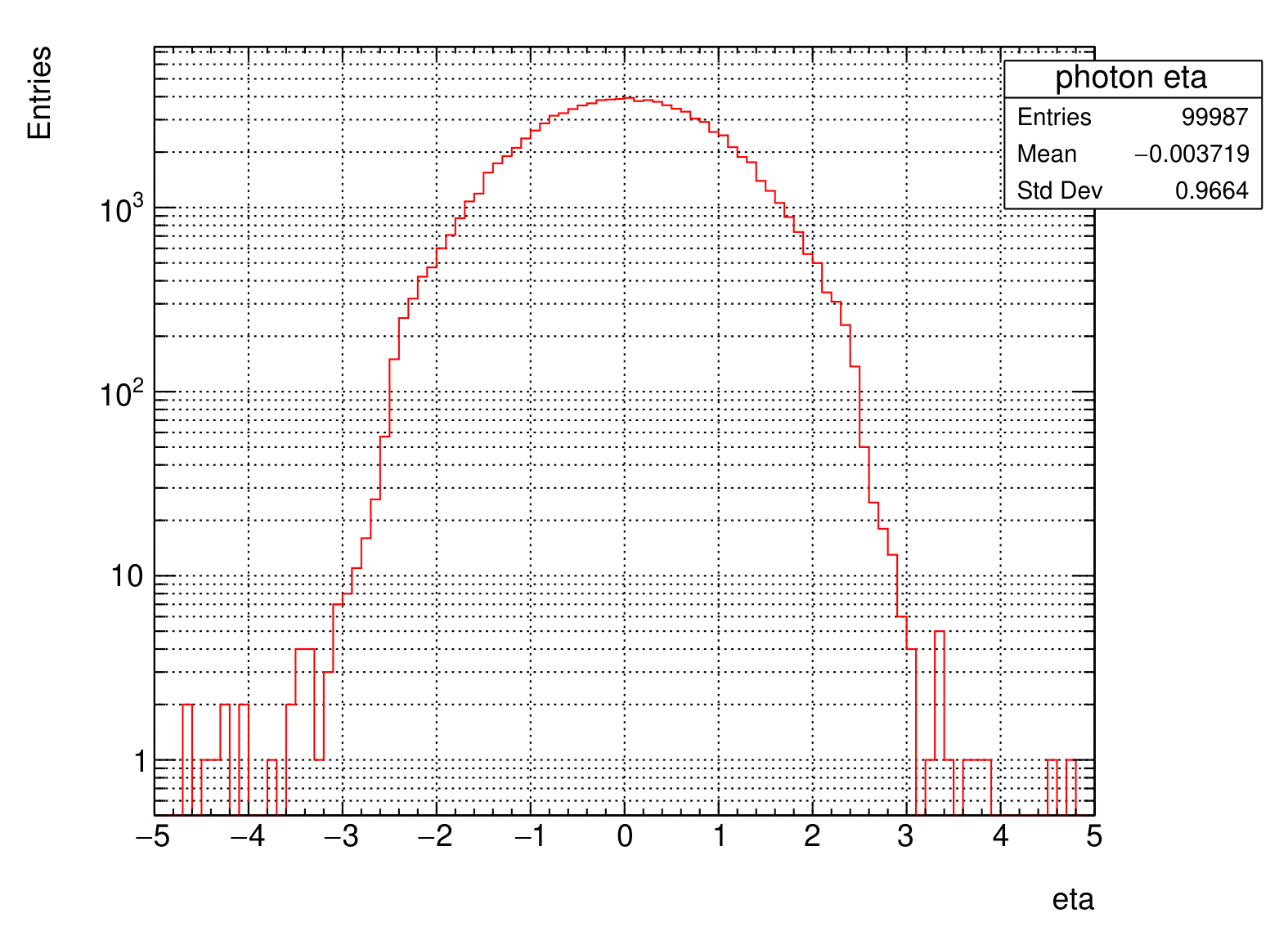}
   \includegraphics[width=0.49\textwidth]{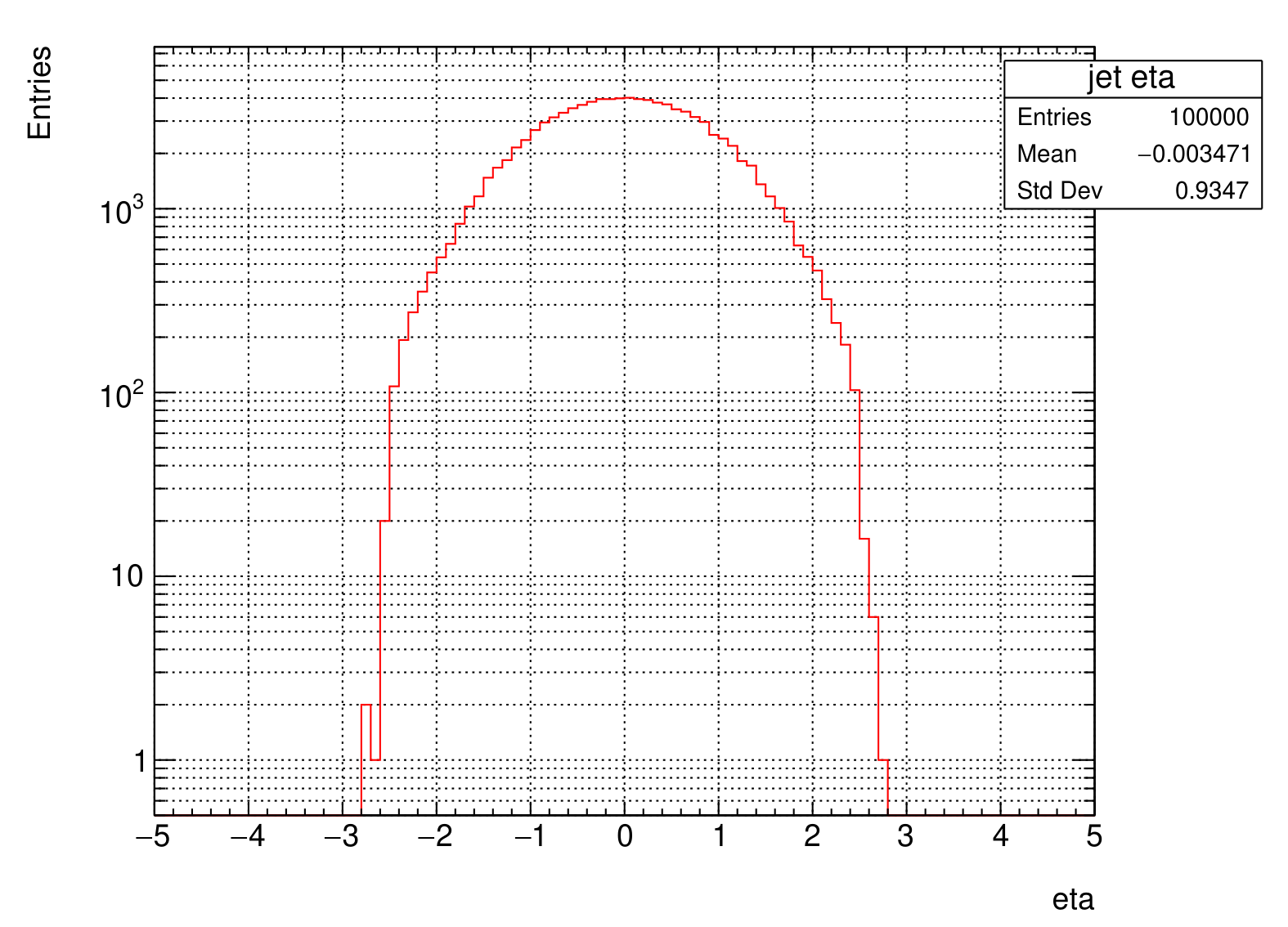}
   \includegraphics[width=0.49\textwidth]{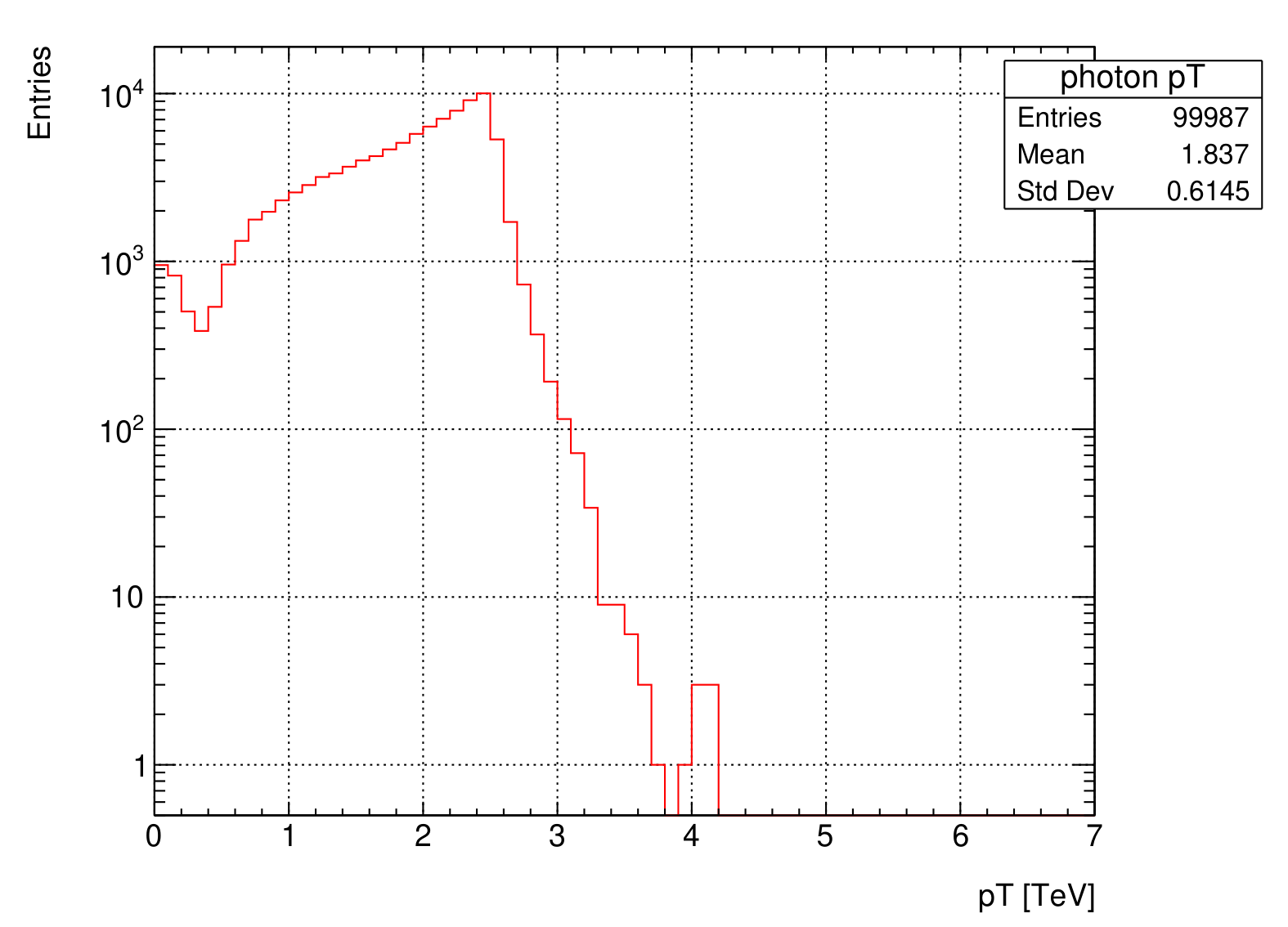}
   \includegraphics[width=0.49\textwidth]{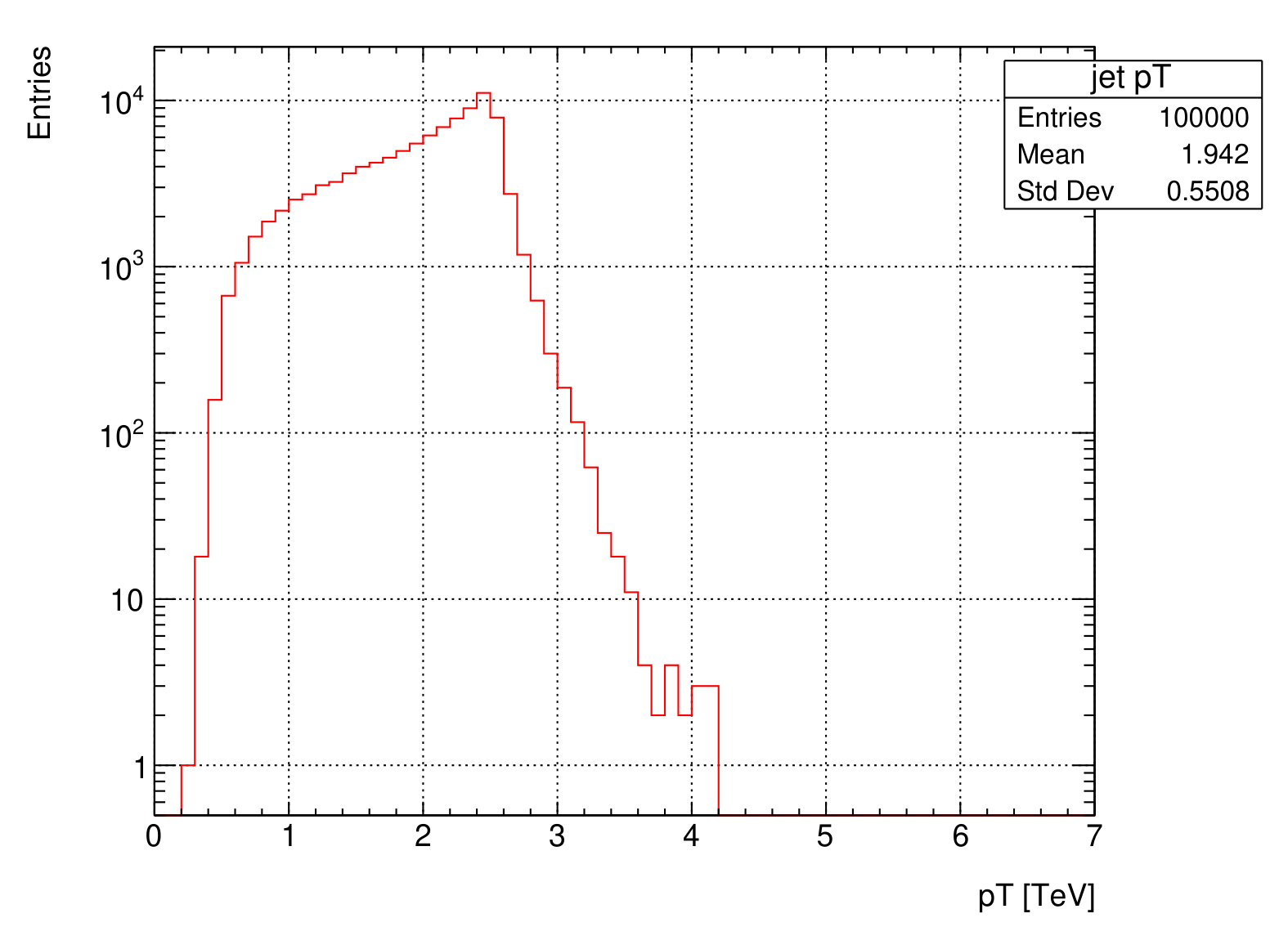}
   \includegraphics[width=0.49\textwidth]{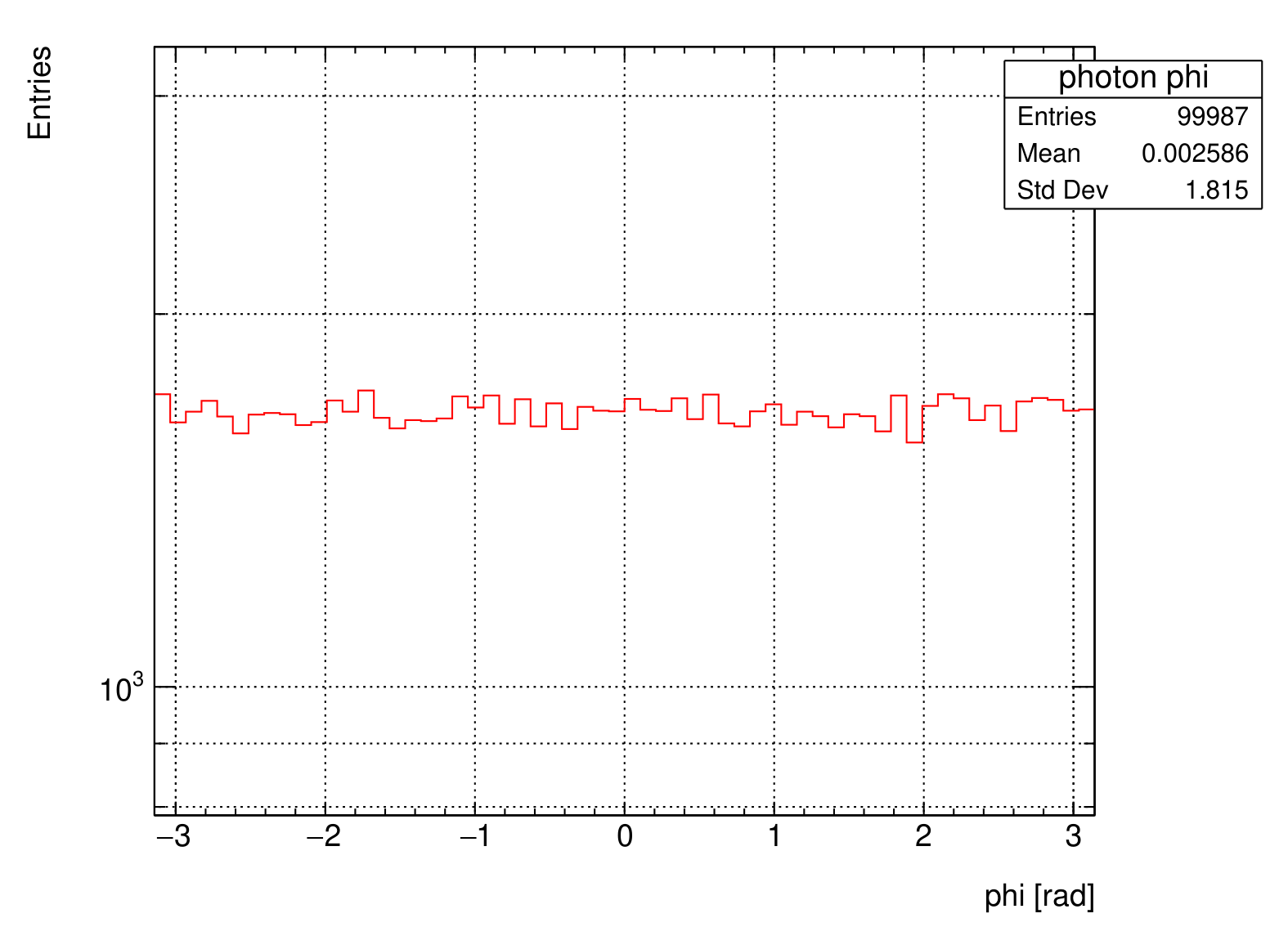}
   \includegraphics[width=0.49\textwidth]{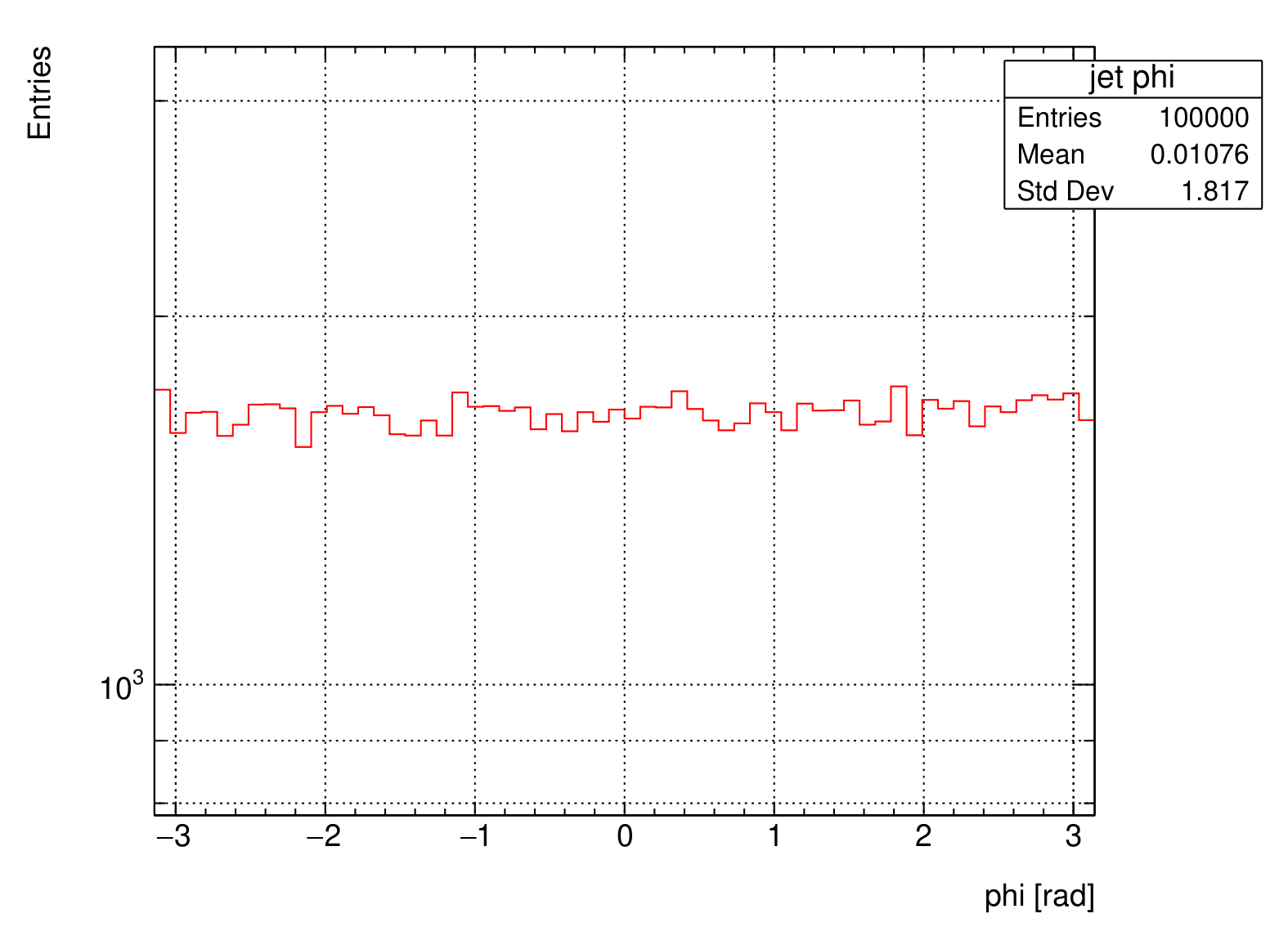}
   \includegraphics[width=0.49\textwidth]{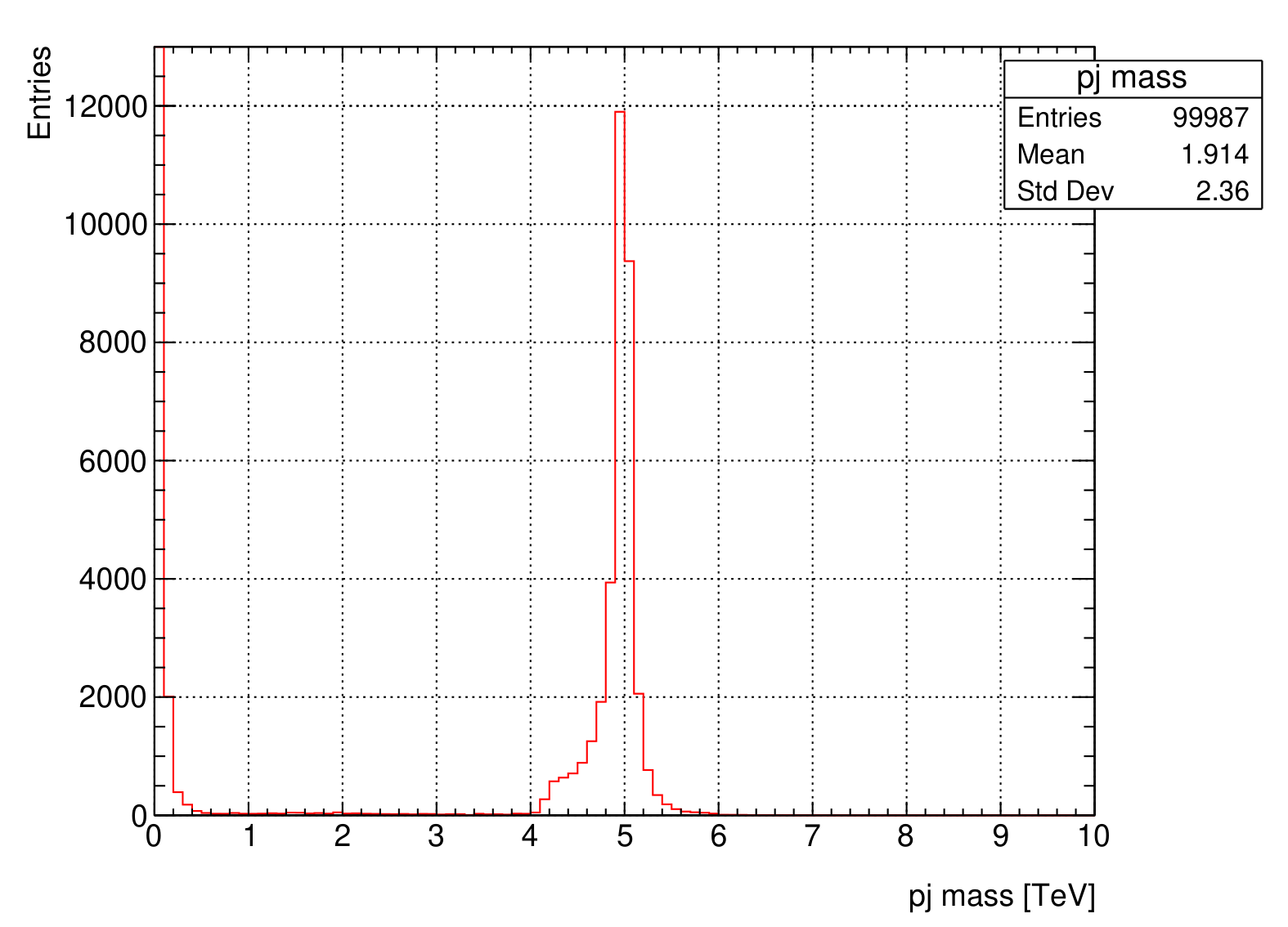}
   \includegraphics[width=0.49\textwidth]{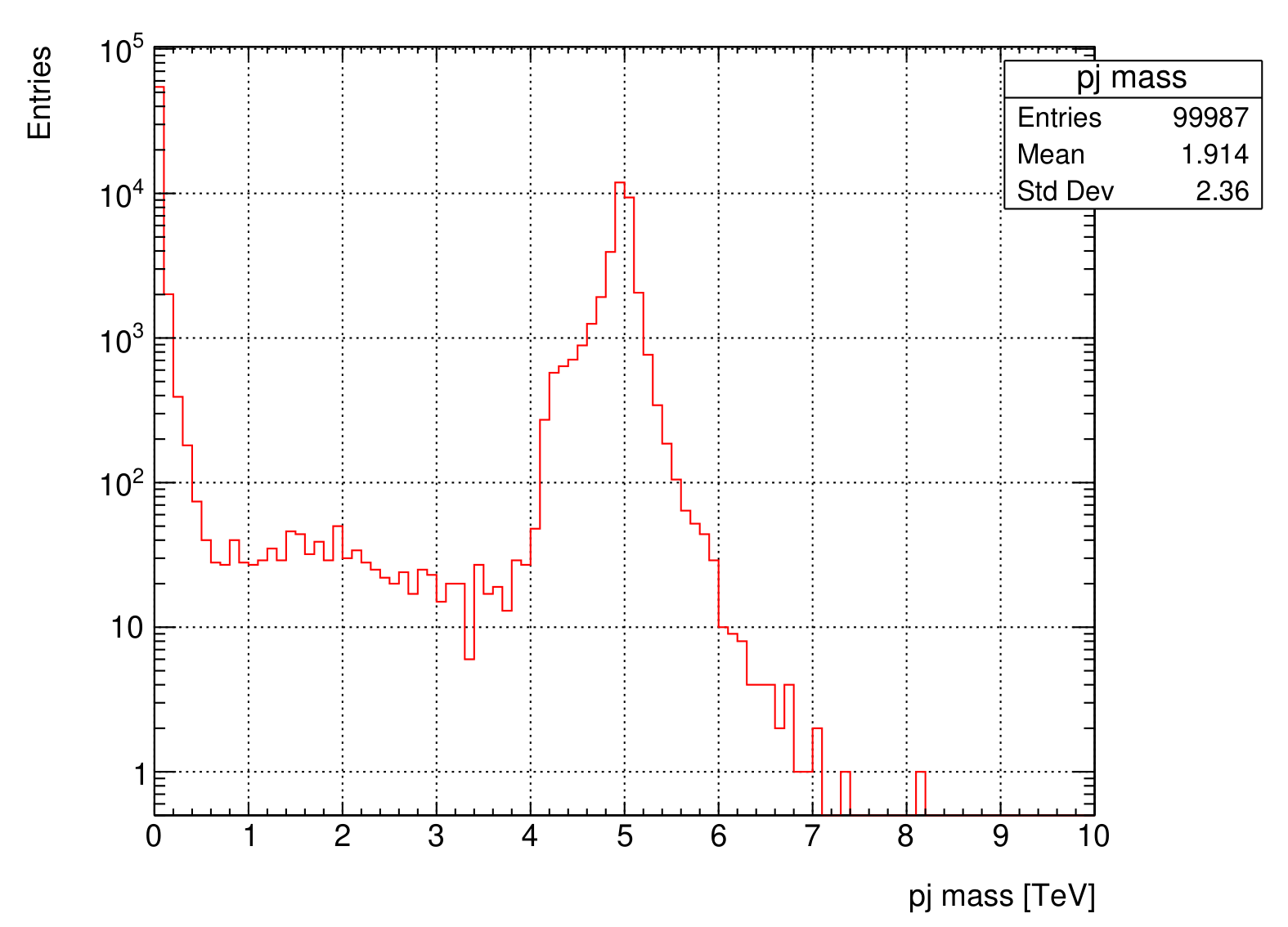}
     \caption{Kinematic data for $M_s = 5$ TeV and $\sqrt{s} = 13$ TeV from Pythia.}
   \label{fig:70}
\end{figure}

\end{document}